\documentclass[useAMS,usegraphicx,usenatbib,fleqn]{mn2e}

\usepackage{times}
\usepackage{graphicx}
\usepackage{grffile}  

\usepackage{xcolor}  
\usepackage[]{amsmath}
\usepackage[]{amssymb}

\newcommand{\simgt}{\,\rlap{\lower 3.5 pt \hbox{$\mathchar \sim$}} \raise
1pt \hbox {$>$}\,}
\newcommand{\simlt}{\,\rlap{\lower 3.5 pt \hbox{$\mathchar \sim$}} \raise
1pt \hbox {$<$}\,}

\usepackage{color}			      
\definecolor{midgray}{gray}{0.4}		
\definecolor{orange}{rgb}{1,0.5,0}    

\usepackage[pdftex,colorlinks=true,citecolor=midgray,linkcolor=midgray]{hyperref}        

\newcommand{\BE}{\begin{equation}}
\newcommand{\EE}{\end{equation}}
\newcommand{\BEA}{\begin{eqnarray}}
\newcommand{\EEA}{\end{eqnarray}}

\title[Spatial Extent of SF in $z\sim1.5$ Mergers]{The Spatial Extent and Distribution of Star Formation in 3D-HST Mergers at $z\sim1.5$}
\author[Schmidt et al. (2013)]
{\parbox{\textwidth}{Kasper B. Schmidt,$^{1,2,}$\thanks{E-mail: kschmidt@physics.ucsb.edu} 
Hans-Walter Rix,$^{1}$ 
Elisabete da Cunha,$^{1}$ 
Gabriel B. Brammer,$^{3}$ 
Thomas J. Cox,$^{4}$
Pieter van Dokkum,$^{5}$
Natascha M. F\"orster Schreiber,$^6$
Marijn Franx,$^7$ 
Mattia Fumagalli,$^7$
Patrik Jonsson,$^{8}$
Britt Lundgren,$^5$ 
Michael V. Maseda,$^1$ 
Ivelina Momcheva,$^{5}$
Erica J. Nelson,$^{5}$
Rosalind E. Skelton,$^{5}$
Arjen van der Wel,$^1$ and
Katherine E. Whitaker$^9$ }
\vspace{0.6cm}\\
\parbox{\textwidth}{$^{1}$Max Planck Institut f\"ur Astronomie, K\"onigstuhl 17, D-69117 Heidelberg, Germany \\
$^{2}$Department of Physics, University of California, Santa Barbara, CA 93106, USA \\
$^{3}$European Southern Observatory, Alonso de C\'ordova 3107, Casilla 19001, Vitacura, Santiago, Chile \\
$^{4}$Carnegie Observatories, 813 Santa Barbara Street, Pasadena, CA 91101, USA\\
$^{5}$Department of Astronomy, Yale University, New Haven, CT 06520, USA\\
$^{6}$Max Planck Institut f\"ur Extraterrestrische Physik, Giessenbachstrasse, D-85748 Garching, Germany\\
$^{7}$Leiden Observatory, Leiden University, Leiden, The Netherlands\\
$^{8}$Harvard-Smithsonian Center for Astrophysics, 60 Garden Street, Cambridge, MA 02138, USA \\
$^{9}$Astrophysics Science Division, Goddard Space Flight Center, Code 665, Greenbelt, MD 20771, USA}}
\begin{document}

\date{Accepted for publication in MNRAS on March 13th 2013}

\pagerange{\pageref{firstpage}--\pageref{lastpage}} \pubyear{2013}

\maketitle

\begin{abstract}
We present an analysis of the spatial distribution of star formation in a sample of 60 visually identified galaxy merger candidates at $z>1$.
Our sample, drawn from the 3D-HST survey, is flux-limited and was selected to have high star formation rates based on fits of their broad-band, low spatial resolution spectral energy distributions. It includes plausible pre-merger (close pairs) and post-merger (single objects with tidal features) systems,Êwith total stellar masses and star formation rates derived from multi-wavelength photometry. Here we use near-infrared slitless spectra from 3D-HST which produce H$\alpha$ or [O{\scshape iii}] emission line maps as proxies for star-formation maps.
This provides a first comprehensive high-resolution, empirical picture of \emph{where} star formation occurred in galaxy mergers at the epoch of peak cosmic star formation rate.
We find that detectable star formation can occur in one or both galaxy centres, or in tidal tails. 
The most common case (58\%) is that star formation is largely concentrated in a single, compact region, coincident with the centre of (one of) the merger components.
No correlations between star formation morphology and redshift, total stellar mass, or star formation rate are found.
A restricted set of hydrodynamical merger simulations between similarly massive and gas-rich objects implies that star formation should be detectable in both merger components, when the gas fractions of the individual components are the same.
This suggests that $z\sim1.5$ mergers typically occur between galaxies whose gas fractions, masses, and/or star formation rates are distinctly different from one another.
\end{abstract}


\maketitle
\label{firstpage}
\section{Introduction}
\label{sec:introEL}

The spatial extent and distribution of star formation in normal, local galaxies is well established \citep[e.g.,][and references therein]{James:2004p11912,Bigiel:2008p22975,Bigiel:2011p24135,Schruba:2011p23762,Calzetti:2012p23653}.
Both self-regulated star formation and merging must be key ingredients in galaxy-formation and evolution and have been studied observationally in detail in the $z\lesssim0.5$ Universe \citep[e.g.,][]{Lambas:2003p13470,Hammer:2005p13460,Barton:2007p13379,Barton:2000p13383,Jogee:2009p13304,Robaina:2009p13298,Robaina:2010p10557} as well as in theoretical simulations \citep[e.g.,][]{Mihos:1996p13561,Barnes:1996p13546,Springel:2000p18417,Cox:2004p13474,Cox:2006p3233,Cox:2008p3243,Lotz:2008p12474,Lotz:2008p24286,DiMatteo:2007p13565}. 
From these studies, it has become evident that rapid star formation can be triggered by tidal interaction in mergers, but the simulations also suggest that mergers trigger both nuclear starbursts and black hole accretion.
Even though galaxy mergers are observed to enhance star formation in galaxies and trigger some of the most violent starbursts known \citep[e.g.,][]{Rieke:1985p21594,Joseph:1985p21589,Melnick:1990p21749,Klaas:1991p21718}, it appears that the net effect of major mergers in the global star formation history of the galaxy population has been relatively modest since at least $z=1$ \citep[e.g.,][]{Robaina:2009p13298,Jogee:2009p13304}. 

Mergers are predicted to play a crucial role in the build-up and formation of massive galaxies \citep[e.g.,][]{Mihos:1996p13561,Springel:2000p18417,Cox:2008p3243,Hopkins:2010p10954}, and therefore, a crucial step towards fully understanding how galaxies have evolved is to study the star formation properties in merging systems at high redshifts. 
Some of the questions that need to be addressed observationally to get a more detailed picture of the star formation history of present-day galaxies are:
Where did stars form in merging galaxies at higher redshift? and Which phase(s) of the merging process seems to trigger most star formation? 

At higher redshift, the interplay between merging and star formation has been investigated in much less detail than in the local Universe. This is mostly due to observational challenges, as resolved observations at $<1"$ resolution are needed. Further, tracing star formation, e.g., through H$\alpha$ at $z\sim1.5$ requires observations in the near-infrared (NIR), and these have been far less feasible than observations in optical wavebands. Nevertheless, the $0.7<z<2$ epoch is immensely important for understanding galaxy-formation, as this is the cosmic time when the majority of the stars we see in galaxies today were formed \citep[e.g.,][]{Hopkins:2006p11948,Karim:2011p12513}. 
Several studies have addressed the impact of (major) galaxy mergers as well as the general galaxy morphology on the amount of star formation at $z\sim1.5$ \citep[e.g.,][]{Swinbank:2004p21851,Law:2007p10887,ForsterSchreiber:2009p8449,Wright:2009p21864,Conselice:2011p26214,Bluck:2012p26210,Bell:2012p26144}. 
However, only a few of these studies investigate three-dimensional (3D) spectroscopy, where both spatial and spectral information is available, which is crucial for investigating the \emph{spatial} extent of star formation.

Until recently, large samples of galaxies, and in particular galaxy
mergers, with rest-frame optical 3D spectroscopic information at high
redshift did not exist. ÊThe largest samples of galaxies with such data
at $1.5<z<2.5$ are from the SINS survey \citep{ForsterSchreiber:2009p8449},
recently expanded with the zC-SINF sample \citep{Mancini:2011p23802},
totalling 110 star-forming galaxies at $1.5 \la z \la 2.5$ observed with SINFONI, and from the MASSIV survey \citep{Contini:2012p23866} which contains 84 star-forming galaxies at  $0.9 < z < 1.8$ also mapped with SINFONI.
Using H$\alpha$ as kinematic and star formation tracer enabled analysis of the spatially-resolved ionised gas kinematics, its distribution,
and the physical properties of these systems \citep{Genzel:2006p23877,Genzel:2008p21878,Genzel:2011p23883,
Shapiro:2008p23480,Cresci:2009p23902,Epinat:2009p23854,Epinat:2012p24404,Queyrel:2009p24405,Queyrel:2012p24406,Newman:2012p23909}. 
High-resolution NIR imaging with \emph{The Hubble Space Telescope's} (HST) Near Infrared Camera and Multi-Object Spectrometer 2 (NICMOS2) for a small subset of the SINS objects provided
additional rest-frame optical morphologies, in agreement with the disc or
merger nature from the H$\alpha$ kinematics \citep{ForsterSchreiber:2011p23940,ForsterSchreiber:2011p23942}.
The selection of these samples was primarily based on integrated photometry or spectroscopic properties, not morphologies, and only a modest fraction of
objects ($\sim$1/3) were kinematically inferred to be (major) mergers.
Roughly comparable fractions were found in other sizeable NIR IFU
samples at $z \sim 1 - 3$, including, e.g., those by \cite{Law:2007p10887,Law:2009p23849,Gnerucci:2011p23872}, and \cite{Wisnioski:2011p9934}. Ê

With the recent 3D-HST slitless grism survey \citep[see][and Section~\ref{sec:3dhstEL}]{Brammer:2012p12977}, much larger samples of objects with NIR 3D emission line spectroscopy have become available, making it possible to address the spatial extent of star formation for extensive samples of galaxy mergers at the peak of cosmic star formation rate density. 
The IFU samples mentioned above have significantly higher spectral resolution, enabling detailed kinematic studies, but AO-assisted IFU observations, which provide angular resolution comparable to HST in the NIR, remain observationallyÊ expensive and suffer from complications due to strong night sky lines. HST grism observations, as the ones taken in the 3D-HST survey, provide more limited kinematic information but allow for unbiased target selection and are much more efficient at detecting and mapping the continuum and line-emission at high angular resolution for all targets within the field-of-view.
3D-HST provides resolved line-emission, enabling studies of the spatial extent of star formation for large samples of galaxies at $z > 1$ in five well-studied cosmological fields. The initial papers use approximately half of the full dataset, as described in \cite{vanDokkum:2011p10254} and \cite{Brammer:2012p12977}.

Using the same 3D-HST data, we explore the spatial extent and distribution of star formation in a 'population snapshot' of presumably \emph{merging} systems at $z\sim1.5$. 
For this sample we make $\sim$0.2$''$ resolution maps of emission lines (H$\alpha$ and [O{\scshape iii}]), which trace the spatial extent of the (unobscured) star formation in these mergers and allow us to study their star formation properties in a statistical and unbiased way.
This is done under the assumption that the H$\alpha$ (for $z\sim0.7-1.5$) and [O{\scshape iii}] (for $z\sim1.2-2.3$) emission of the systems trace the star formation. This has been shown to be a fair assumption for both H$\alpha$ \citep{Kennicutt:1983p9397,Gallagher:1984p17584,Kennicutt:1994p11885,Kennicutt:1998p9250,Kennicutt:1998p17963} and [O{\scshape iii}] \citep{Kennicutt:1992p18181,Teplitz:2000p18120,Hippelein:2003p18273}, even though using [O{\scshape iii}] as a quantitative indicator of star formation rate (which is not what we aim to do here) includes several complicating factors \citep{Teplitz:2000p18120}.

To help interpret our observations in a theoretical context, we create a sample of pseudo-observations from state-of-the-art smoothed-particle hydrodynamic (SPH) simulations of individual mergers, which we compare to the 3D-HST data. 
The simulations predict a merger sequence and star formation picture with centrally-concentrated triggered starbursts at final coalescence, enhanced star formation in tidal features, and black hole growth and accretion.
The goal is to understand the observational results from 3D-HST by making a direct comparison with the predictions from the simulations.
These comparisons will help explore which parameters, e.g., viewing angle, merger phase, gas fraction, mass ratio, etc. play a crucial role in determining, for example, star formation rates from observations. 

In Section~\ref{sec:3dhstEL} we describe the 3D-HST survey from which the merger sample was selected. 
We then describe the selection of our sample in Section~\ref{sec:sampleEL} and the procedure used to map the spatial extent of the star formation in Section~\ref{sec:ELmapEL}. 
In Section~\ref{sec:resultsEL} we split the sample into four different morphological types and find that most mergers exhibit star formation in only one component.
In Section~\ref{sec:SimspecEL} we compare numerical merger simulations to the observed 3D-HST spectra and find that in simulations star formation most commonly occurs in both components, before we summarise and conclude in Section \ref{sec:concEL}.

\section{The 3D-HST Survey Data}\label{sec:3dhstEL}

To construct the $0.2''$ resolution emission line maps, the proxies for star formation maps, we take advantage of the NIR 3D spectroscopy survey possibilities that the \emph{Wide Field Camera 3} (WFC3) on HST brings. 
The 3D-HST survey is a 248 orbit NIR spectroscopic Hubble treasury program (Cycles 18 and 19, PI~van~Dokkum).
It provides NIR imaging with the F140W filter and grism spectroscopy with the G141 grism over well-studied extragalactic survey fields (AEGIS, COSMOS, GOODS-S, GOODS-N, and UKIDSS/UDS).
The grism spectroscopy is slitless so both spatial and spectroscopic information is available for every single object in the survey fields. The 3D-HST survey provides rest-frame optical spectra for a sample of $\sim$7000 galaxies at $1<z<3.5$ \citep{vanDokkum:2011p10254,Brammer:2012p12977}. 

As of August 2011 the survey had observed 68 pointings over the GOODS-S, GOODS-N\footnote{The GOODS-N data were taken as part of the HST program GO-11600 (PI B. Weiner).}, COSMOS and AEGIS fields \citep{vanDokkum:2011p10254}. The present work is based on 30 of these 68 pointings, where the extensive ancillary data available enables robust spectral energy distribution (SED) modelling necessary for our sample selection as described in Section~\ref{sec:sampleEL}.

\subsection{The 3D-HST Grism Spectroscopy}\label{sec:grism}

The WFC3 G141 grism used in 3D-HST disperses the light over the wavelength range from 1.05~$\mu$m to 1.7~$\mu$m with a low spectral resolution of $R\sim130$. 

Since the grism spectroscopy is slitless, the WFC3 G141 grism basically produces an emission line image that is superimposed onto a sequence of dispersed monochromatic continuum images, and some of the key features of slitless spectroscopy therefore need to be taken into account. 
First of all, the width of emission/absorption lines in the dispersion direction in slitless spectroscopy is not only caused by velocity broadening (which is negligible for the low resolution 3D-HST spectra) and the intrinsic broadening of the wavelength dispersion: 
as slitless spectroscopy produces shifted monochromatic images, the spatial extent of the dispersed emission line image reflects the spatial distribution of the line emission both along and perpendicular to the dispersion direction.
We will take advantage of this 'morphology-broadening' (which can be seen in the third panel of Figure~\ref{fig:ELmapill}), to map the spatial extent of star formation as described in Section~\ref{sec:ELmapEL}. Figure~\ref{fig:ELmapill} will be explained in more detail here.

As with multi-slit spectroscopy, the differing wavelength coverage of the spectra is an issue. Since the detector onto which the field-of-view is dispersed has a finite size, approximately 10\% of the spectra are cut off on the edge of the detector. 

Lastly, `contamination' is an important property of slitless spectroscopy. Since the focal plane is not blocked out with a slit or a mask as is usually done in standard spectroscopy, all the light from a given object, and all other objects in the observed field, are dispersed onto the detector. Hence, spectra will often overlap and therefore `contaminate' each other as explained in \cite{Brammer:2012p12977}.  

For more information on the data reduction methods, the data products of the 3D-HST survey, and the survey itself, we refer to \cite{Brammer:2012p12977}.

\section{Selecting Merger Candidates}\label{sec:sampleEL}

We select our sample of merger candidates from the first 68 pointings obtained as part of the 3D-HST survey based on three different inputs:
(i) The 3D-HST survey catalogue, (ii) SED modelling, and most importantly (iii) visual inspection of NIR (F140W) morphologies. The first two selection steps are performed to define an initial sample of systems with sufficient spectral coverage and to minimise the number of objects to visually inspect. We will describe each of these three steps below.

To obtain robust star formation rate estimates via the SED fitting described in Section~\ref{sec:SEDfit}, we require extensive ancillary photometric catalogues. We therefore focus on the 30 pointings of data available in GOODS-S (6) and COSMOS (24), where the photometric data in the FIREWORKS \citep{Wuyts:2008p9535} and the NEWFIRM medium band survey \citep[NMBS;][]{Whitaker:2011p9509} catalogues are available, respectively. Hence, this work is performed on approximately 1/5 of the final 3D-HST data product. 

\subsection{Grism Catalogue Cuts}

The first step in defining our merger sample is to select a well-defined sample of objects based on the data products of the 3D-HST survey. We ensure that at least 75\% of each spectrum in the sample falls on the detector. Since we are looking for merging objects we do not put any constraints on the contamination of the individual spectra, as spectra of close pairs will always have a high level of contamination. We rely on the visual inspection (Section~\ref{sec:visclasEL}) to remove cases with heavy contamination from objects that are not part of the potentially merging system.

Each individual object in the 3D-HST catalogue has been matched to the available ancillary photometric catalogues. Since the 3D-HST catalogue is selected from the deep \citep[$H_\textrm{F140W} \approx 26.1$;][]{Brammer:2012p12977} high-resolution NIR HST F140W images, and the photometric catalogues are ground-based and shallower, not all 3D-HST objects can be matched to an object in the photometric catalogues.
We only selected objects with a counterpart (within $0.3''$) in ground-based photometric catalogues.
Faint objects have the risk of being assigned to a bright(er) counterpart's photometry, have low signal-to-noise (S/N), and less reliable redshifts, and we therefore restrict ourselves to objects with $m_\textrm{F140W}\leq23.5$.

Last but not least, the 3D-HST catalogues provide a redshift estimate for the objects based on the extracted grism spectra. The catalogue grism redshifts, $z_\textrm{grism}$, are obtained by collapsing the 2D grism spectrum into a 1D spectrum, combining it with available photometry, and then estimating the redshift with an updated version of the EAZY code \citep{Brammer:2008p13280}. The redshift range where H$\alpha$ and/or [O{\scshape iii}] emission fall in the G141 grism wavelength range is $0.7<z<2.3$ \citep[see Figure~1 in][]{Brammer:2012p12977}. We are interested in tracing the star formation in the merging systems via either H$\alpha$  or [O{\scshape iii}] emission, and therefore use $z_\textrm{grism}$ to select objects in this particular redshift range.

Applying these five initial cuts (listed in the top half of Table~\ref{tab:samplecut}) reduces the full sample of 21460 detections in the 30 GOODS-S and COSMOS pointings to 1542 objects.

\setcounter{table}{0}
\begin{table}
\centering{
\caption[ ]{Grism (Top) and SED (Bottom) Selection Criteria}
\label{tab:samplecut}
\begin{tabular}[c]{|ccccc|}
\hline
0.75		 &  $\leq$ 	& 	Spectral coverage							& 	 	&   \\
		 &	 	& 	Photometric match							& $\leq$ 	& $0.3''$  \\
		 &  	 	&	$m_\textrm{F140W}$						& $\leq$	& 23.5  \\
0.7		 &  $\leq$ 	& 	$z_\textrm{grism}$							& $\leq$ 	& 2.3  \\
\hline
\multicolumn{5}{c}{  } \\
\hline
9.0		 &  $\leq$ 	& 	$\log\left(\frac{M_*}{[M_\odot]}\right)$			& $\leq$ 	& 12.0  \\
-9.5		 &  $\leq$ 	& 	$\log\left(\frac{sSFR}{[\textrm{yr}^{-1}]}\right)$		&  		&   \\
1.0		 &  $\leq$ 	& 	$\log\left(\frac{SFR}{[M_\odot/\textrm{yr}]}\right)$	& 	 	&   \\
\hline
\end{tabular}}
\end{table}

\subsection{Fitting SEDs to Photometry}\label{sec:SEDfit}

We select star-forming systems that are expected to have significant emission line features based on their star formation rates (SFR), specific SFR (sSFR), and stellar masses ($M_*$). We obtain the SFR, sSFR, and $M_*$ of each individual object from modelling the SED based on the ancillary photometric catalogues using a \cite{Chabrier:2003p23945} IMF with the code MAGPHYS presented in \cite{daCunha:2008p9413}.
We use the 37 NMBS bands \citep{Whitaker:2011p9509} for the COSMOS objects and the 17 FIREWORKS bands \citep{Wuyts:2008p9535} for the GOODS-S objects. Both catalogues span from the far-UV to MIPS 24~$\mu$m.
The 3D-HST catalogue redshift $z_\textrm{grism}$ is used as a prior when fitting the ancillary photometric data for each object. 

Although the photometric measurements are a blend of two or more components in many cases, i.e., only one photometric ID corresponds to each merger whereas several components are clearly distinguishable in the high-resolution HST imaging, selecting the high-SFR objects based on SED fits to the photometry is still very effective in selecting objects with strong emission line features.

As shown in the bottom part of Table~\ref{tab:samplecut} we select objects with SFR~$> 10 M_\odot/\textrm{yr}$, sSFR~$>10^{-9.5} \textrm{yr}^{-1}$ and $10^9M_\odot < M_* < 10^{12}M_\odot$. 
A SFR of $10 M_\odot/\textrm{yr}$ roughly corresponds to an emission line flux $F_{\textrm{H}\alpha}\sim10^{-16}$~erg/s/cm$^2$ at $z=1.5$ which corresponds to a (collapsed) emission line S/N of $\sim$8 at $1\times10^{-16}$~erg/s/cm$^2$ \citep{Brammer:2012p12977}. Hence, concentrated star formation, i.e., emitted from a modest amount of pixels of this order, should be well-detected in the 3D-HST spectra. On the other hand, if the total  emission line flux $F_{\textrm{H}\alpha}$ is spread over a larger area, the S/N per pixel might become too low for clear detection (see Section~\ref{sec:resultsEL}).

We find 352 of the initial 1542 objects in the $0.7<z<2.3$ range that satisfy these SED criteria.

\subsection{Visual Inspection of NIR Morphology}\label{sec:visclasEL}

Previous studies have argued both for visual \citep{Robaina:2009p13298} and algorithmic merger identification \citep[e.g.,][]{Lotz:2008p12474,Conselice:2008p22641}. For this pilot study, which is focussed on the \emph{emission line morphology} not on the merger rates, we have, as discussed below, decided to use a visual classification. The visual inspection of the remaining 352 objects is the crucial final step in the merger sample selection process.

The visual inspection is based on the NIR F140W morphology of the objects from the 3D-HST direct imaging. 
The NIR images show the rest-frame optical emission at the redshifts of our galaxies. We assume that the observed NIR (i.e., rest-frame optical) morphology traces the distribution of the (intermediate-age) stellar component of the galaxies, and that it is therefore different from the stars being formed (current star formation) as traced by the emission lines. A caveat to such an assumption is that if a galaxy does not have a dominating intermediate-age stellar population and has a high SFR, the morphology in the rest-frame optical will to some extent reflect the distribution of young stars as well.
The criteria used to select the merger candidates from the 352 objects are that 
(i) they should show a morphology that differs from the bulk of the `normal' isolated galaxies, i.e., a disturbed irregular/asymmetric morphology, and 
(ii) they have to show several distinct components in the continuum image, either multiple objects within the $\sim3\times3$~arcsec  F140W thumbnails or pronounced tidal features extending from the main continuum emission component.
It should be noted that because our merger selection is based only on this visual classification, and the low resolution of the G141 grism ($R\sim130$) does not offer any kinematic information of the individual companions, it is impossible to address whether or not the systems are gravitationally bound. Our merger sample therefore consists of \emph{potentially} merging systems. 

A fraction of the more widely separated merger pairs could therefore be potential low- or high-redshift interlopers which would artificially enhance the number of mergers found. The distances between the majority of the individual merger components in the candidates selected here are of the order $1''$, corresponding to roughly 8.5~kpc at $z\sim1.5$.
\cite{Law:2012p23503} estimates that $7^{+1}_{-1}$\% of galaxies have projected false pairs within 16~kpc. This serves as an upper limit on the expected false pair fraction for our sample. As we also include late-stage mergers and not only widely separated pairs, based on this a more realistic estimate of the amount of interlopers would probably be $\sim4$\%.

Furthermore, several studies have shown that the morphology of isolated star-forming galaxies at higher redshifts is often clumpy and irregular \citep{Bournaud:2007p26648,Genzel:2008p21878,Genzel:2011p23883,Law:2009p23849,Law:2012p23503,Elmegreen:2009p24419,Kriek:2009p26672,ForsterSchreiber:2011p23940,ForsterSchreiber:2011p23942,Wuyts:2012p23967,Wisnioski:2011p9935}. 
This potentially biases our visual classification as inclusion of such systems will artificially enhance the number of selected mergers and hence the estimated fraction of mergers in our sample. As described below both parametric and visual classification schemes will be subject to this bias.
Hence, what is described as a merger in the present selection might in fact be a galaxy with a clumpy and irregular light distribution appearing like a merger remnant. This caveat should be kept in mind when evaluating the merger candidates presented here and elsewhere. 
In the remainder of the paper we will therefore use the shorthand \emph{merger} for \emph{likely merger candidates}.

The visual inspection is not only used to select morphologically disturbed systems. 
Inspecting the full grism spectra, as well as the one-dimensional collapsed grism spectra, of the individual objects, we are able to discard objects without any emission line features. Without emission line features creating a star formation map as described in Section~\ref{sec:ELmapEL} is impossible.
Assuming that the estimated SFR, mass, and redshift are correct, the fact that these objects lack emission line features make them very interesting in themselves, as they might be highly dust-obscured systems blocking the star formation emission. However, these objects are ignored for the present study. 
If an emission line feature on the other hand is observed, i.e., the (collapsed) emission line flux is roughly larger than $\sim10^{-16}$~erg/s/cm$^2$ (see Section~\ref{sec:SEDfit}), the object is included in our sample and we attempt to create an emission line map. However, as noted in Section~\ref{sec:SEDfit} and as we will see in Section~\ref{sec:resultsEL} this does not necessarily mean that the S/N per pixel in the full grism spectrum is good enough to create an emission line map.

Visually inspecting the grism spectra also ensures that any strong contamination is due to the different components of the merging system and not due to interlopers.

From the parent sample of 352 catalogue-selected objects, the visual inspection discards 292 objects as they appear to be isolated `normal' galaxies (252/292), or having high contamination not stemming from the merging components (13/292), or having no obvious emission line features in their spectra (24/292 corresponding to $7^{+3}_{-2}$\% of the 352 catalogue-selected objects). 
Hence, our final sample of (potential) mergers from 6 GOODS-S and 24 COSMOS 3D-HST pointings consists of 60 systems, corresponding to $\sim17^{+4}_{-4}$\% of the 352 catalogue-selected objects, or approximately 10-13\% if we correct for the expected fraction of false pair interlopes. 
Here the confidence intervals are the 95\% quantiles of the beta distribution following \cite{Cameron:2011p26280}.
This merger candidate fraction is on the high side compared to what is generally found in the literature (see, e.g., Figure~1 in \citealp{Lotz:2011p24053} and \citealp{Williams:2011p23620}), again suggesting that a fraction of the visually identified mergers may be single objects with a blotchy distribution of young stars which increases the apparent merger fraction. The obtained merger fraction is subjective as it relies on a visual assessment of disturbance.
As described above, also potential galaxy interlopers which have passed the subjective visual inspection could be biasing our sample towards higher merger fractions. 
\cite{Williams:2011p23620} used a mass-selected sample ($\log(M/ M_{\odot}) > 10.5$) of galaxy pairs when estimating a major merger fraction of $\sim$5\%. This could also account for part of the discrepancy, as we are not only looking at distinct pairs of galaxies and use a flux limit as opposed to a mass limit in the selection process. 
With an average merger fraction just below 10\% at $z\sim1$ the merger fractions summarised in Figure~1 of \cite{Lotz:2011p24053} which includes samples selected on both mass and luminosity cuts is somewhat closer to the $17^{+4}_{-4}$\% ($\sim$10-13\% if corrected for interlopes) reported here.

The selection cuts in Table~\ref{tab:samplecut} positions all 60 3D-HST mergers in the blue cloud of star-forming galaxies \citep{Bell:2004p24172,Strateva:2001p24138} in colour magnitude diagrams, which implies (as expected) that the selected mergers are mainly gas-rich `wet' mergers. We note however, that this may be very different for mass-selected samples where no limits have been imposed on the overall SFR of the objects.

\begin{figure}
\centering{
\includegraphics[width=0.45\textwidth,bb=30 0 680 660]{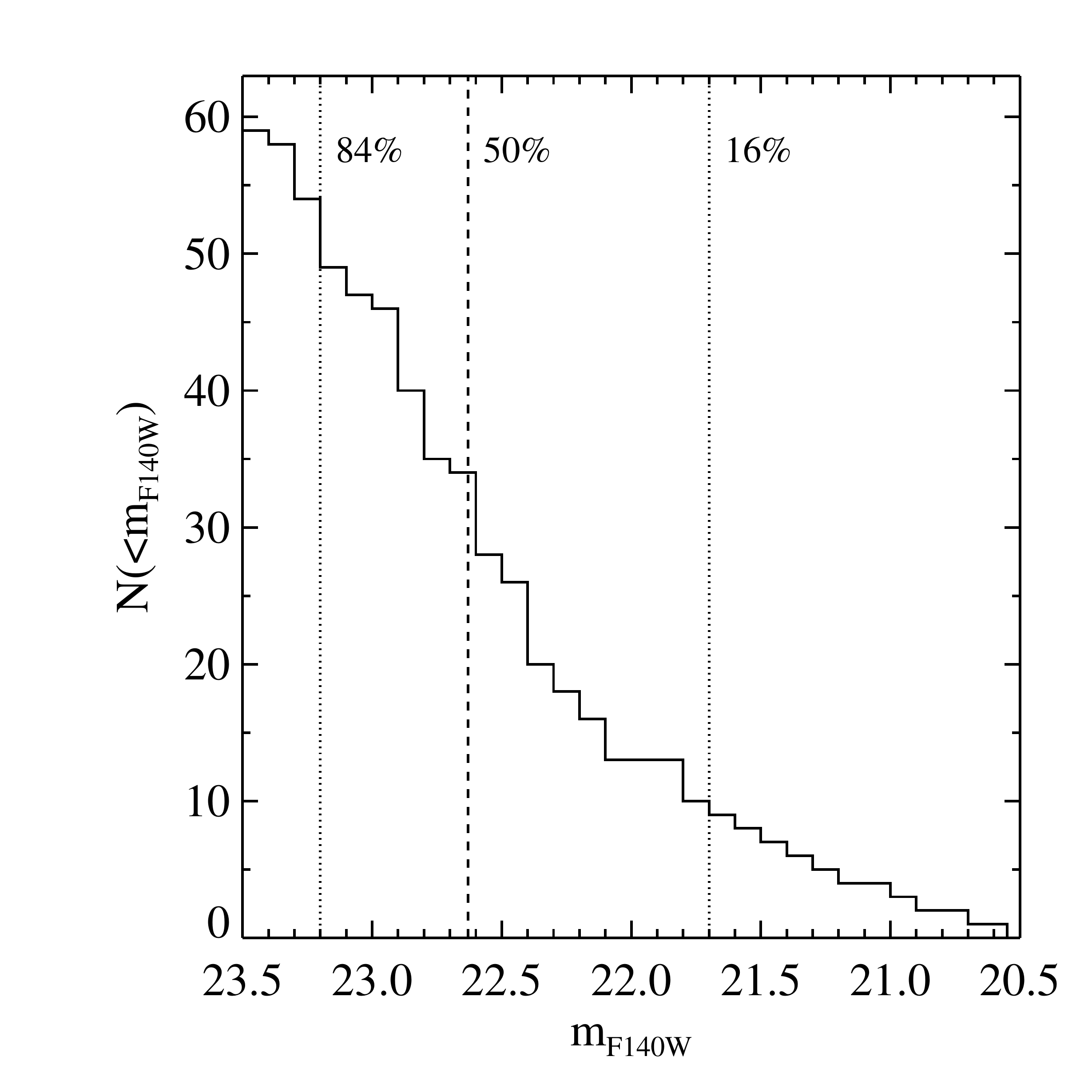}
\caption{The cumulative distribution of $m_\textrm{F140W}$ magnitudes for the 60 3D-HST mergers. The dotted lines indicate the 16th and 84th percentiles of the distribution whereas the dashed line shows the distribution median. The 16th and 84th percentile values are used when defining the parameter space for the simulated grism spectra in Section~\ref{sec:SimspecEL}.}
\label{fig:mhist}}
\end{figure} 

\begin{figure}
\centering{
\includegraphics[width=0.45\textwidth,bb=30 0 680 660]{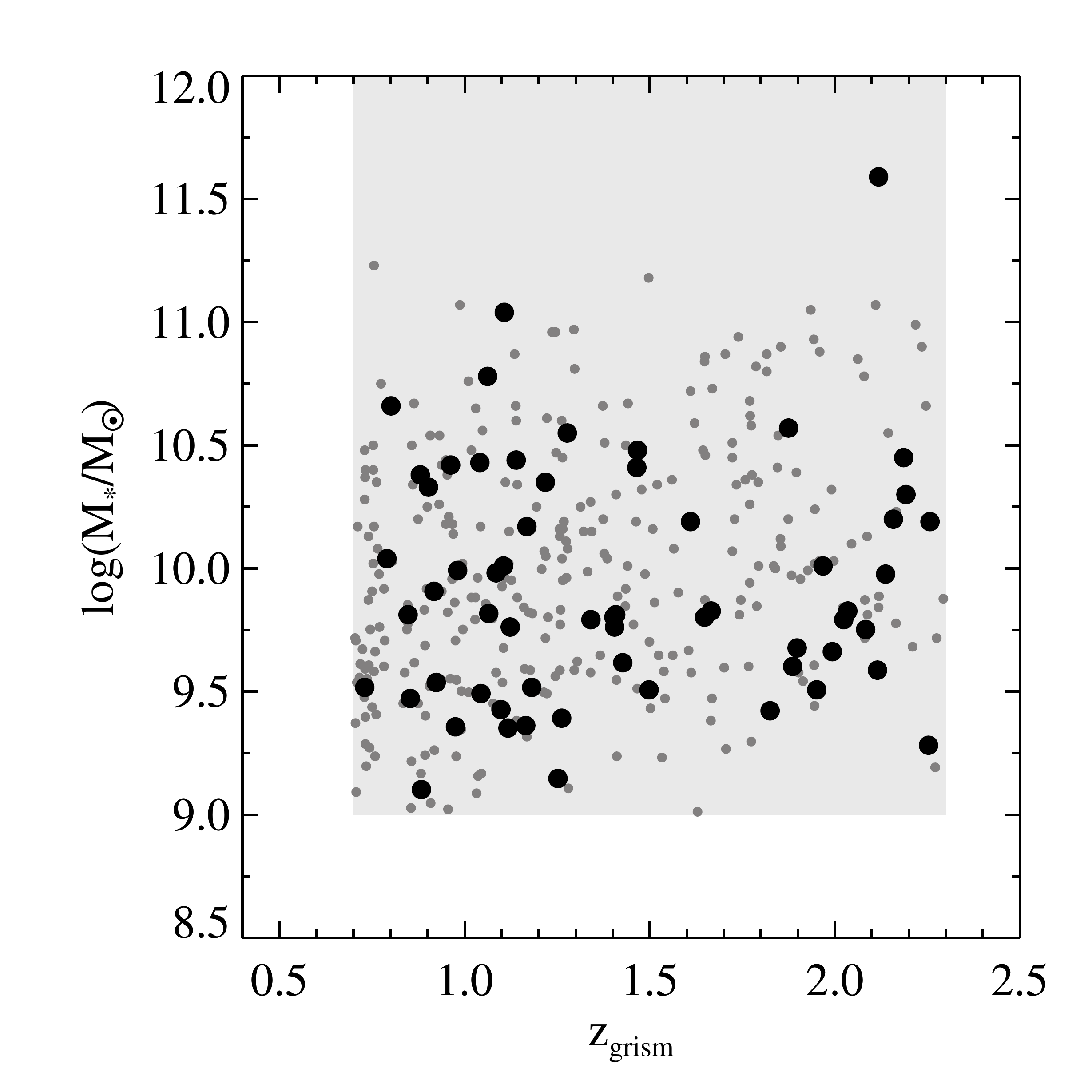}
\caption{The distribution of inferred stellar mass as a function of redshift for the 60 3D-HST merger candidates (large points). The small grey points represent the 292 objects discarded by the visual inspection and the grey shaded region shows the selection region from Table~\ref{tab:samplecut}.}
\label{fig:Massz}}
\end{figure} 

\begin{figure*}
\centering{
\includegraphics[width=0.9\textwidth]{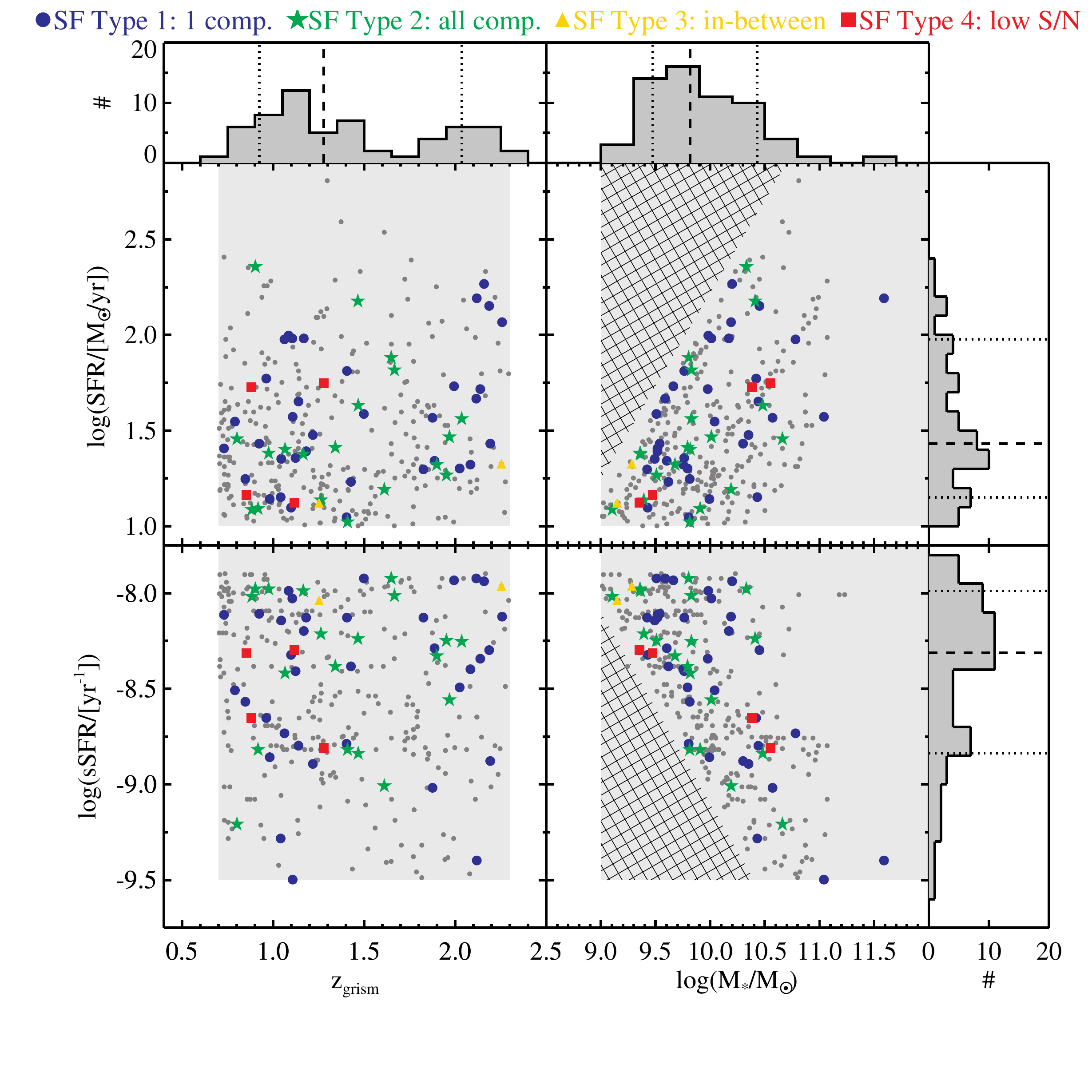}
\caption{The 60 3D-HST mergers (large symbols) plotted in the main region of the catalogue selection space given in Table~\ref{tab:samplecut} (grey shaded regions). The small grey points represent the 292 objects discarded by the visual inspection. The hashed regions are empty due to our selection criteria. The histograms attached to the scatter plots show the distribution of $z_{grism}$, $M_*$, SFR, and sSFR for the 60 mergers. The dotted and dashed lines in the histograms indicate the 16th and 84th percentiles and the median of the distributions, respectively. These values are used to define the parameter space of the simulated grism spectra described in Section~\ref{sec:SimspecEL}. 
The different symbols (and colours) indicate the morphology of the emission line maps (SF type) described in Section~\ref{sec:resultsEL} as indicated above the panels.
No obvious trends are found between the SF morphology and the SFR, sSFR, redshift, or stellar mass.}
\label{fig:SFRvsM}}
\end{figure*}

In Figure~\ref{fig:mhist} we show the $m_\textrm{F140W}$ magnitude distribution of the 60 3D-HST merger candidates.
In Figure~\ref{fig:Massz} and \ref{fig:SFRvsM} we plot them together with the selection regions from Table~\ref{tab:samplecut} (grey shaded regions) as large solid symbols. 
The small grey points represent  the 292 objects discarded by the visual classification, i.e., the general galaxy populations satisfying the selection cuts in Table~\ref{tab:samplecut}.
In Figure~\ref{fig:SFRvsM} the distributions of SFR, sSFR, $z_\textrm{grism}$, and $M_*$ for the 60 merger candidates are shown as histograms on the axes of the scatter plots. 
In both the histograms in Figure~\ref{fig:mhist} and in Figure~\ref{fig:SFRvsM}, the dotted lines correspond to the 16th and 84th percentiles of the distributions, and the dashed lines show the median values. In Section~\ref{sec:SimspecEL} we will use these values to determine the parameter space to sample when simulating 3D-HST grism spectra. 

Visual inspection is a more subjective way of selecting mergers than for instance empirically established parametric merger classification schemes like for instance the Gini $(G)$, $M_{20}$ and $CAS$ selections \citep[e.g.,][]{Lotz:2004p12689,Lotz:2006p17552,Lotz:2008p12474,Lotz:2008p24286,Conselice:2003p26534,Conselice:2008p22641,Conselice:2009p12404,Papovich:2005p26615,Scarlata:2007p26618}.
Both parametric and visual classification schemes have advantages and disadvantages. Determining a merger population based on parameters ensures that the selection is done in a consistent and uniform way for all objects. However, at redshifts where the NIR morphology is not fully understood and where galaxies look clumpy and irregular  as described above, and where disturbed morphologies are prominent without necessarily being part of a recent merger, a parametric classification scheme might fall short of a visual classification. On the other hand visual classification is potentially biased by the subjectivity of the classifier. Nevertheless, the human eye is known to be excellent at detecting and distinguishing features in noisy images and spectra and arguably minimises the bias of the clumpy irregular morphology of high-$z$ systems. 

To ease comparison with the extensive literature using parametric merger classifications we have estimated the $G$, $M_{20}$, $C$, $A$, and $S$ morphological parameters for the 60 visually selected merger candidates and the 292 visually discarded objects as shown in Appendix~\ref{sec:GM20CAS}. The selected candidates partially satisfy the empirical parametric merger selection but in general seems to be an average subset of the parent distribution, i.e., not distinguishing itself clearly from it. Assuming that the 60 merger candidates are reliable therefore speaks in favour of using visual classification when the NIR morphology is complicated, as a parametric $GM_{20}CAS$ classification would not be able to clearly distinguish the 60 merger candidates from the parent population.

Despite the limitations of a visual selection of mergers and the only partial agreement with the empirical parametric classifications (Appendix~\ref{sec:GM20CAS}), we believe that the advantages of the visual classification scheme  described above outweighs a `blind' parametric selection for a study like the one presented here.
Furthermore, we are probing an unexplored regime (NIR at high redshift) and the current parametric methods might not be appropriately calibrated and tested here and we have therefore chosen to use the visual selection in the reminder of this paper.

\section{Emission Line Mapping}
\label{sec:ELmapEL}

\begin{figure}
\includegraphics[width=0.49\textwidth]{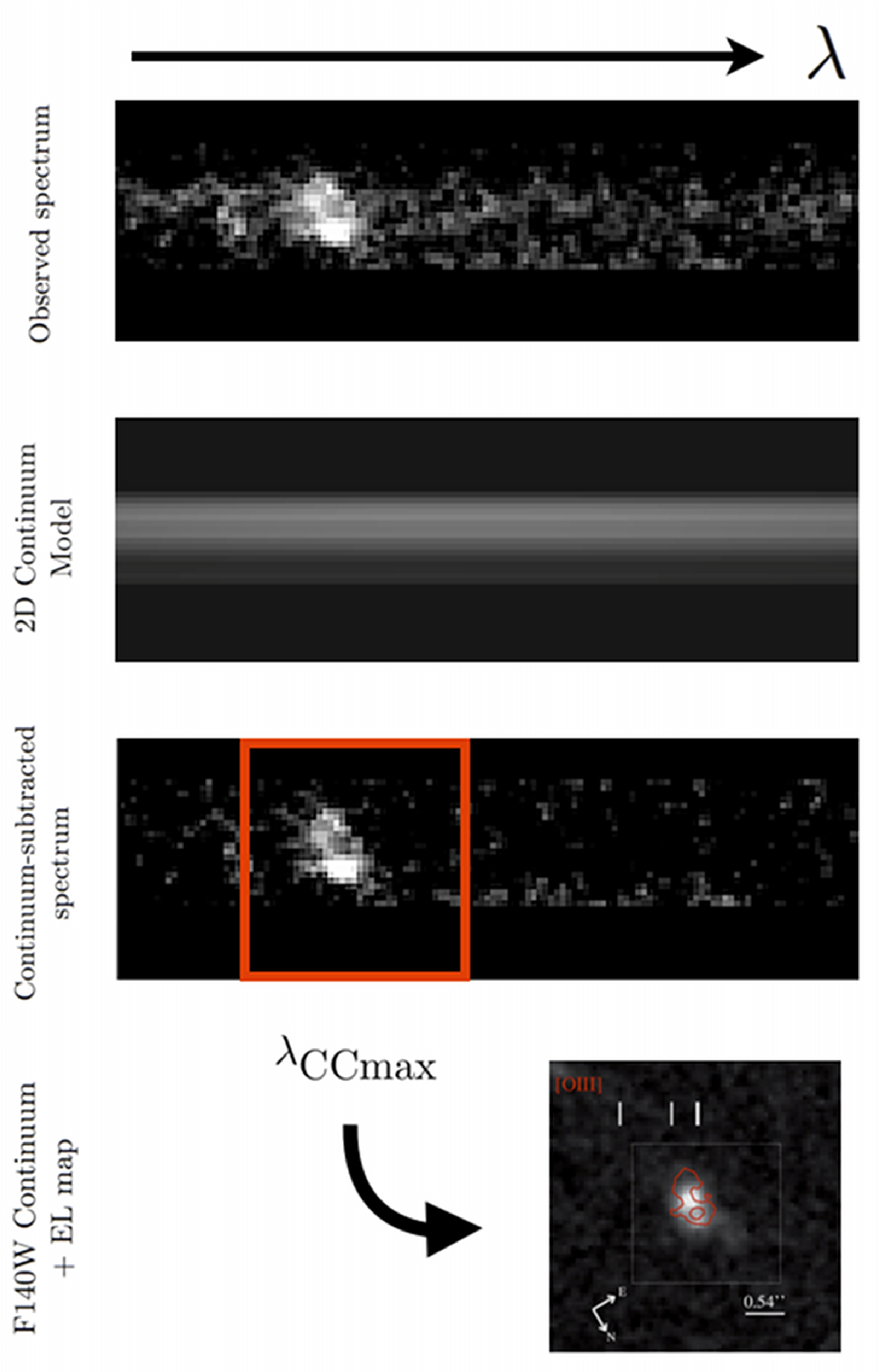}
\caption{Steps performed to obtain the emission line (star formation) maps as described in Section~\ref{sec:ELmapEL}. The top panel shows a standard 3D-HST spectrum with a prominent emission line feature. The 2D continuum model, a set of polynomial fits to the observed spectrum when excluding the emission line feature (Section~\ref{sec:2dfit}), is shown just below that. The third panel shows the spectrum from the top panel after subtraction of the 2D continuum model. The red square marks the emission line map cut-out at $\lambda_\textrm{CCmax}$ obtained by cross-correlating the NIR F140W thumbnail of the object (grey scale bottom panel) with the continuum-subtracted spectrum. The bottom panel illustrates how this emission line map cut-out is mapped back (red contours) onto the continuum image.}
\label{fig:ELmapill}
\end{figure}

To quantify the extent of (unobscured) star formation in the 60 3D-HST mergers described in the previous section, we rely on the spatial information of the H$\alpha$ and [O{\scshape iii}] emission lines that the slitless grism spectroscopy provides. From the grism spectra we create emission line maps by subtracting a model of the continuum light in the grism spectra such that only the probed emission line feature is remaining. This can then be mapped back onto the NIR continuum light-distribution of the object. In practice we:
\begin{enumerate}
\item Create a 2D continuum model for the grism spectrum.
\item Subtract this continuum model from the 2D grism spectrum itself.
\item Cross-correlate the F140W thumbnail with the continuum subtracted spectrum to map the (cut-out) continuum-subtracted thumbnail back onto the F140W thumbnail.
\end{enumerate}
Each of these steps is described in detail in the following subsections and are illustrated in Figure~\ref{fig:ELmapill}.

\subsection{Continuum Modelling and Subtraction}\label{sec:2dfit}

The 2D continuum models we subtract from the 3D-HST spectra are based on a third order polynomial fit to a one-dimensional spectrum. The 1D spectra are obtained from a weighted sum of the individual lines in the 2D spectra. 
The polynomial fit to the 1D spectrum is turned back into a 2D continuum model by concatenating rows with the 1D polynomial form weighted by a `slit-profile' obtained from the columns blue-ward and red-ward of the probed emission line feature in the full 2D spectrum. 
Subtracting this model from the 3D-HST spectrum returns a two-dimensional emission line map as illustrated in Figure~\ref{fig:ELmapill}, where all that is left is the emission line feature.
This approach is similar to the one used in \cite{Nelson:2012p12947}. 
Often when dealing with slitless spectroscopy, and the 3D-HST grism spectra in particular, the goal is to remove contamination in a systematic manner. However, we relied on visual inspection to remove badly contaminated objects, since mergers per definition are contaminated.
Remaining contamination not affecting the continuum flux of the merger was masked out when modelling the continuum before subtraction. 

\subsection{Constructing Emission Line Maps}
\label{sec:zgrismEL}

We used a simple cross-correlation between the NIR F140W thumbnail continuum image of each object and the corresponding full 2D emission line map in order to map the emission line map back onto the NIR image. In practice we calculate 
\BE
\textrm{CC}(\lambda) = \sum_i^{N_\textrm{width}}\sum_j^{N_\textrm{width}} f_{i,j,\textrm{F140W}}\times f_{i,j,\textrm{ELmap}}
\EE
for each of the first $k=N_\textrm{2D}-N_\textrm{width}$ columns in the full 2D emission line map, where $N_\textrm{2D}$ is the number of columns in the 2D emission line map and $N_\textrm{width}$ is the width of the F140W thumbnail. The $f_{i,j,\textrm{F140W}}$ and $f_{i,j,\textrm{ELmap}}$ is the flux in the pixel $(i,j)$ for the F140W thumbnail and 2D emission line map cut-out (indicated by the red box in Figure~\ref{fig:ELmapill}), respectively.

The maximum of the cross-correlation function, CC, indicates the wavelength, $\lambda_\textrm{CCmax}$, where there is the largest overlap between the NIR light distribution of the object and the $k$th cut-out of the full emission line map. The $\lambda_\textrm{CCmax}$ corresponds to the $k$th column of the 2D emission line map plus $N_\textrm{width}/2$.

\begin{figure*}
\centering{
\hspace{0.037\textwidth}
\includegraphics[width=0.38\textwidth]{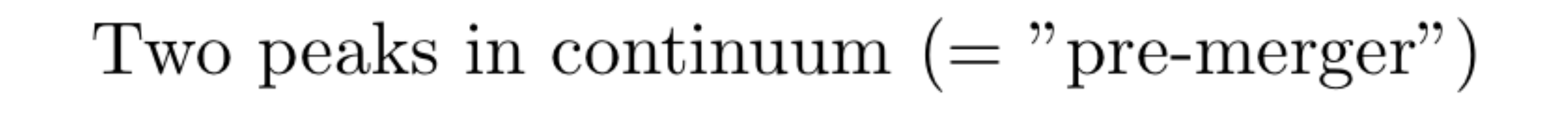}\hspace{1.5cm}
\includegraphics[width=0.38\textwidth]{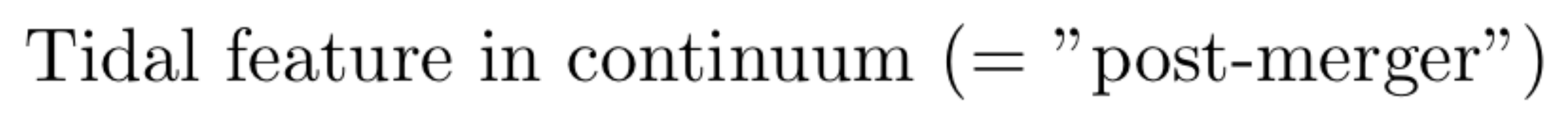}\\
\includegraphics[width=0.022\textwidth]{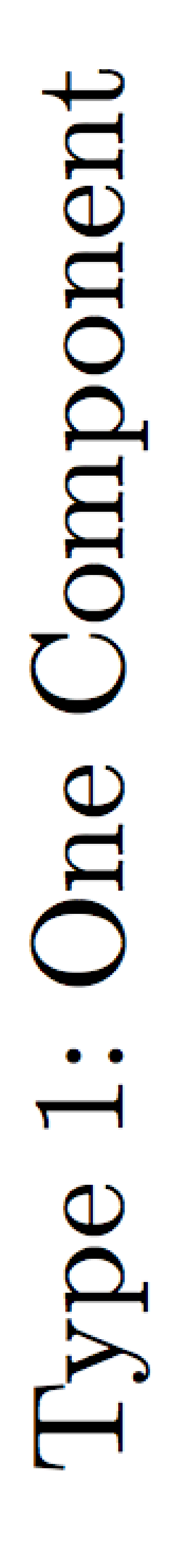}\hspace{0.01\textwidth}
\includegraphics[width=0.22\textwidth]{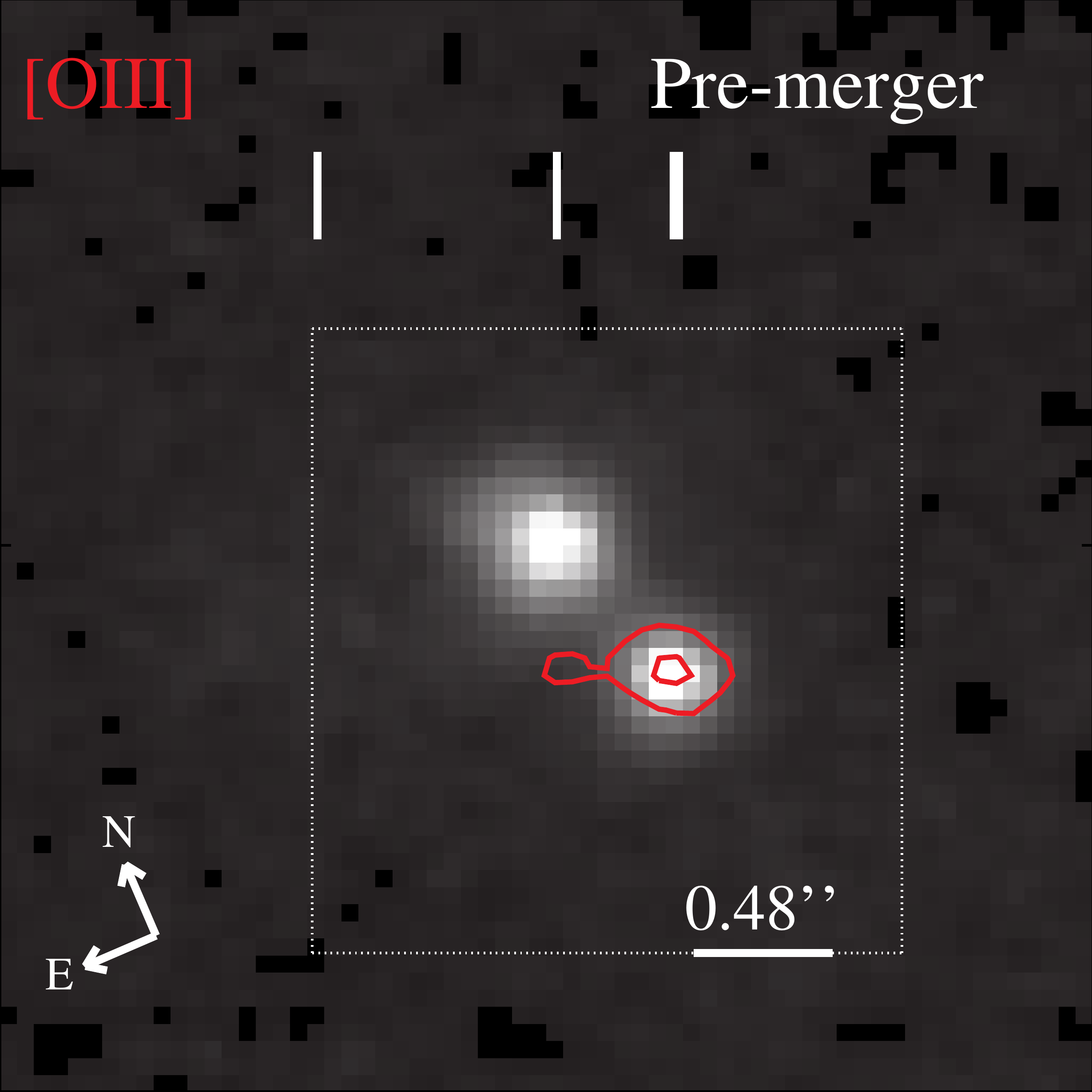} 
\includegraphics[width=0.22\textwidth]{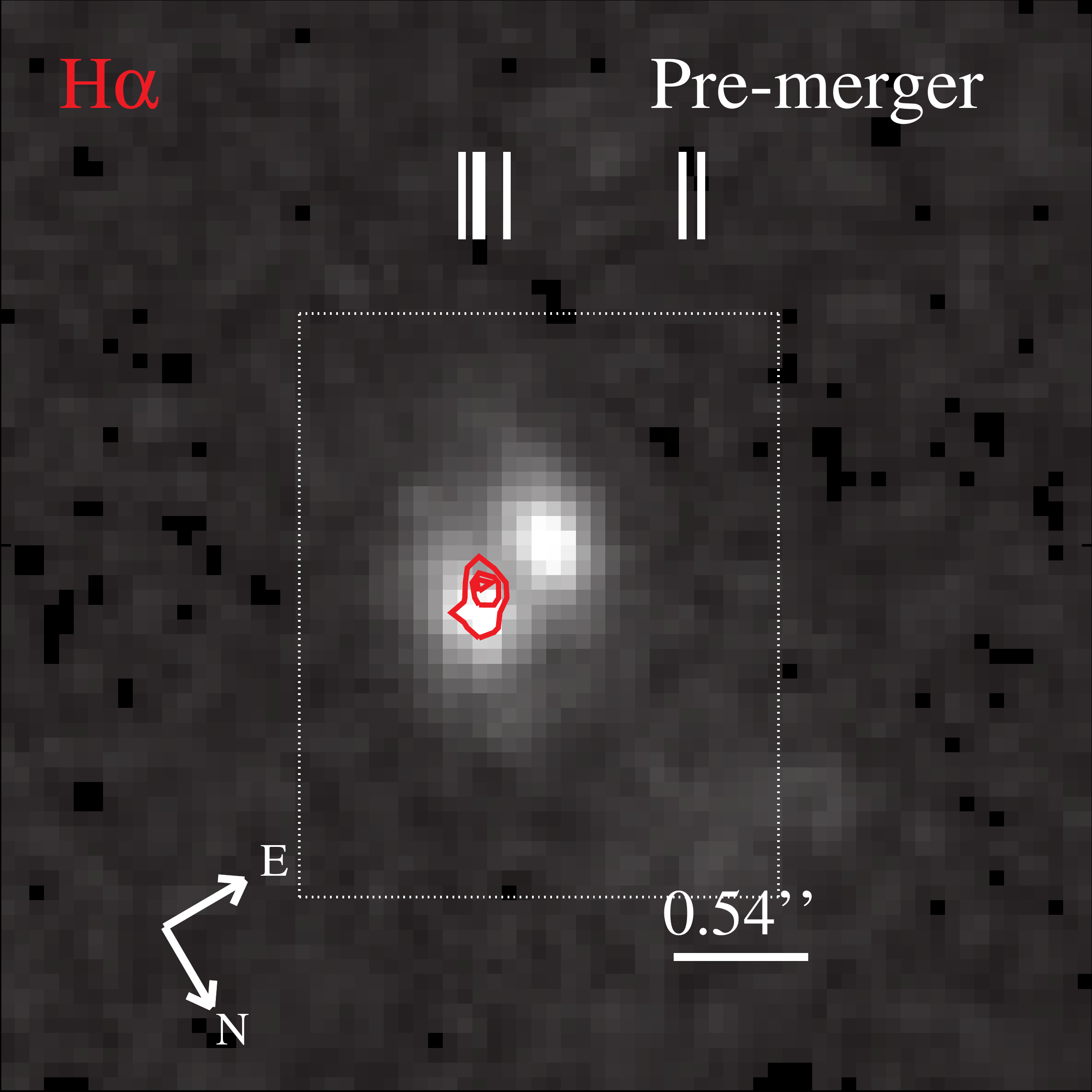} \hspace{0.3cm}
\includegraphics[width=0.22\textwidth]{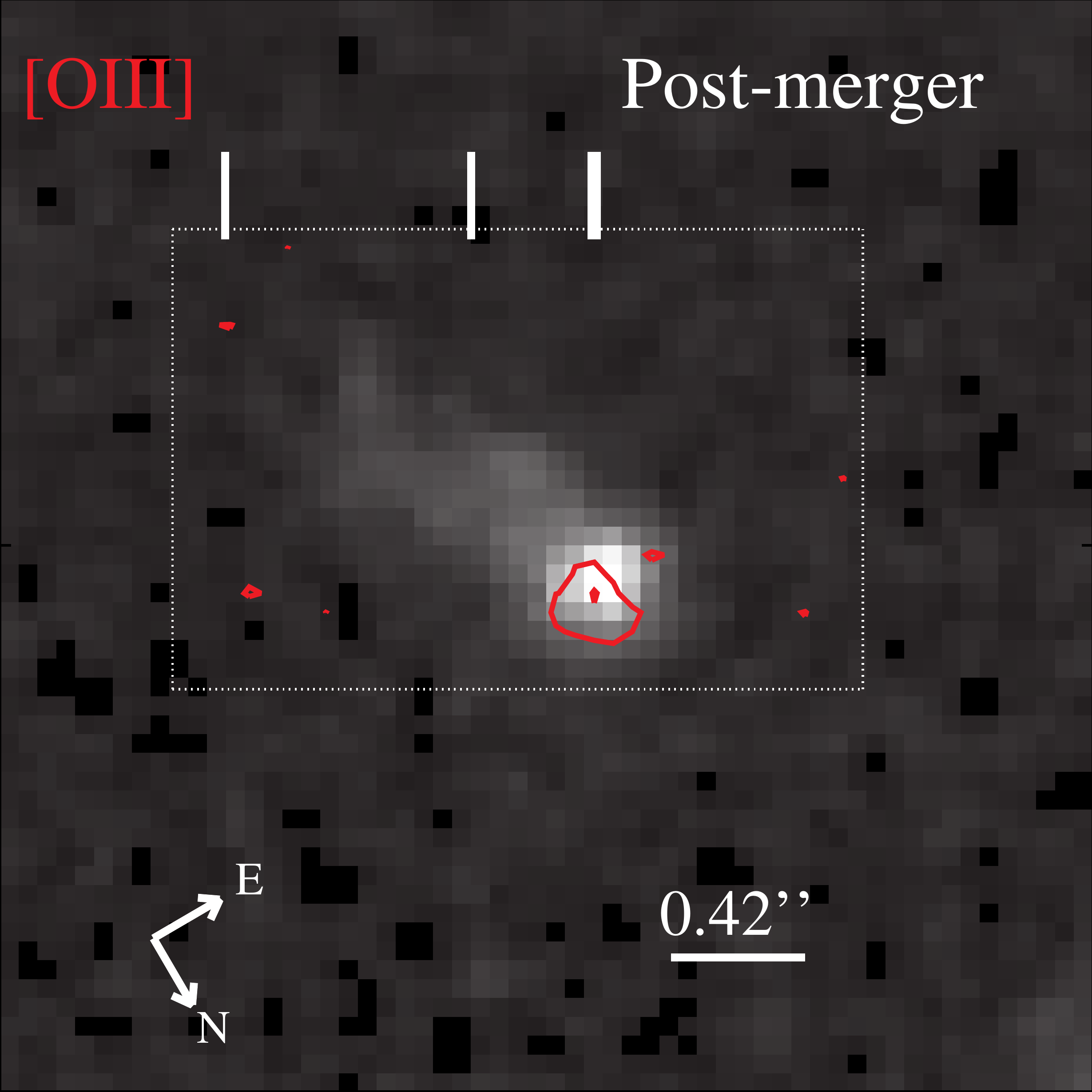} 
 \includegraphics[width=0.22\textwidth]{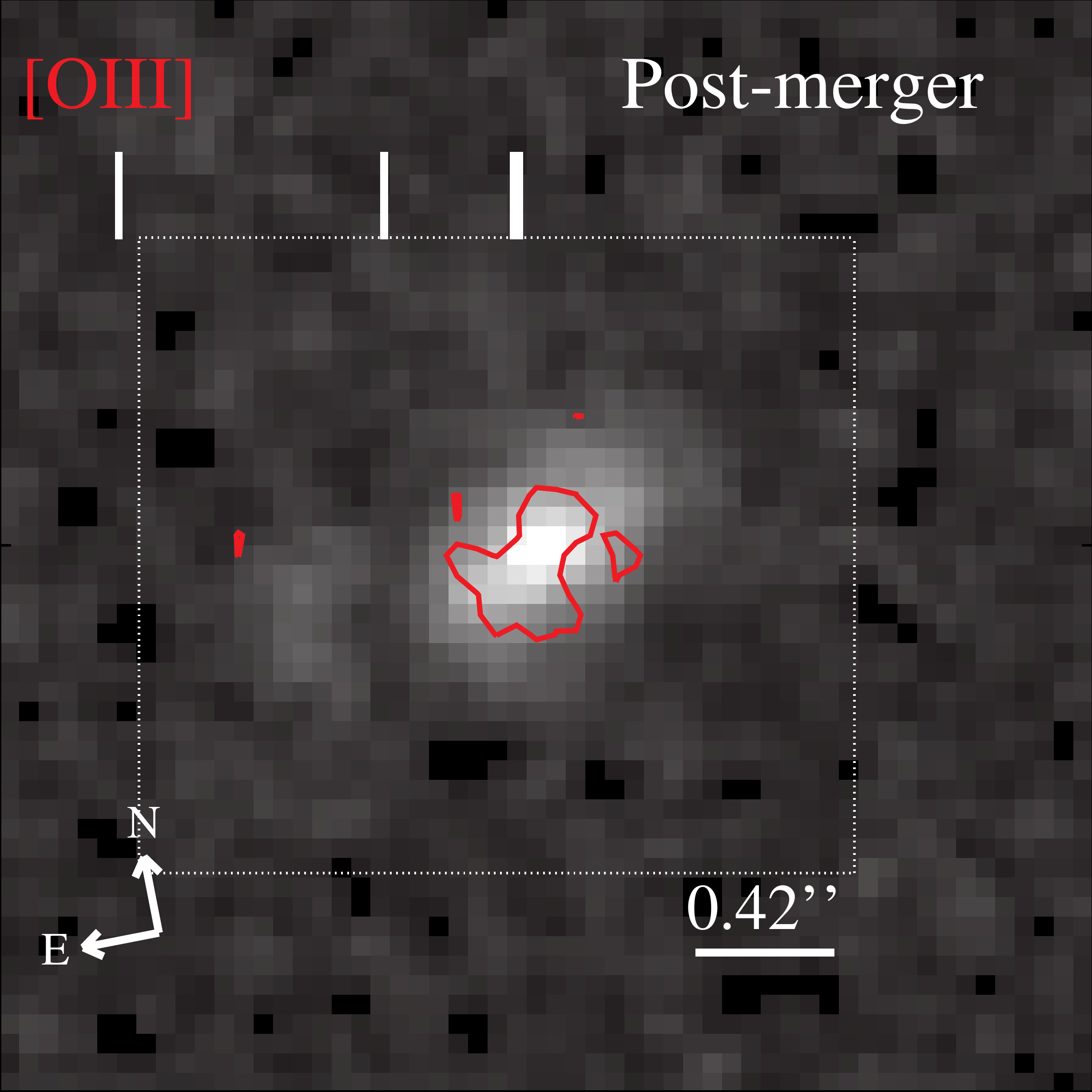} 
 \\ \vspace{0.1cm}
\includegraphics[width=0.022\textwidth]{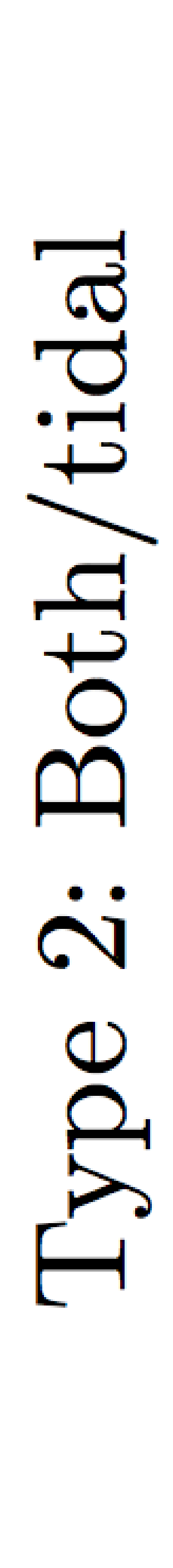}\hspace{0.01\textwidth}
 \includegraphics[width=0.22\textwidth]{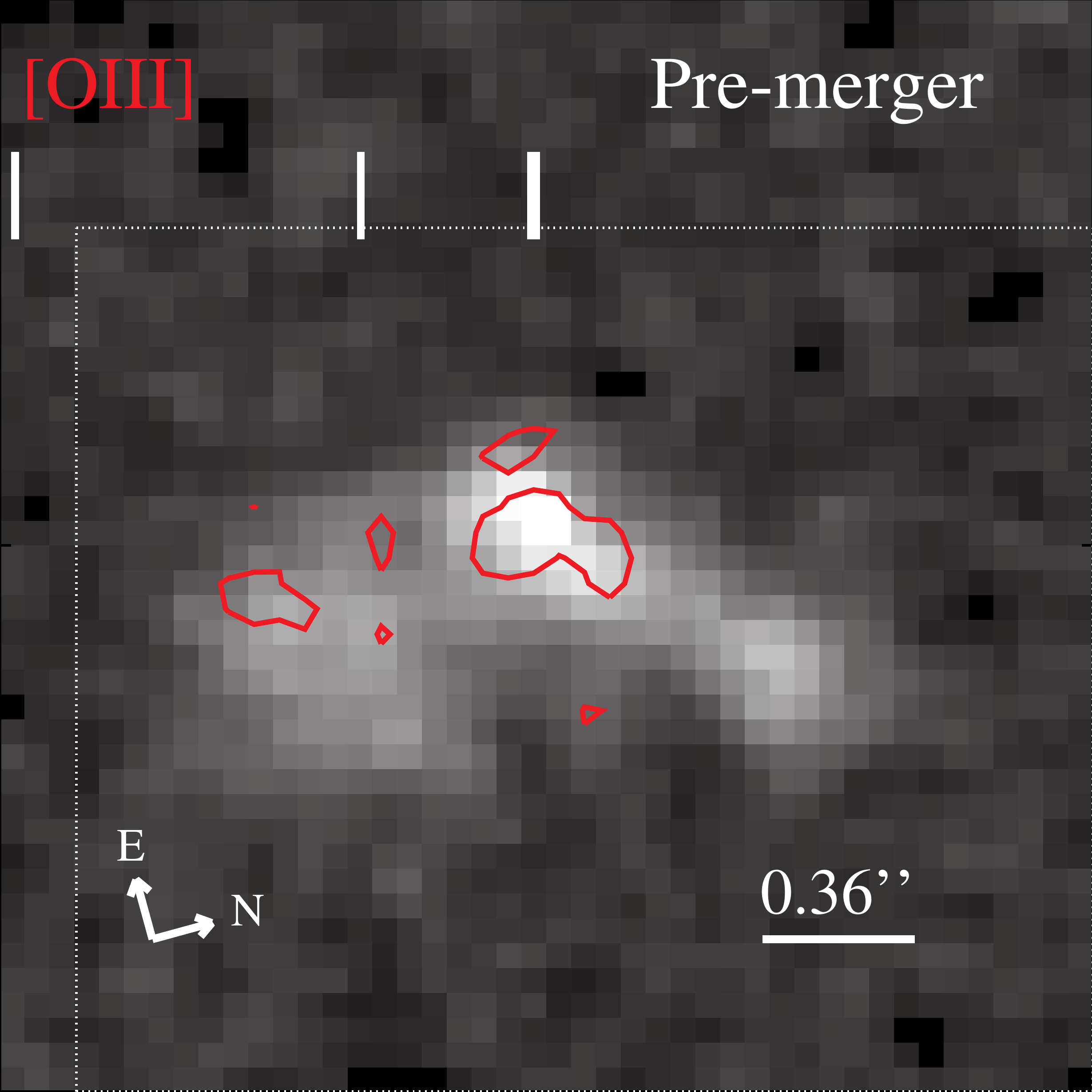} 
 \includegraphics[width=0.22\textwidth]{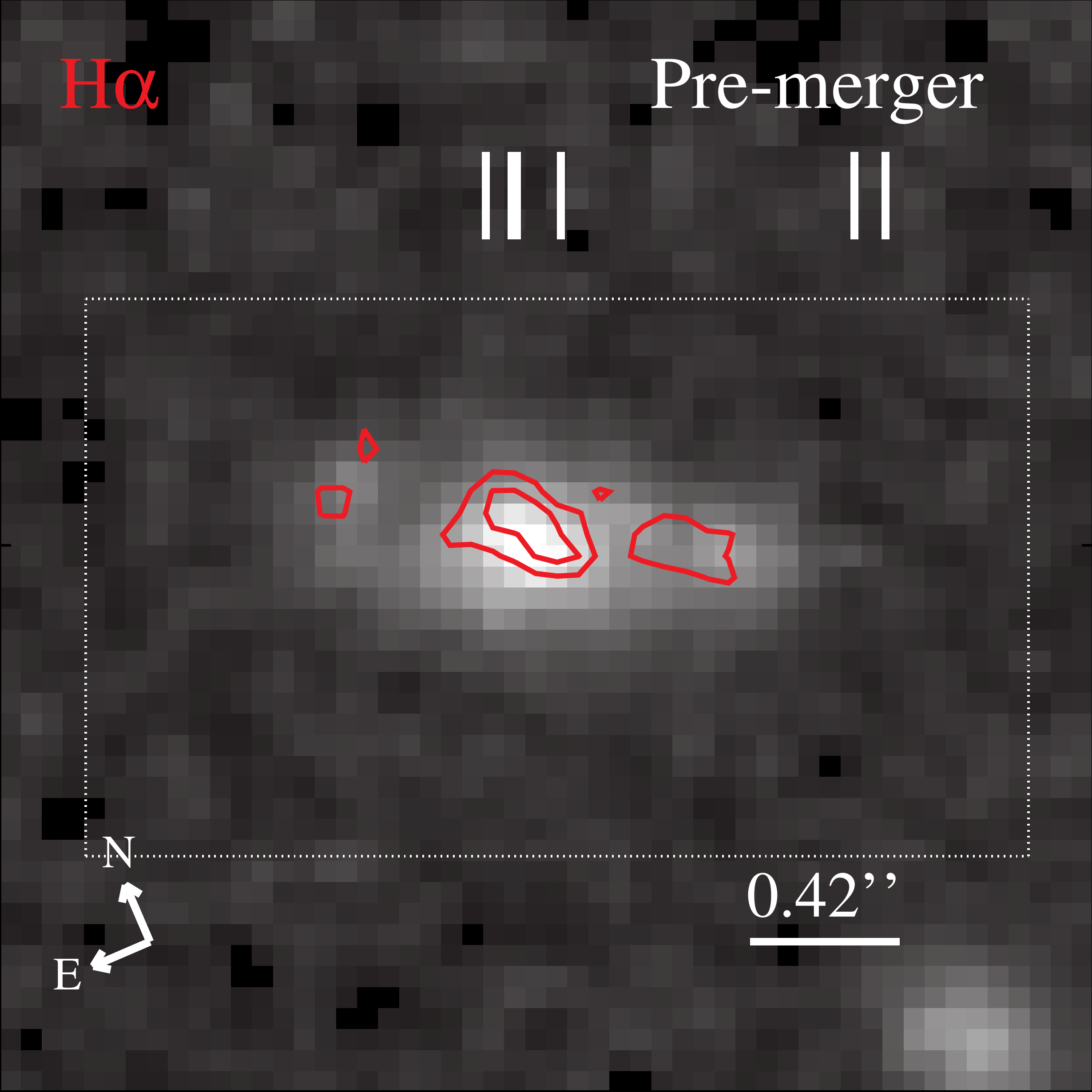} \hspace{0.3cm}
 \includegraphics[width=0.22\textwidth]{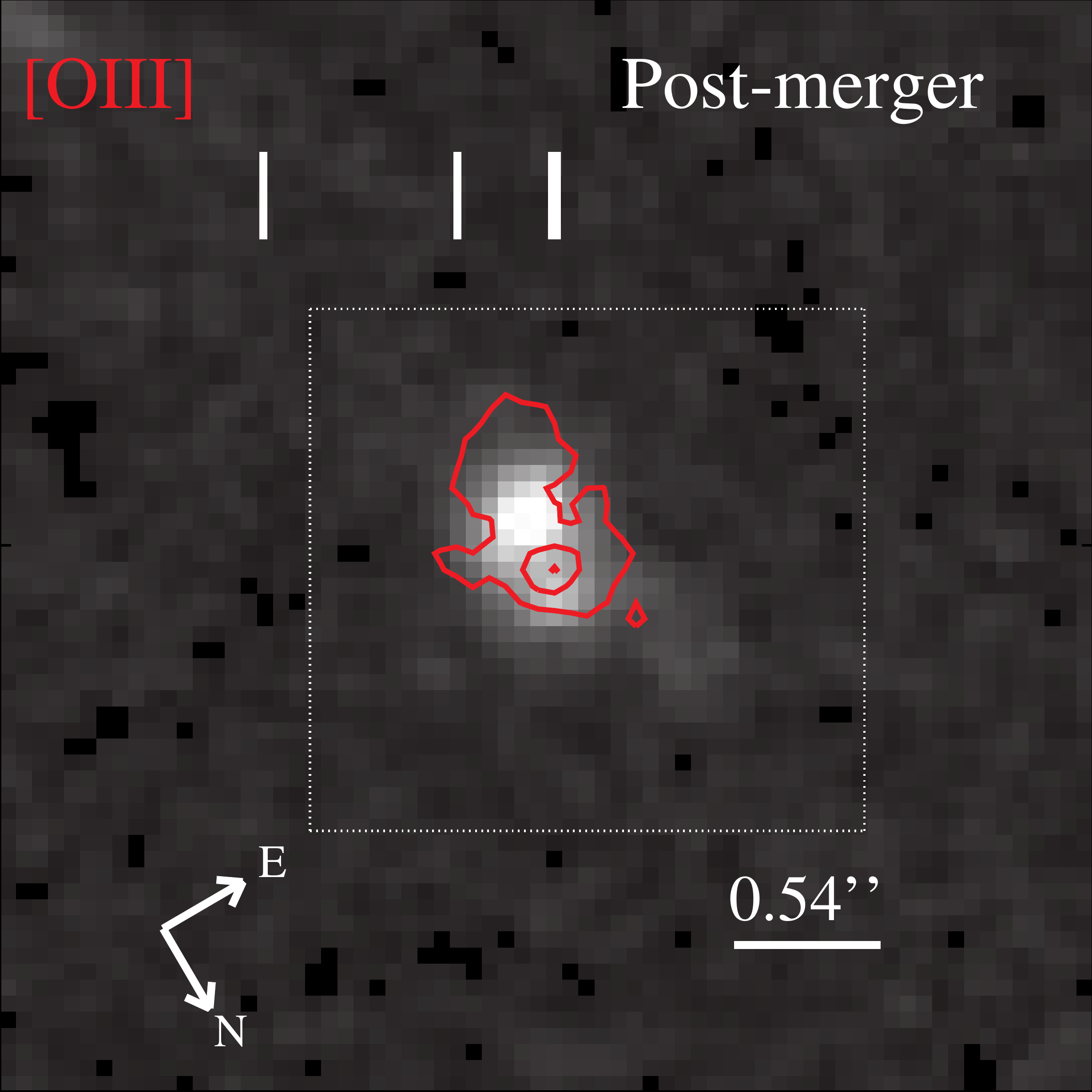} 
 \includegraphics[width=0.22\textwidth]{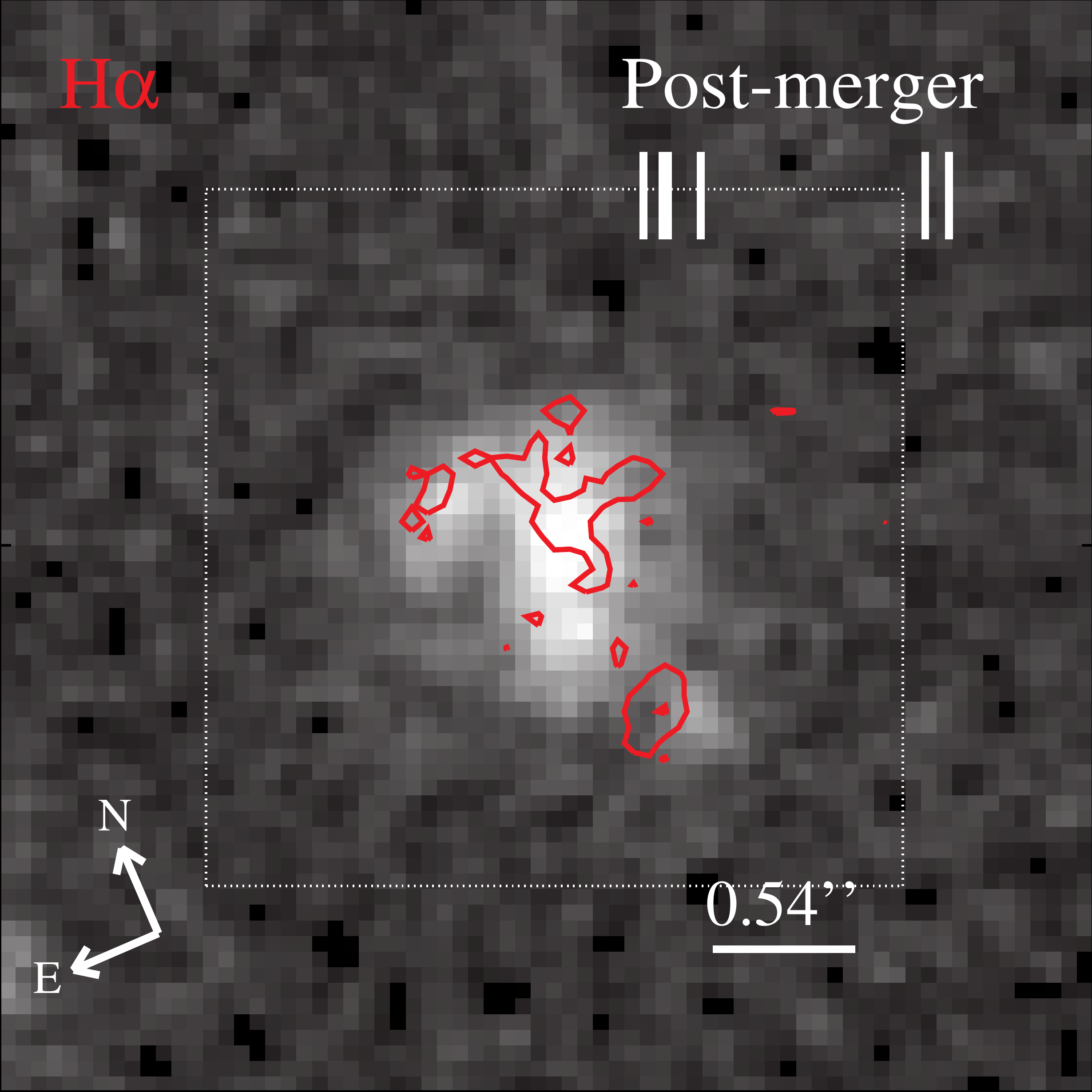} 
\\ \vspace{0.5cm}
\includegraphics[width=0.019\textwidth]{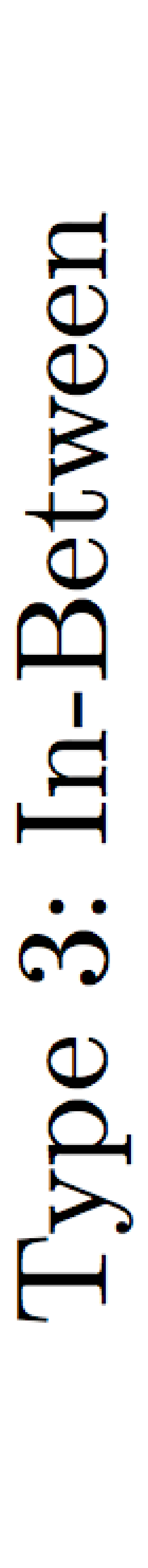}\hspace{0.01\textwidth}
 \includegraphics[width=0.22\textwidth]{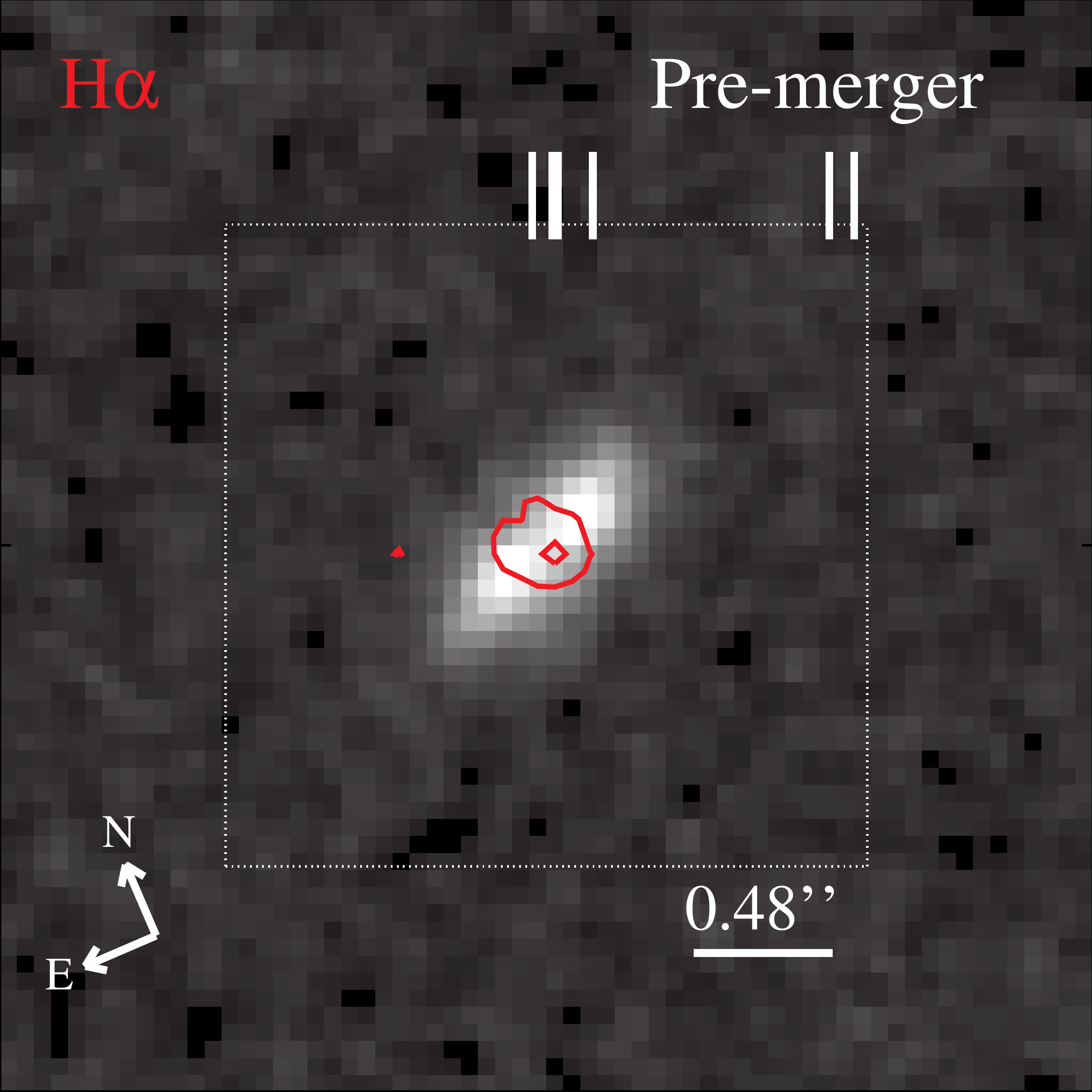}
 \includegraphics[width=0.22\textwidth]{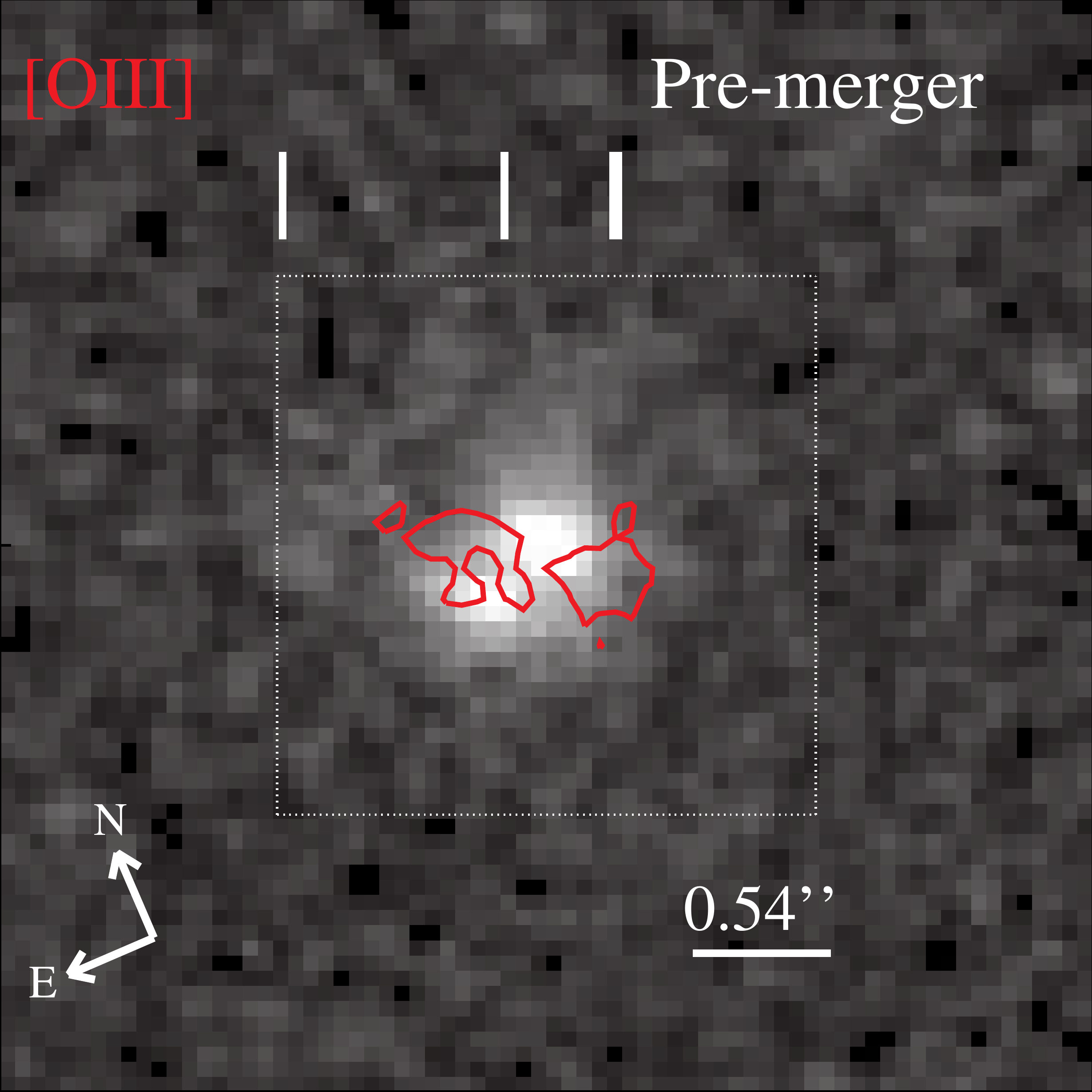}\\ \vspace{0.1cm}
\includegraphics[width=0.022\textwidth]{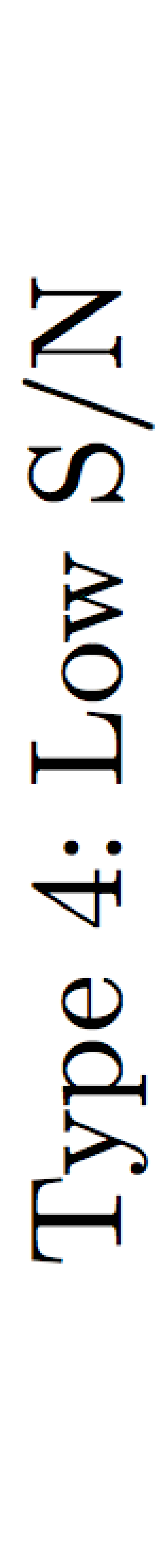}\hspace{0.01\textwidth}
 \includegraphics[width=0.22\textwidth]{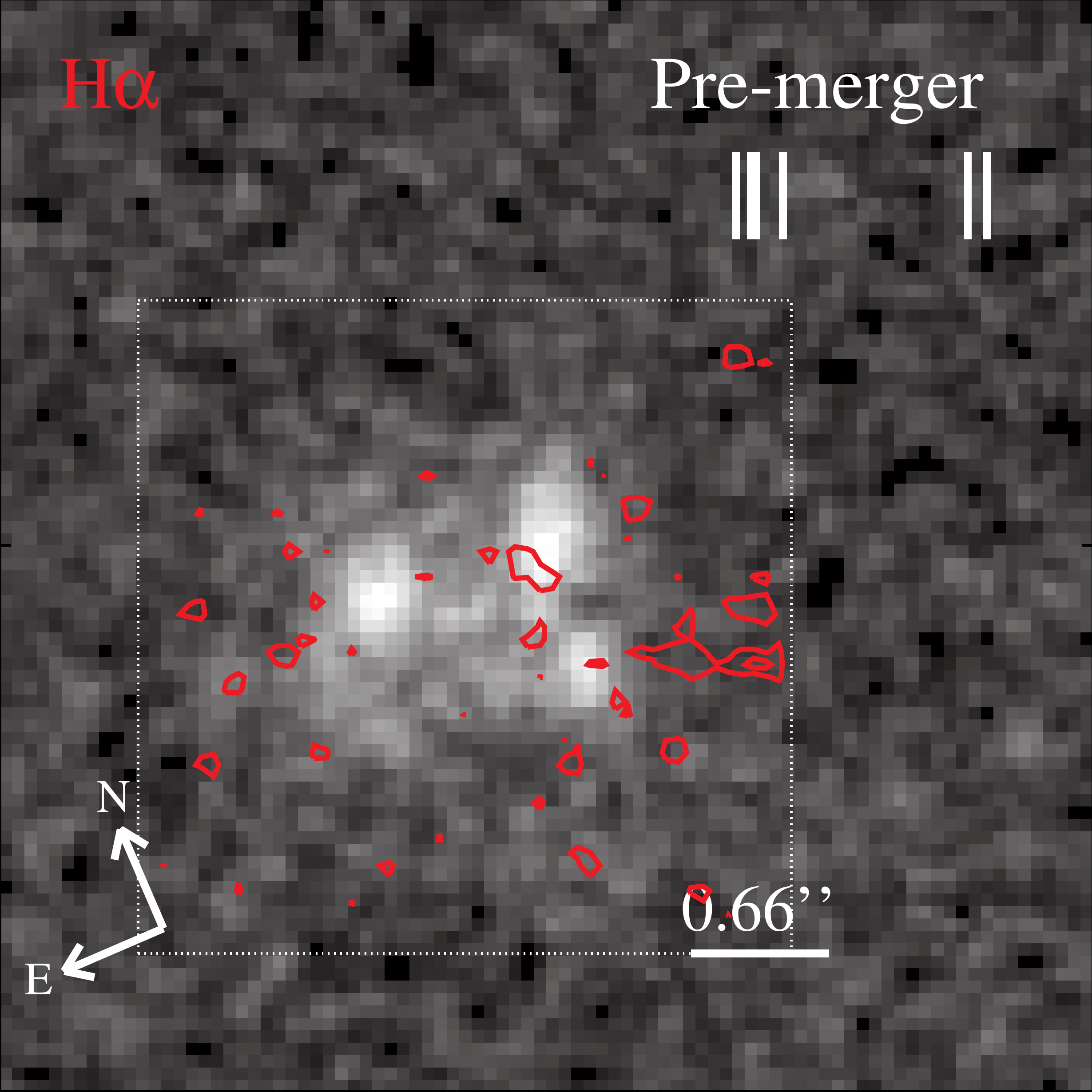}
  \includegraphics[width=0.22\textwidth]{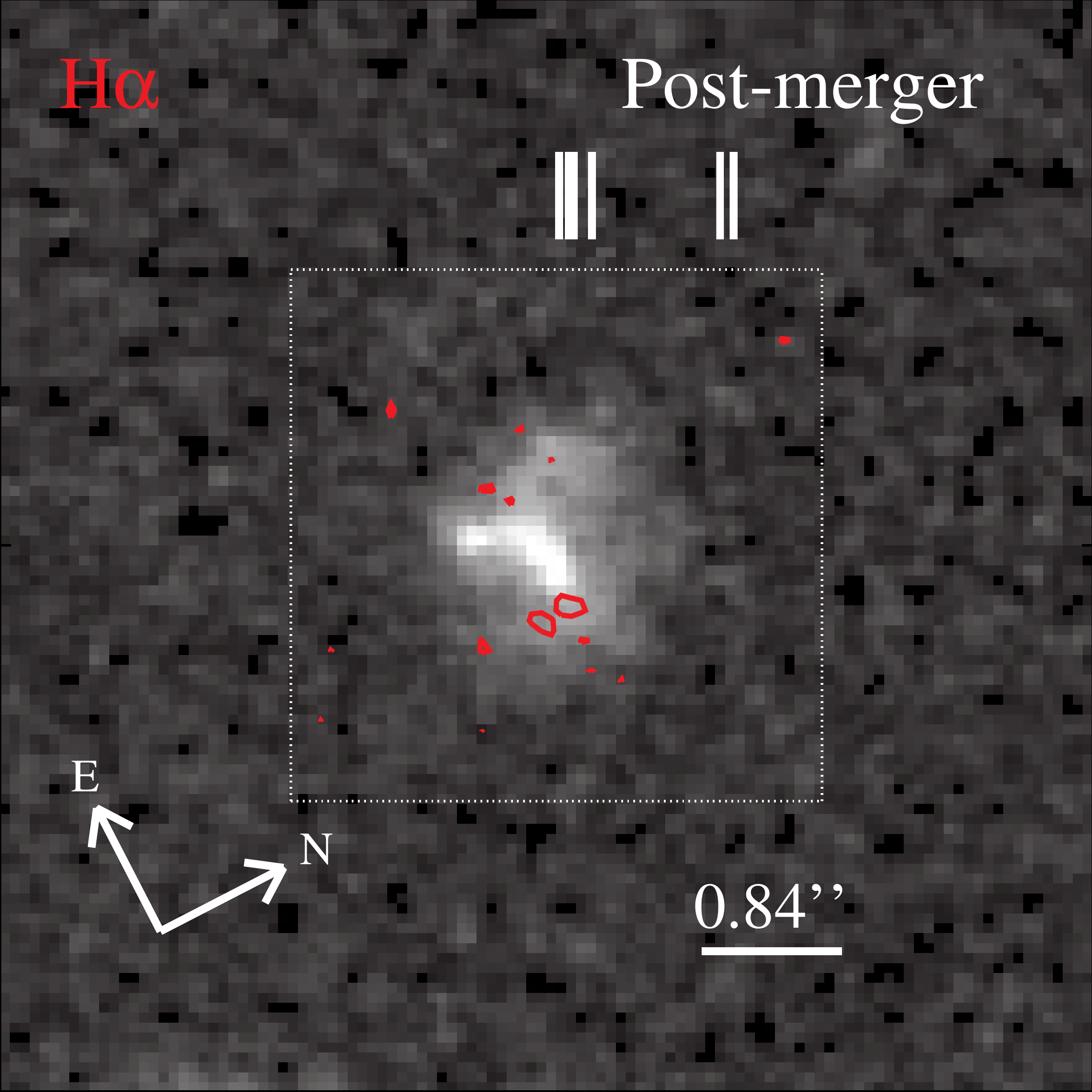}
 \includegraphics[width=0.22\textwidth]{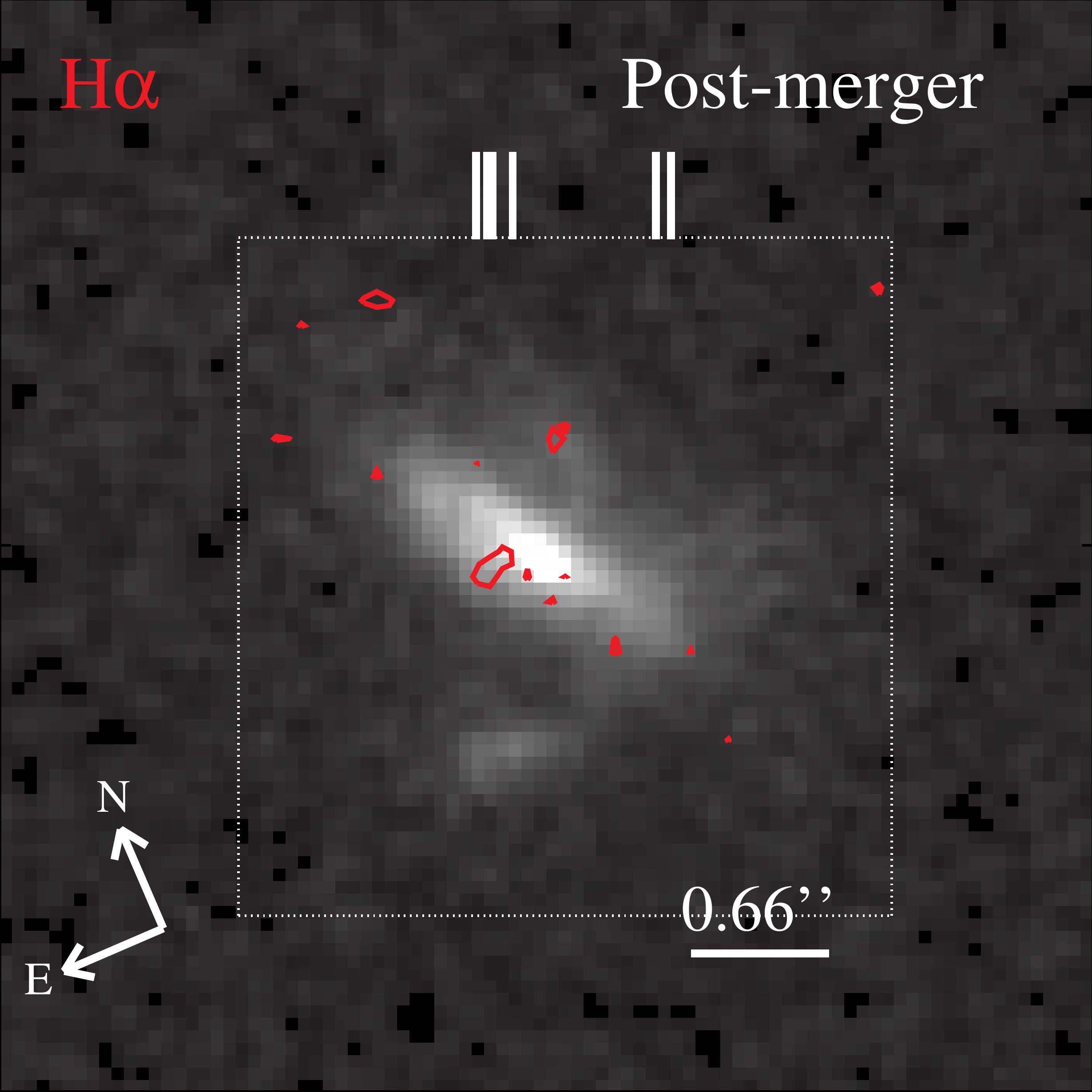}
 \includegraphics[width=0.22\textwidth]{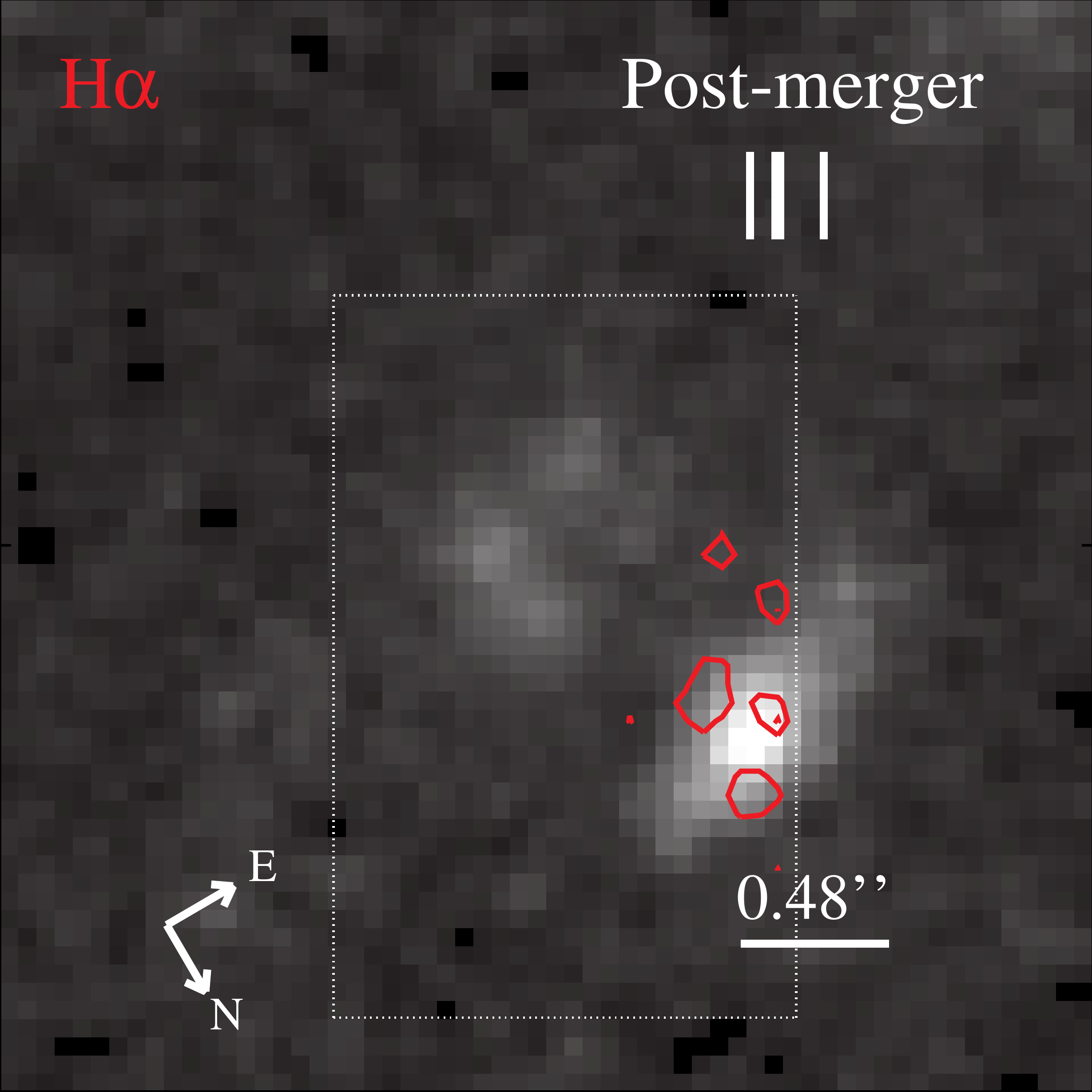}
\caption{Examples of H$\alpha$ and [O{\scshape iii}] emission line maps (red contours) plotted on the WFC3 F140W thumbnails (grey scale) for a sub-subsample of the 60 3D-HST mergers. The dashed boxes indicate the mapped region. Each row represents one morphological type of star formation distribution as described in Section~\ref{sec:selfctEL}. From top to bottom each row show
maps of SF type 1 (`one component' maps), SF type 2 (`both (all) component' maps), SF type 3 (`in-between' maps), and SF type 4 (`low S/N' maps). The Type 1 and 2 maps (first two rows) have been divided into into pre- (left) and post-mergers (right). The vertical white lines indicated the relative distance between [N{\scshape ii}] $\lambda\lambda$6548,6583, H$\alpha$ $\lambda$6563, [S{\scshape ii}] $\lambda\lambda$6716,6730 and H$\beta$ $\lambda$4861, [O{\scshape iii}] $\lambda\lambda$4959,5007 for the H$\alpha$ and [O{\scshape iii}] maps, respectively. 
The full sample of emission line maps is shown in Appendix~\ref{sec:maps}.}
\label{fig:ELmap}}
\end{figure*}

In Figure~\ref{fig:ELmap} we show a collection of emission line maps (red contours) from our 3D-HST merger sample. The individual maps correspond to the region at $\lambda_\textrm{CCmax}$ that has been mapped back onto the NIR continuum image as illustrated in Figure~\ref{fig:ELmapill}. 
In each map we have marked the relative location of [N{\scshape ii}] $\lambda\lambda$6548,6583, H$\alpha$ $\lambda$6563, [S{\scshape ii}] $\lambda\lambda$6716,6730 and H$\beta$ $\lambda$4861, [O{\scshape iii}] $\lambda\lambda$4959,5007 for the H$\alpha$ and [O{\scshape iii}] maps, respectively. In some cases the [O{\scshape iii}] doublet is marginally resolved as seen in the upper left emission line map in Figure~\ref{fig:ELmap}. The H$\alpha$-[N{\scshape ii}] composite is however not resolved in the 3D-HST grism resolution. We do not attempt to de-convolve these emission lines when creating the emission line maps. Assuming that $F_\textrm{[O{\scshape iii}] $\lambda$5007} = 3\times F_\textrm{[O{\scshape iii}] $\lambda$4959}$ the redshift uncertainty imposed by ignoring the [O{\scshape iii}] doublet is only 0.0024. This is less than the quoted average 3D-HST redshift precision of 0.0034$(1+z)$ \citep{Brammer:2012p12977} and is therefore not affecting the conclusions of this study.

The different `morphologies' of the emission line maps in Figure~\ref{fig:ELmap} will be addressed in Section~\ref{sec:resultsEL}.
In Appendix~\ref{sec:maps} we show the full sample of 3D-HST merger emission line maps. 

\section{Results: The Spatial Extent of Star Formation in High-$z$ Mergers}\label{sec:resultsEL}

As noted in the introduction, star formation and mergers are important parts of understanding how high-$z$ galaxies evolved into the galaxies we observe in the low-$z$ Universe. In the previous sections we have described how we select the mergers from the 3D-HST data, and subtract the continuum light in the spectra to create the emission line (star formation) maps. In this section we characterise the morphological \emph{type} of the spatial extent of the star formation in the 60 3D-HST mergers. However, we first divide our sample into two different kinds of mergers: the `pre-mergers' and the `post-mergers'. By pre-mergers we mean objects that show multiple clearly-distinct and pronounced continuum peaks in the NIR images, i.e.,  the optical continuum emission comes from multiple objects in the process of merging or about to merge. Examples of those are shown to the left in the two top panels of Figure~\ref{fig:ELmap}. 
The post-mergers, on the other hand, are systems that have (presumably) undergone merging and now appear to be dominated by a nuclear feature in the continuum with pronounced tidal features surrounding it. Examples of these systems are shown to the right in the two top panels in Figure~\ref{fig:ELmap}. Dividing the 60 mergers into these two sub-samples return 32 pre-mergers and 28 post-mergers.  
We note that the distinction between pre- and post-mergers is (operatively) our distinction to guide-the-eye when classifying and inspecting the emission line maps of the individual objects as described below. The identification with the merger phases is plausible, but will need more modelling. 

\begin{figure}
\centering{
\includegraphics[width=0.49\textwidth]{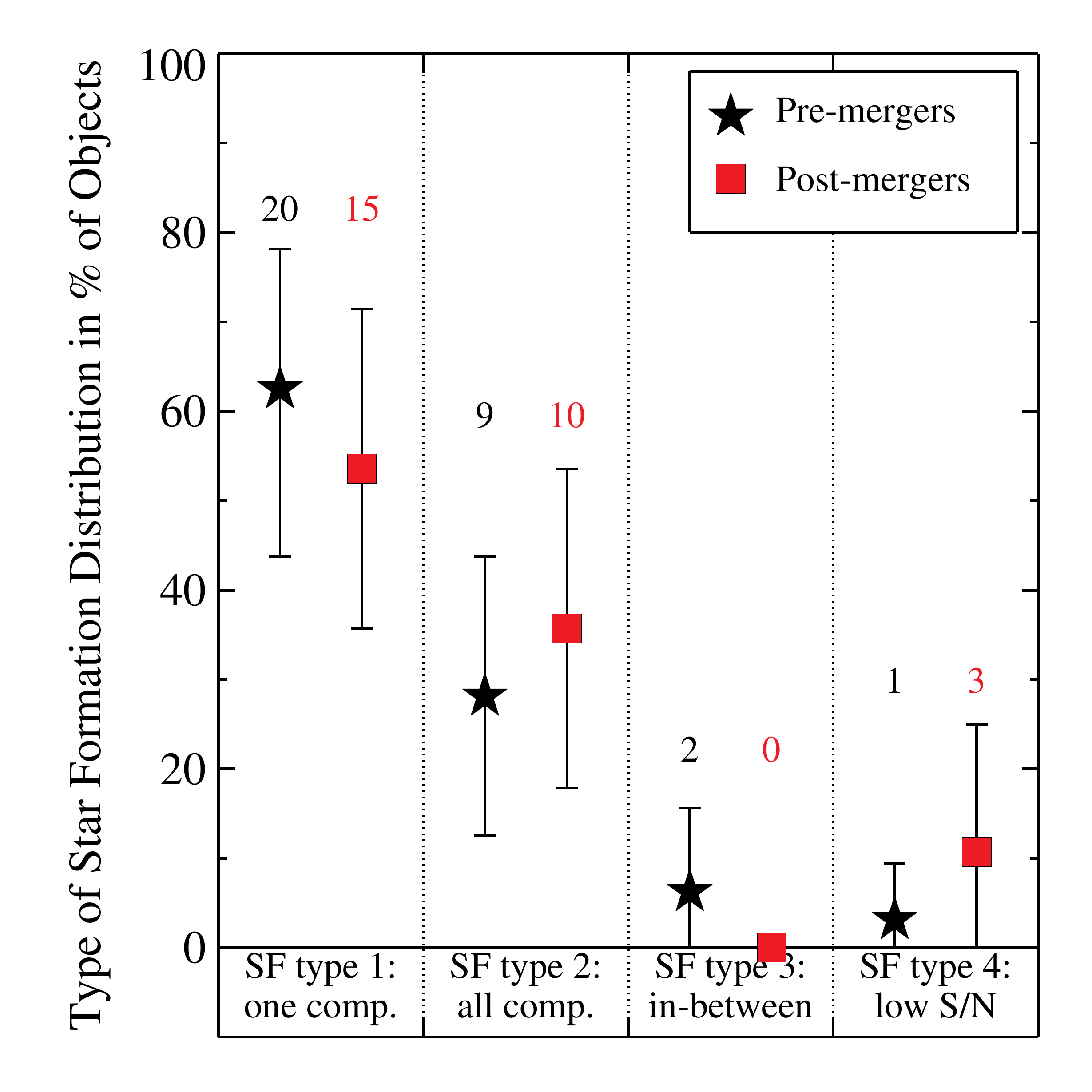}
\caption{The morphological types of the star formation distribution (Section~\ref{sec:resultsEL}) observed in the 60 3D-HST merger candidates split into `pre-mergers' of multiple individual objects (32 objects, black stars) and `post-mergers' of systems with a nucleus and tidal features (28 objects, red squares). The number of objects with a given SF type is indicated above each point. The error-bars are obtained via bootstrapping as described in the text. The two samples have similar star formation distributions. About 20\% of the objects show star formation in all merger components (individual objects or nucleus and tidal feature) whereas $\sim$60\% of the systems only show star formation in one component.}
\label{fig:SFsplit}}
\end{figure}

To characterise the location and the spatial extent of the star formation in the 3D-HST mergers we categorise the star formation maps into the following four morphological types (SF type):
\begin{enumerate}
\item[\textbf{1)}] \textbf{One Component:} The star formation in the primary emission line feature is significantly stronger than any secondary emission line feature. 
The threshold used is $F_\textrm{p} > 2.5 \times F_\textrm{s}$, where $F_\textrm{p}$ and $F_\textrm{s}$ is the estimated aperture flux of the primary and secondary emission line feature, respectively.
For the pre-merging systems the primary emission line feature corresponds to one of the multiple objects and for the post-mergers it refers to either the nuclear region or a tidal feature.
\item[\textbf{2)}] \textbf{Both (All) Components:} The star formation is pronounced/detected in all (or the majority if more than two) components of the system, i.e., $F_\textrm{p} < 2.5 \times F_\textrm{s}$.
\item[\textbf{3)}] \textbf{In-between:} The mapped star formation appears to be emerging from in-between the merging components. None of the post-mergers show this feature, so indeed this means in-between clearly distinguishable objects.
\item[\textbf{4)}] \textbf{Low S/N:} The S/N per pixel of the emission line features in the 2D grism spectrum is too low to produce a convincing emission line map. This may be the case for extended star formation as pointed out in Section~\ref{sec:SEDfit}. It can also be due to very dust-enshrouded star formation making the emission lines very weak.
\end{enumerate}
Each of the four rows of star formation maps in Figure~\ref{fig:ELmap} show examples of these four SF types.
The results from characterising the two classes of mergers with these four SF types are shown in Figure~\ref{fig:SFsplit}. The error-bars are obtained by bootstrapping the results, i.e., by randomly drawing 60 SF types from the results 1000 times and then using the 2$\sigma$ width of the resulting SF type distributions as error-bars (hence no error-bar on the post-merger SF type 3 in Figure~\ref{fig:SFsplit}, as none were found). 
For both the pre- and post-mergers, the star formation is most prominent in just one of the components (SF type 1) for roughly 3/5 of the objects. In roughly 1/3 of the objects, star formation was detected in all components (SF type 2). Hence, the distribution of the spatial extent of star formation among the pre- and post-mergers is consistent. In Table~\ref{tab:res} we have listed the fractions for all 60 3D-HST mergers resulting from the classification of the emission line maps.

The difference in the rates of objects with prominent star formation in just one component (SF type 1) and and mergers with star formation of type 2 might be a consequence of dust obscuration. 
We are only able to probe the unobscured star formation, so in cases where one component (or the tidal feature) is much more dust-obscured than the other we would end up with star formation maps of type 1.
The discrepancy between the rates could also be due to different SFRs in the different components. 
Since the mergers are selected based on morphology and we do not have any kinematic information, the fraction of SF type 1 objects might be biased by chance superpositions of objects on the sky at different redshifts, such that we only see line emission from one object in the NIR. 
As described in Section~\ref{sec:visclasEL} we expect approximately 4-7\% of such objects due to the modest distances of $\sim$8.5~kpc involved.

The results indicate that most mergers happen between objects of different gas fractions and/or different SFR, i.e., two merging components with significantly different properties. We will show below that this is backed up by an initial comparison with simulated mergers. 

In Figure~\ref{fig:SFRvsM} the different SF types are represented by different symbols to look for dependencies between the morphology of the star formation maps and SFR, sSFR, $z_\textrm{grism}$, and $M_*$. These quantities are obtained on the \emph{total} photometry of the merging components as only a single photometric ID was assigned to each merger and hence includes the flux of the total system. 
The absence of correlations seen in Figure~\ref{fig:SFRvsM} suggests that all star formation morphologies occur at all redshifts irrespective of SFR and mass.

\begin{table}
\centering{
\caption[ ]{The Spatial Extent of Star Formation}
\label{tab:res}
\begin{tabular}[c]{|l|rr|rr|}
\hline
  \multicolumn{1}{|c|}{SF type} &
  \multicolumn{2}{c|}{3D-HST} &
  \multicolumn{2}{c|}{Simulations} \\
\hline
1) One Comp.			& 58$^{+12}_{-13}$\%	& (35/60) 	& 28$^{+5}_{-5}$\% 	& (83/296) \\
2) Both (All) Comp.		& 32$^{+12}_{-12}$\% 	& (19/60) 	& 59$^{+6}_{-6}$\% 	& (175/296) \\
3) In-Between Comp.	& 3$^{+5}_{-3}$\% 	& (2/60) 	& 0$^{+0}_{-0}$\% 	& (0/296) \\
4) Low S/N per pixel			& 7$^{+8}_{-5}$\% 	& (4/60)	& 13$^{+4}_{-4}$\% 	& (38/296)\\
\hline
\multicolumn{5}{l}{Uncertainties are obtained via bootstrapping as described in the text.}\\
\multicolumn{5}{l}{See Figure~\ref{fig:comp} for a plot of these values.}
\end{tabular}}
\end{table}

\section{Simulating 3D-HST Spectra}\label{sec:SimspecEL}

With the exceptional data of the 3D-HST merger sample presented above, we can perform comparisons with the star formation produced in simulated mergers at high redshift. 
Current high-resolution merger simulations are able to predict the spatial distribution of star formation in mergers and produce simulated images by including the emission by young, newly-formed stars and the transfer of starlight through gas and dust \citep{Cox:2006p3233,Cox:2008p3243,Jonsson:2010p3216}.
\cite{Lotz:2008p12474,Lotz:2008p24286,Lotz:2010p24059,Lotz:2010p24041} used these and similar simulations to describe the correlations between the rest-frame optical morphology of mergers, their mass ratios, total star formation, projected size, gas fractions, and merger time-scales, enabling comparisons with observations.

Simulations have shown examples of star formation triggered by direct galaxy interaction, star formation originating in the central cores of the individual merging components, and star formation appearing in tidal features. In the nearby Universe, many examples of such features have been found, e.g., in the Antennae Galaxies \citep{Wang:2004p21685}. Whether the predictions of simulations are also representative of the star formation in mergers at  $z\sim1.5$, 
i.e., when the Universe was only 3-4 Gyr old, have not been tested yet.
With 3D-HST, the sample size of data with both rest-frame optical morphology \emph{and} star formation morphology is becoming large enough that we can start looking at a \emph{population} (snapshot) of mergers instead of individual case studies, and hence comparisons with predictions from simulations become feasible. In this section we make an initial illustrative attempt at comparing state-of-the-art merger simulations with the observational results from 3D-HST. 
In the following we describe how we create simulated WFC3 G141 grism spectra from a small initial set of high resolution simulated mergers.

\subsection{SPH Merger Simulations}\label{sec:DC}

Knowledge about both the spatial and spectral extent of the objects is crucial for simulating grism spectra.
As input for our grism simulations we use three-dimensional data cubes of simulated mergers. These input data cubes are based on the simulations described in \cite{Younger:2008p24037} and are designed in the spirit of the \cite{Cox:2006p3233,Cox:2008p3243} simulations.
The simulations are $N$-body/SPH simulations from the GADGET code \citep{Springel:2001p12527} from which simulated observations are generated with the radiative transfer code SUNRISE \citep{Jonsson:2006p3306,Jonsson:2010p3216}. For our initial comparison of simulations with actual data, we use the three models listed in Table~\ref{tab:sims}. The three simulations are all of mass ratio 2:1 and the two merging objects in each simulation both have a gas fraction ($f_\textrm{gas}$) of 40\%, making them `wet' mergers like the 3D-HST systems and mimicking the suggested gas fraction of $z>1$ galaxies \citep{Daddi:2010p24304,Tacconi:2010p24301}. The three mergers happen at three different relative orbital inclinations ($\Theta$) of the two merging components. Hence, what we test here is the effect of orientation on the detected star formation distribution. We further have a series of time steps or snapshots of each galaxy ($N_t$) and a set of viewing angles ($N_\textrm{view}$) that we can simulate spectra for. Thus, the simulation parameter space sampled here is spanned by time, orientation of the two mergers, and viewing angle. As described below, we combine this with a set of parameters determined by the 3D-HST sample to define the total parameter space to be covered by the simulated spectra.
%

\begin{table}
\centering{
\caption[ ]{The Simulated 3D Input Data Cubes}
\label{tab:sims}
\begin{tabular}{|l|r|r|c|c|r|}
\hline
  \multicolumn{1}{|c|}{Name} &
  \multicolumn{1}{c|}{$N_t$} &
  \multicolumn{1}{c|}{$N_\textrm{view}$} &
  \multicolumn{1}{c|}{Mass ratio$^\star$} &
  \multicolumn{1}{c|}{$f_\textrm{gas}$} &
  \multicolumn{1}{c|}{$\Theta$} \\
\hline
Sim. 1   & 6	& 3 	& (Sb) 2:1 (Sc) 		& 0.4:0.4	& $30^\circ$ 	\\ 
Sim. 2   & 5 	& 3	& (Sb) 2:1 (Sc)		& 0.4:0.4	& $90^\circ$ 	\\ 
Sim. 3   & 3 	& 3	& (Sb) 2:1 (Sc)		& 0.4:0.4	& $150^\circ$ 	\\ 
\hline
\multicolumn{6}{l}{$^\star$ Parenthesis give disc model \citep{Younger:2008p24037}.} 
\end{tabular}}
\end{table}

\subsection{Simulating 3D-HST grism Spectra}
\label{sec:selfctEL}

Simulating the grism spectra from the merger simulation output data cubes presented above is fairly straightforward.
The mock grism spectra are created by dividing the data cube into wavelength slices, i.e., images corresponding to a `filter' of width $\Delta\lambda$. Offsetting or dispersing this sequence of images and co-adding the fluxes results in a grism spectrum. Using this approach, we turn the input data cubes into  a sequence of simulated WFC3 G141 grism spectra. We use a pixel scale of $0.06''$ and a spectral resolution of $\Delta\lambda=22\textrm{\AA}$ according to the 3D-HST grism spectra characteristics described in \cite{Brammer:2012p12977}. For the HST PSF we use a Tiny Tim PSF\footnote{http://www.stsci.edu/hst/observatory/focus/TinyTim} and for the system throughput we used the G141 sensitivity curve.

We use the parameter space of the observed merger sample from 3D-HST to define the parameter space of the simulations.
By re-scaling each data cube we simulate grism spectra corresponding to the 16th and 84th percentiles of the magnitude ($m_\textrm{F140W}$), redshift ($z_{grism}$), and SFR distributions of the 3D-HST data as indicated by the dashed lines in the corresponding histograms in Figure~\ref{fig:mhist} and \ref{fig:SFRvsM}. 

To adjust the SFR of the data cubes before turning them into grism spectra, we assume that the SFR to H$\alpha$-flux conversion follows the empirical relation of \cite{Kennicutt:1994p11885}:
\BE
 \frac{\textrm{SFR}}{[M_\odot/\textrm{yr}]} = \frac{1}{1.26\times10^{41}}\; \frac{L_{H\alpha}}{[\textrm{erg/s}]}   \; .
\EE
This relation assumes a \cite{Salpeter:1955p24401} initial mass function and is applied to the emerging SFR, i.e., after the effects
of extinction are taken into account by SUNRISE.
The $m_\textrm{F140W}$ is obtained by scaling the flux (including emission lines) integrated over the F140W passband independently. 
Combining the values from Table~\ref{tab:sims} with these data-determined parameters, we end up with a sample of 336 simulated spectra spanning the parameter space $(N_t,N_\textrm{view},\Theta,m_\textrm{F140W},z,\textrm{SFR})$. 

Added to the raw mock grism spectra are Poisson noise and noise terms corresponding to the read noise and dark current quoted on the WFC3 website, as well as the background sky levels presented in \cite{Brammer:2012p12977}.

\begin{figure*}
\centering{
\includegraphics[width=0.35\textwidth]{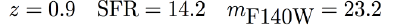}\\
\includegraphics[width=0.15\textwidth]{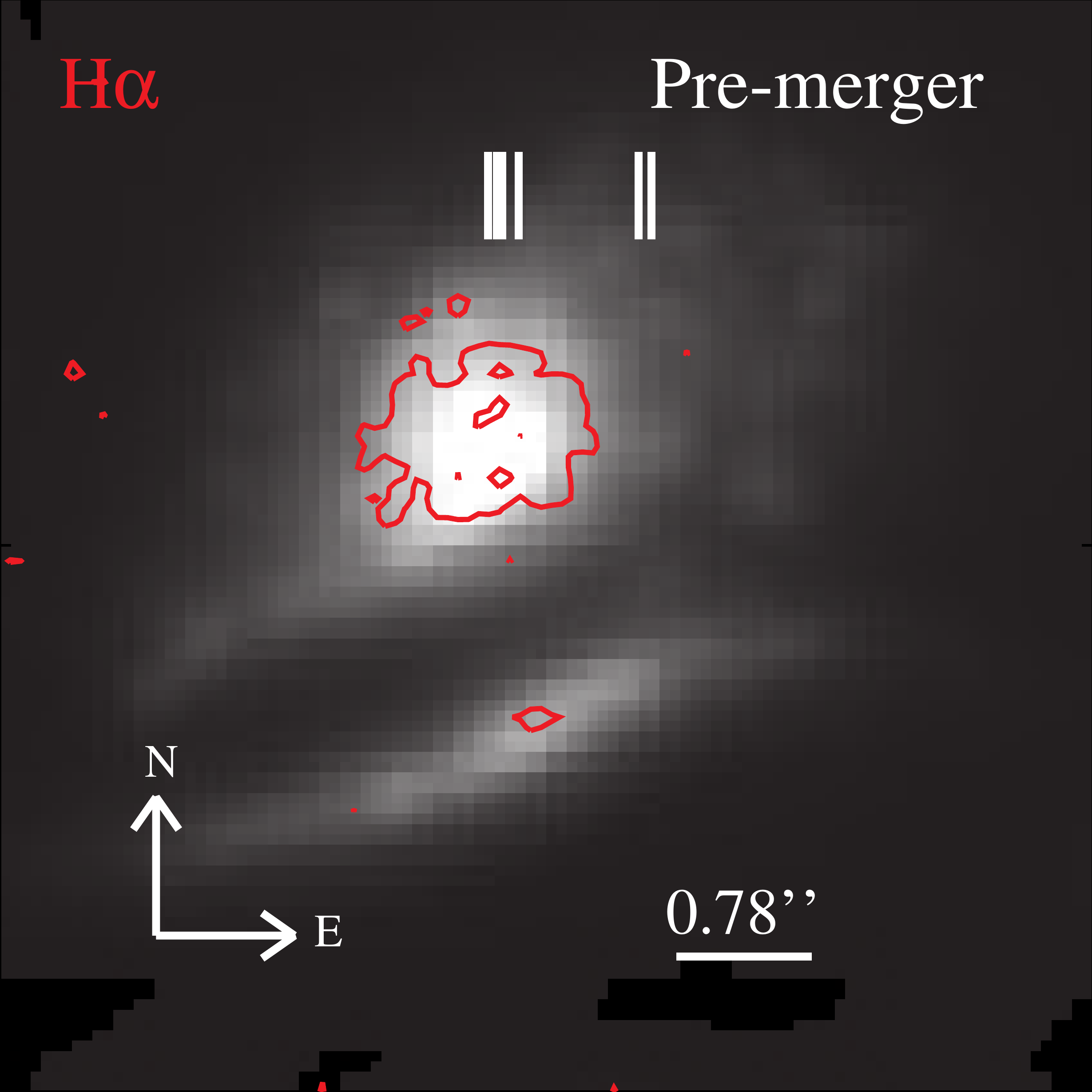}
\hspace{0.25cm} 
\includegraphics[width=0.61\textwidth,bb=30 25 258 88,clip]{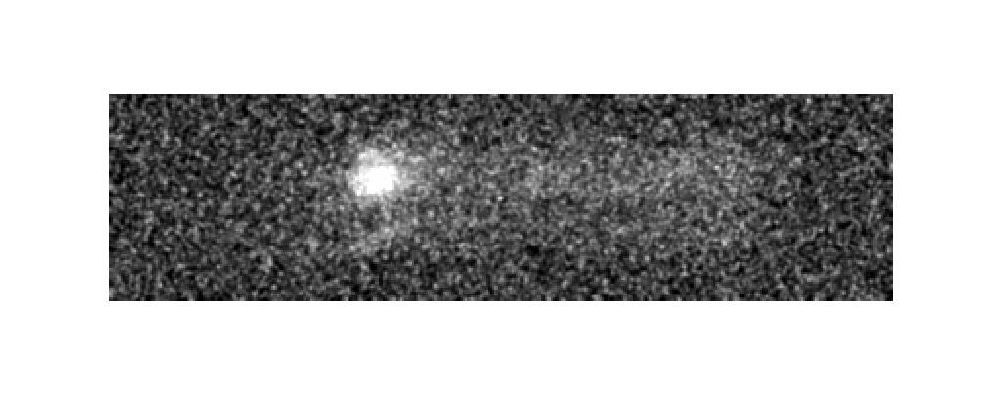} \\ 
\vspace{0.4cm} 
\includegraphics[width=0.35\textwidth]{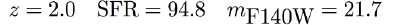}\\
\includegraphics[width=0.15\textwidth]{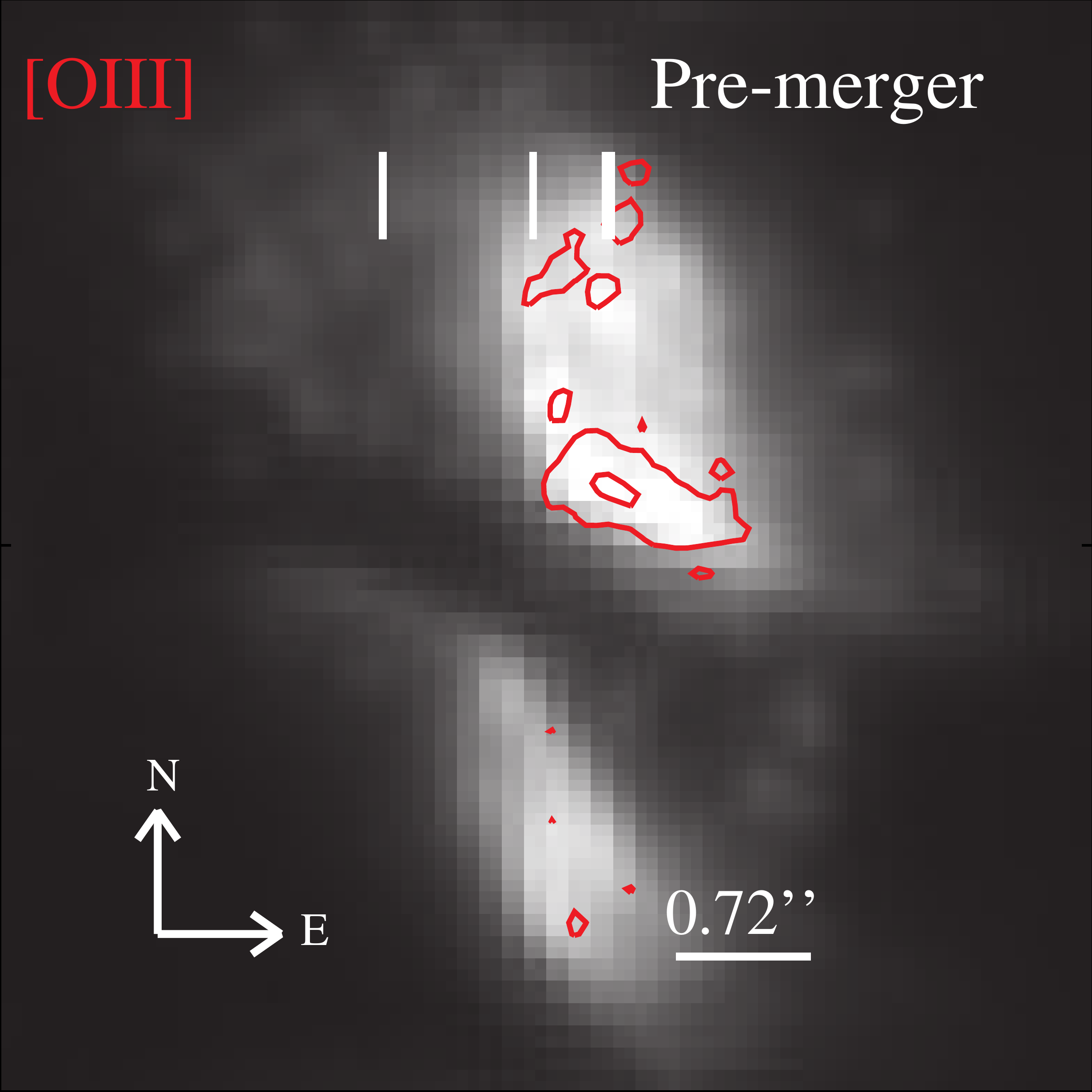}
\hspace{0.25cm}
\includegraphics[width=0.61\textwidth,bb=30 25 258 88,clip]{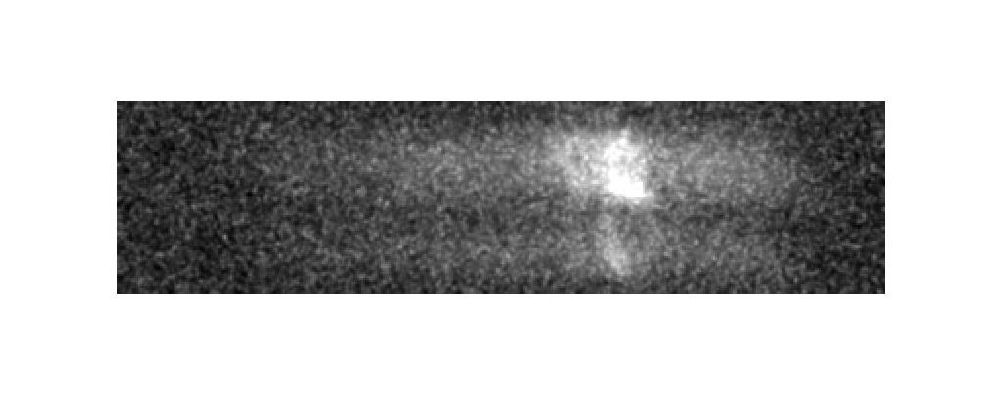}\\ 
\vspace{0.4cm} 
\includegraphics[width=0.35\textwidth]{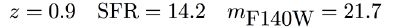}\\
\includegraphics[width=0.15\textwidth]{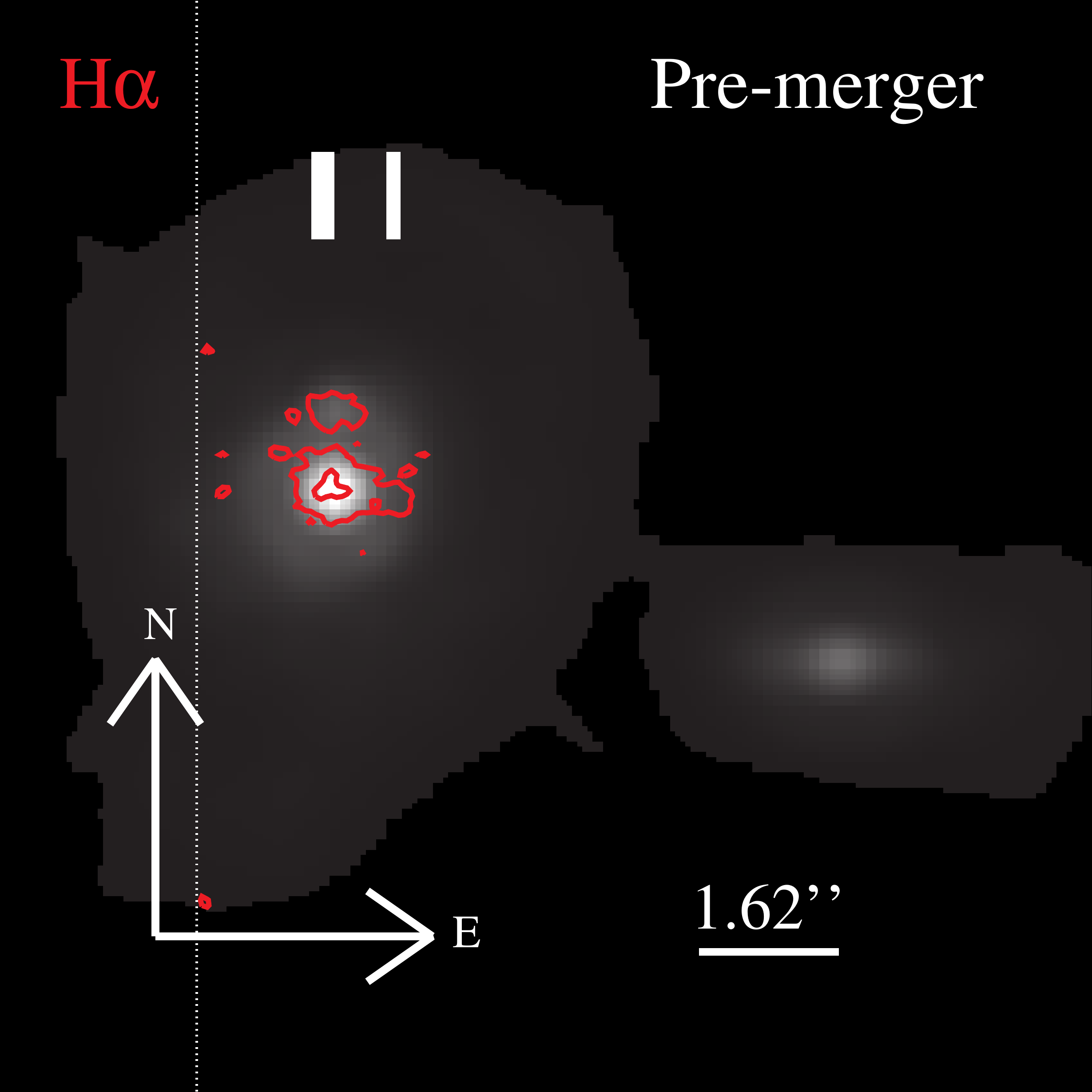}
\hspace{0.25cm}
\includegraphics[width=0.40\textwidth]{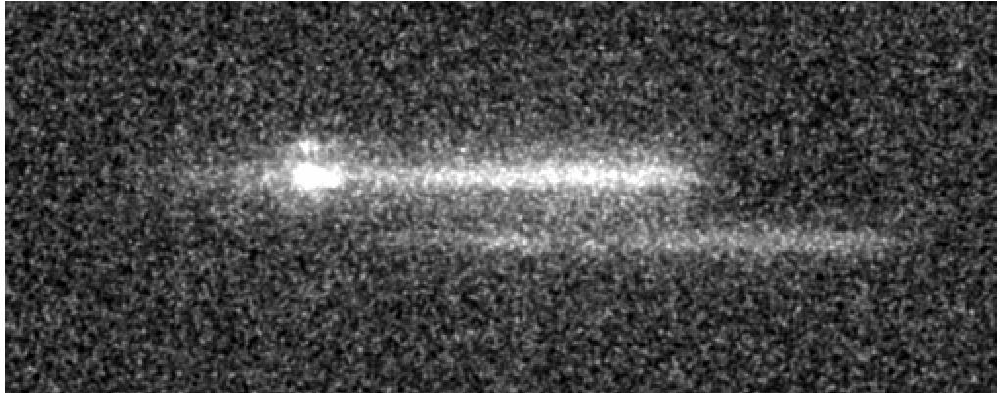} \\
\vspace{0.4cm} 
\includegraphics[width=0.35\textwidth]{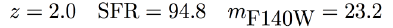}\\
\includegraphics[width=0.15\textwidth]{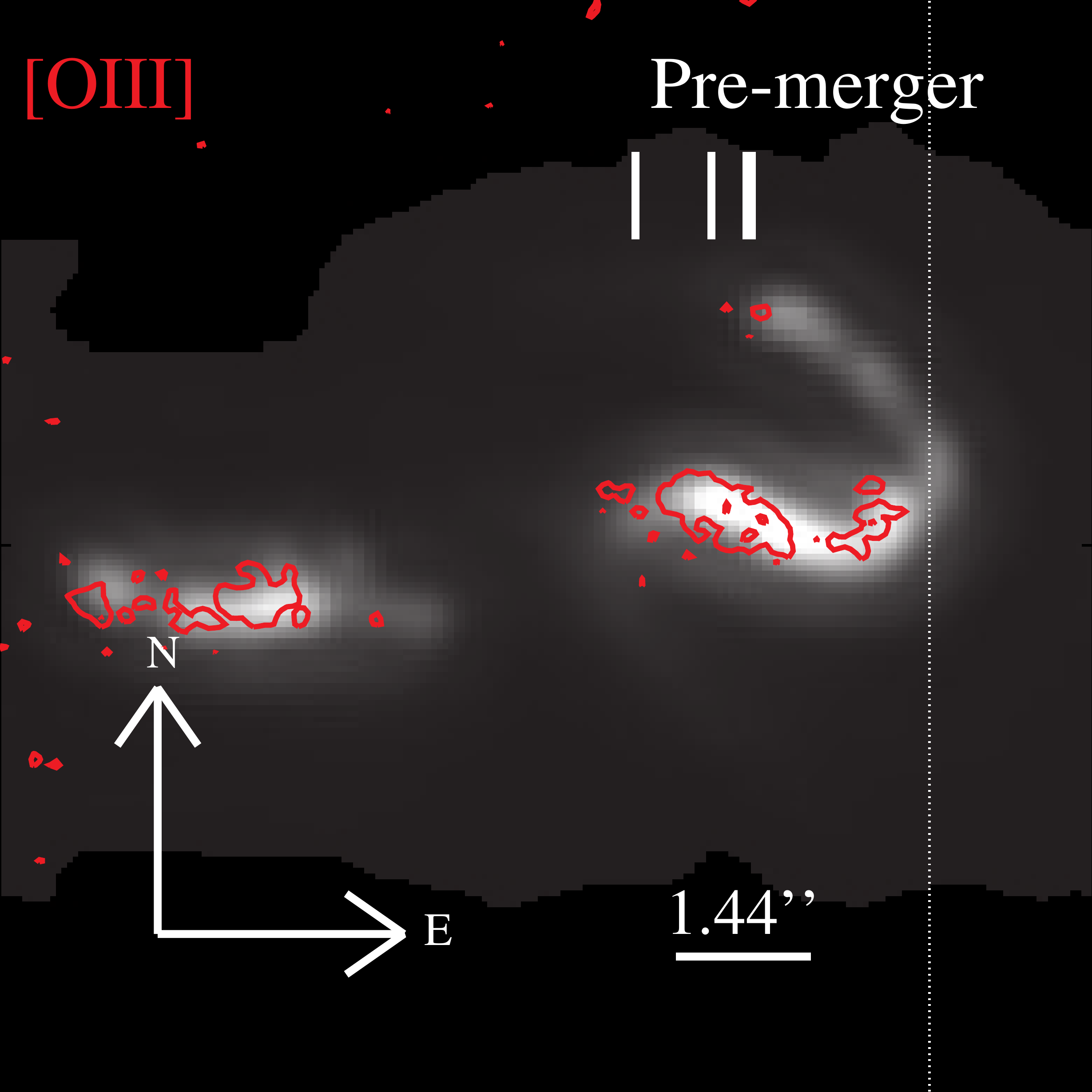}
\hspace{0.25cm}
\includegraphics[width=0.40\textwidth]{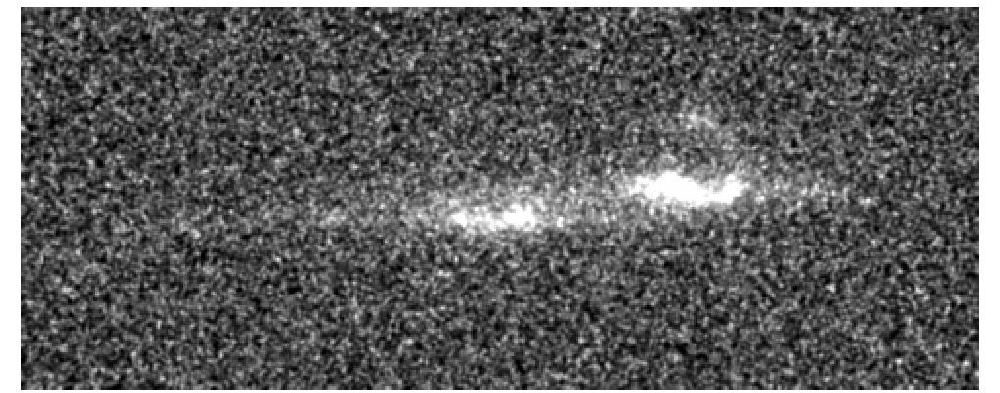}\\ 
\vspace{0.4cm} 
\includegraphics[width=0.35\textwidth]{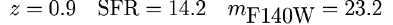}\\
\includegraphics[width=0.15\textwidth]{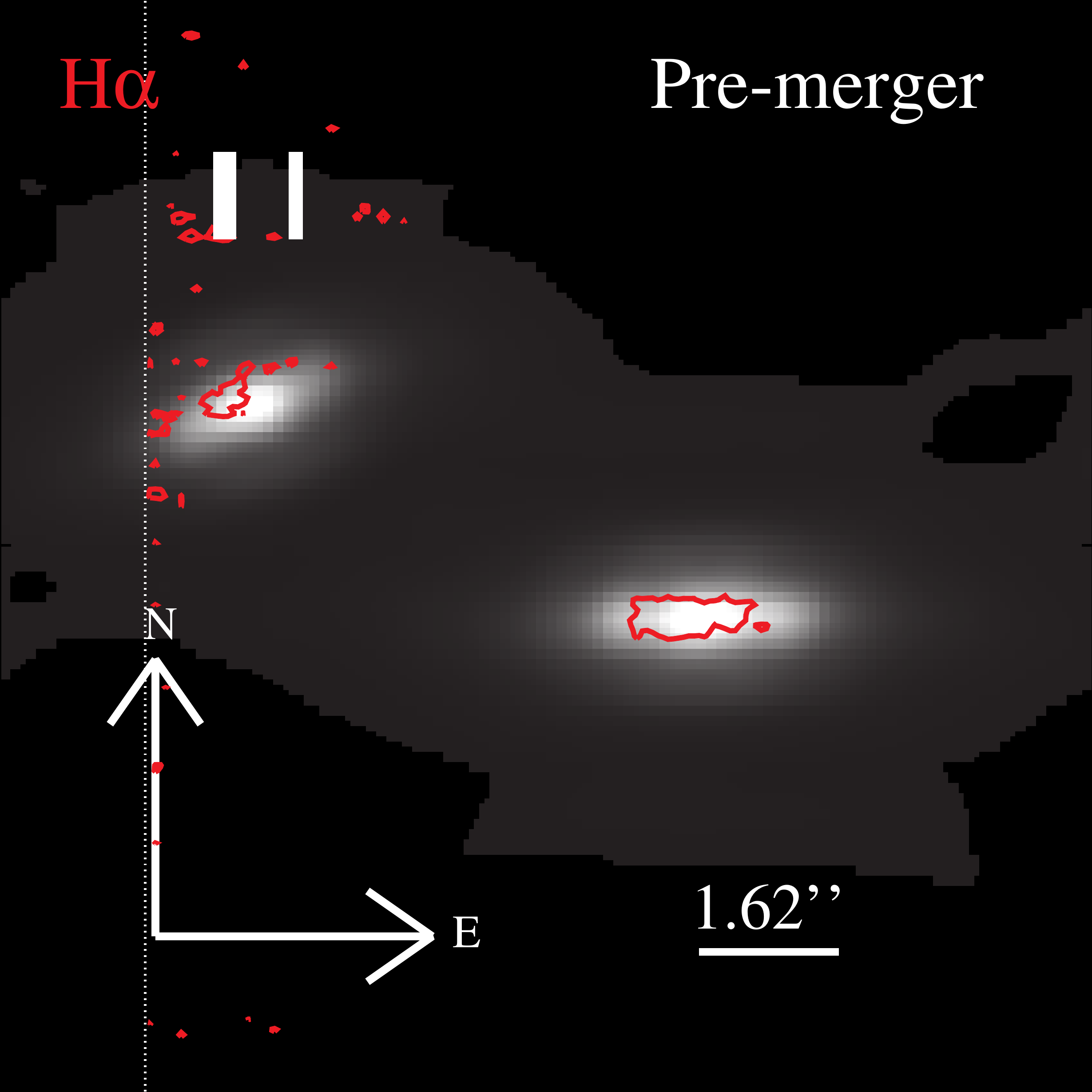}
\hspace{0.25cm}
\includegraphics[width=0.40\textwidth]{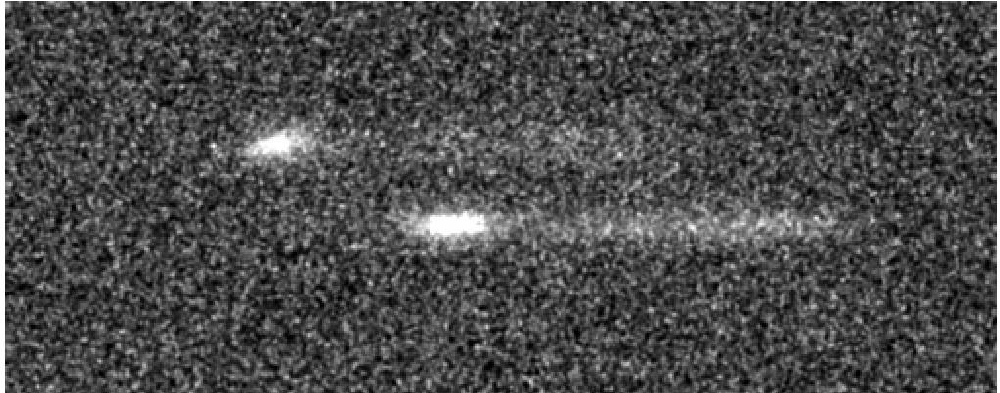}\\ 
\vspace{0.4cm} 
\includegraphics[width=0.35\textwidth]{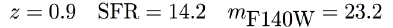}\\
\includegraphics[width=0.15\textwidth]{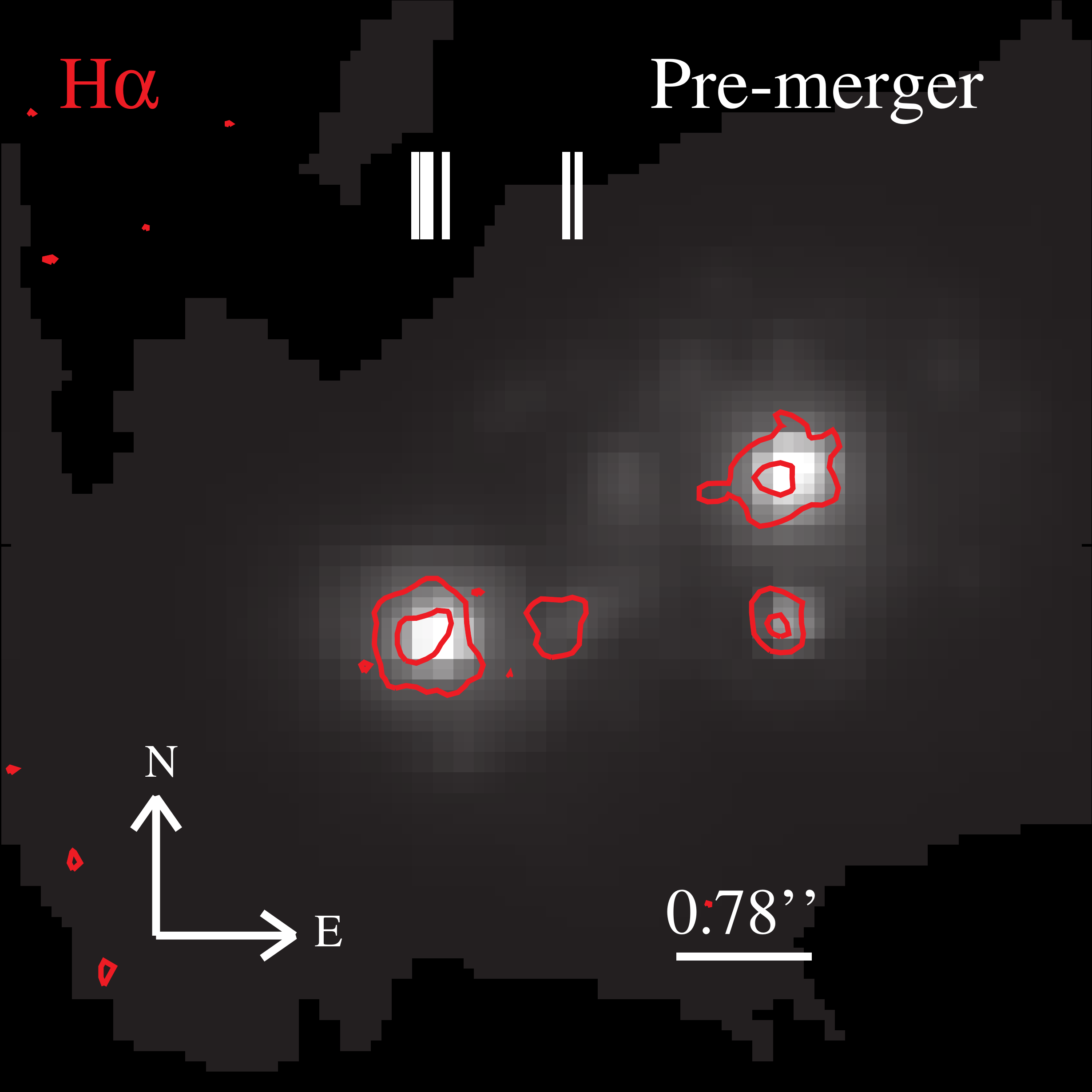}
\hspace{0.25cm}
\includegraphics[width=0.61\textwidth,bb=30 25 258 88,clip]{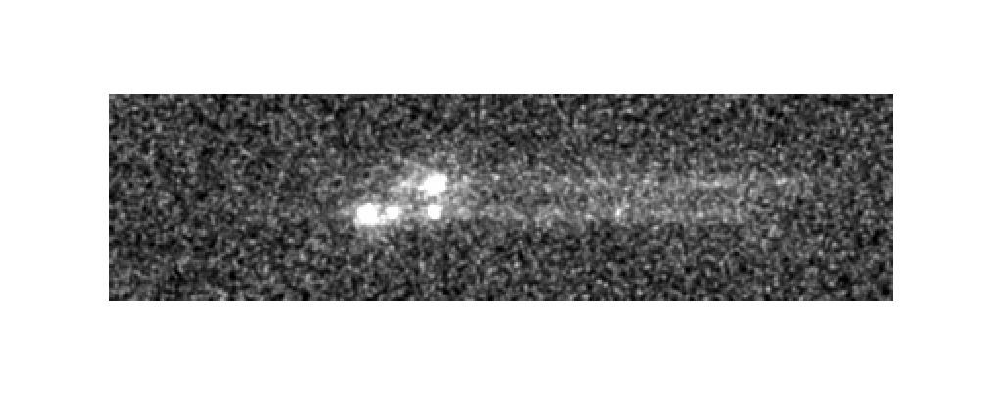} 
\caption{Examples of simulated 3D-HST data based on simulated observations of N-body/SPH mergers generated with the radiative transfer code SUNRISE. The thumbnails (left) are emission line maps similar to the ones shown for the 3D-HST data in Figure~\ref{fig:ELmap}. The first three spectra (right) result in star formation maps of SF type 1, while the last three result in SF type 2 star formation maps.
Various combinations of SFR, $m_\textrm{F140W}$, $z$, time-step in the merging of the components ($N_t$), viewing angles ($N_\textrm{view}$), and initial relative orbital inclination of the two merging components ($\Theta$) are shown.
}
\label{fig:simspec}}
\end{figure*}

In Figure~\ref{fig:simspec} we present a sequence of simulated G141 grism spectra. The spectra shown have various combinations of $N_t$, $N_\textrm{view}$, $\Theta$, $m_\textrm{F140W}$, $z$, and SFR.
The thumbnails on the left are the noise-free simulated F140W images of the objects. 

\subsection{Creating Star Formation Maps of Simulated Spectra}

To create the star formation maps for the simulated spectra, we treat them in exactly the same way as the actual 3D-HST spectra (see Section~\ref{sec:ELmapEL}). First, we visually inspect the spectra to make sure that an emission line feature is available; for some of the faint magnitude, low-SFR combinations the emission lines do not show up in the noise-added spectra. Having eliminated spectra without emission line features, as well as cases of viewing angle and time step were the two merging objects could not be distinguished, the sample of 336 simulated spectra is reduced to 296. For these 296 spectra, we create emission line maps by the method described in Section~\ref{sec:ELmapEL}.

As for the actual 3D-HST data, we characterise the 296 emission line maps of the simulated spectra according to the four SF types described in Section~\ref{sec:resultsEL}. The results from this classification are shown in Table~\ref{tab:res} together with the results for the 3D-HST mergers.
The simulated spectra in Figure~\ref{fig:simspec} show examples of both spectra resulting in star formation maps of type 1 (first three spectra) and type 2 (last three spectra).

\subsection{Comparing the 3D-HST Data with Simulations}
\label{sec:resultsCOMP}

\begin{figure}
\centering{
\includegraphics[width=0.49\textwidth]{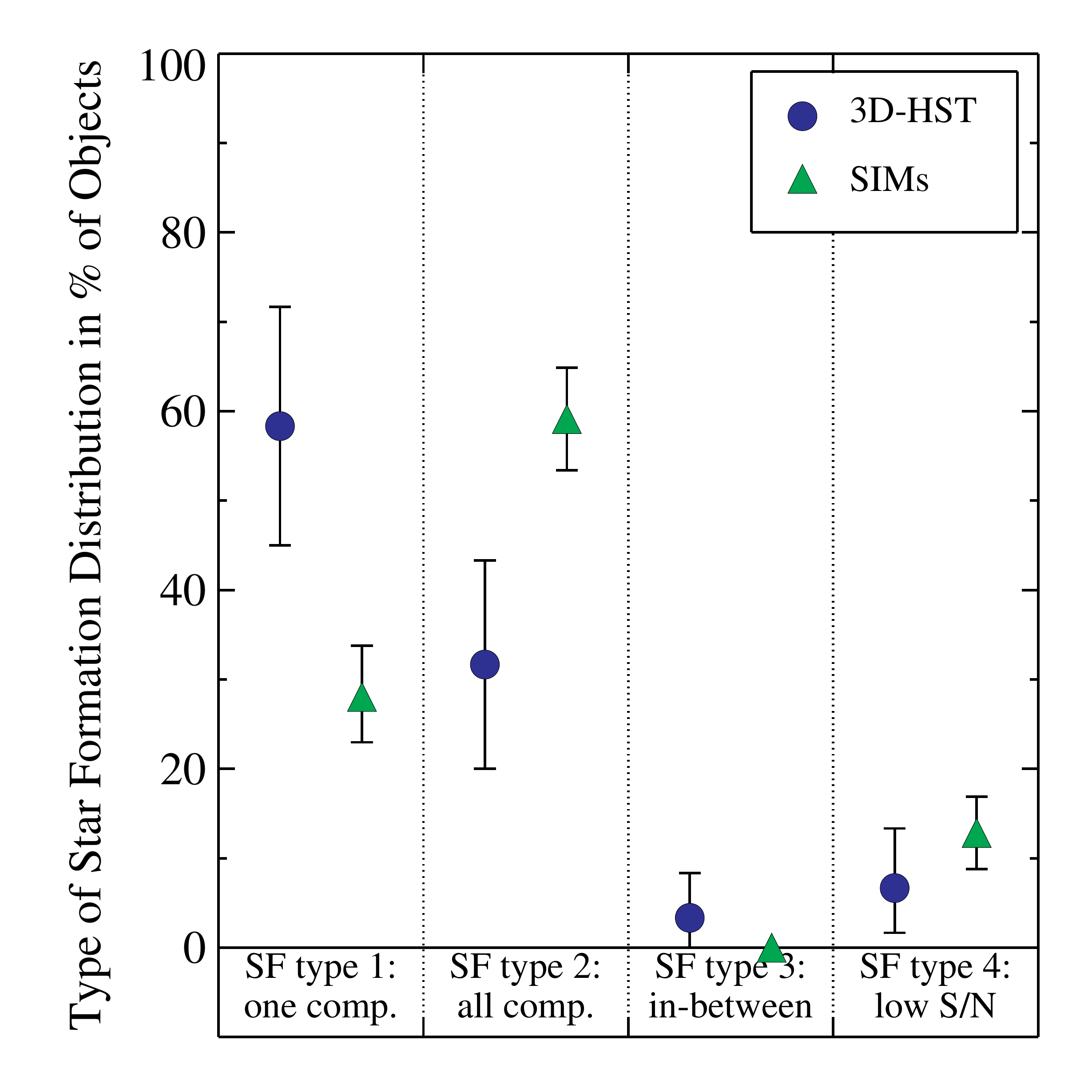}
\caption{Comparing the spatial extent of star formation (SF type; Section~\ref{sec:resultsEL}) in the 60 3D-HST mergers (blue circles) with the 296 simulated spectra (green triangles).  Each point has been assigned an error-bar obtained via bootstrapping. Direct comparisons should be done with care due to the limited size of the parameter space the simulations span. Nevertheless, it is evident that the simulations show many more cases where star formation is seen in both the merging components than the 3D-HST data do, indicating that the majority of mergers have different gas fractions and/or mass ratios prior to merging.}
\label{fig:comp}}
\end{figure}

We compare the 3D-HST SF types we inferred in Section~\ref{sec:resultsEL} with the simulated SF types in order to investigate how, e.g., SFR and viewing angle influences the frequencies of different types of mergers from the  characterisation of the extent of star formation in mergers. 
In Figure~\ref{fig:comp} the SF types of the full sample of 60 3D-HST mergers (blue circles) are shown together with the SF types of the 296 simulated spectra of the 3 mergers from Table~\ref{tab:sims} (green triangles). The plotted percentages are given in Table~\ref{tab:res}. The error-bars are again obtained via bootstrapping.
The comparison needs to be done with care, as the sample of simulated mergers from Table~\ref{tab:sims} does not span the gas fraction and mass ratio dimensions of the parameter space, expected to influence the `observability' of star formation \citep{Lotz:2010p24059,Lotz:2010p24041,Lotz:2011p24053}, as they are fixed at 40\% and 2:1, respectively (see Section~\ref{sec:DC}). Nevertheless, our analysis shows how the 3D-HST survey enables a direct comparison of predictions from simulations with an extensive sample of high-$z$ mergers.

Despite the limitations of he spanned simulation space, it seems that the fraction of spectra where the S/N per pixel is too low to create an actual emission line map is fairly consistent. Likewise, the fraction of cases where the star formation seems to emerge from in-between the merging components (SF type 3) is comparable: in fact we do not find any objects in this category among the simulated spectra.  
Looking at the two SF type 3 cases we find for the data (third row of Figure~\ref{fig:ELmap}) could indicate that the reason we are not seeing star formation from the components themselves is due to dust-obscuration rather than more pronounced star formation in-between the components. To find similar trends in the simulated emission line maps we would need to investigate a range  of gas fractions rather than just the $f_\textrm{gas}=40\%$ cases presented here. This is however beyond the scope of this initial comparison.

From Figure~\ref{fig:comp} it is also evident that the simulated spectra, on average, have twice as many star formation maps where both components have pronounced star formation (SF type 2), i.e., nearly 2/3 of the simulated maps, as compared to $\sim$1/3 for the 3D-HST star formation maps. 
Correspondingly, the fraction of single-component emission line maps is significantly lower for the simulations than for the observations. 
As the simulations assume fairly similar properties between the two merging components (same gas fraction, similar mass in the form of major mergers as opposed to minor mergers, same dust content, and total SFR), this suggests that the majority of mergers do not happen between such galaxies. According to the 3D-HST data, only $\sim$1/3 of real mergers seem to happen between galaxies of similar properties. 
For example, different gas fractions or a larger gap between the masses (minor mergers) could change this picture, in agreement with the results presented in Section~\ref{sec:resultsEL}. 

Also, the distribution of orbits and orientations of the merger components could play a significant role here. The presented simulations consist of both prograde-prograde (Sim. 1), prograde-polar (Sim. 2) and prograde-retrograde (Sim. 3) mergers. We find a deficit of SF type 1 mergers among the Sim. 1 emission line maps, in good agreement with the notion that prograde-prograde mergers generally result in more symmetric star formation \citep{DiMatteo:2007p13565}. 
Hence, several factors might play a role in the difference found between the 3D-HST and simulated emission line (star formation) maps.

\section{Conclusion}\label{sec:concEL}

We have presented a sample of galaxy merger candidates with three-dimensional (R.A., Dec., and $\lambda$) spectroscopy at redshift $z\sim1.5$. The sample consists of 60 morphologically selected mergers from the Hubble treasury slitless grism survey 3D-HST with total masses and star formation rates for the systems derived from multi-wavelength photometry. From the slitless grism spectroscopy we created emission line maps of the rest-frame optical emission lines H$\alpha$ and/or [O{\scshape iii}] as a proxy for the \emph{spatial} extent of the unobscured star formation. 
Our results go towards a comprehensive empirical picture of where star formation happens in galaxy mergers at $z\sim1.5$ when the cosmic star formation and merger rate were at their peak. 

We have also carried out an illustrative comparison of the 3D-HST mergers to recent SPH simulations of galaxy mergers which include star formation and dust extinction. 
This small sample of simulations points the way toward more extensive simulation programs aimed at spanning the full parameter space of the observations.

The main conclusions of the present study are:
\begin{itemize}
\item The spatial distribution of star formation in $z\sim1.5$ candidate mergers shows a broad range of morphologies.Ê It is often concentrated in a single, compact region, but can also be located in tidal tails or in-between the main stellar bodies of the progenitors.
\item In the majority (58$^{+12}_{-13}$\%) of the early-stage, pre-coalescence mergers, the star formation is significantly more pronounced in just one of the merging components.ÊThis is likely due to different gas fractions of the progenitors. Alternatively, dust content may differ in quantity or configuration.
\item The star formation morphologies show no clear correlations with the estimated SFR, sSFR, $z_\textrm{grism}$, and $M_*$, suggesting that all star formation morphologies are present at all epochs irrespective of star formation rate and mass.
\item Simulated mergers among galaxies with similar masses and similar gas fractions typically predict similarly intense star formation in both merger components, at odds with the observations: as opposed to 32$^{+12}_{-12}$\% of the observed mergers, as many as 59$^{+6}_{-6}$\% of the simulated mergers show detectable star formation in both components. Together with the large fraction of early-stage, pre-coalescence mergers with star formation in just one component, this discrepancy supports the notion that $z\sim1.5$ mergers typically occur between galaxies with distinctly different properties such as gas fraction, mass and/or SFR.
\end{itemize}

\section*{Acknowledgments}
We acknowledge funding from ERC grant HIGHZ no. 227749.
This work was funded in part by the Marie Curie Initial Training Network ELIXIR of the European Commission under contract PITN-GA-2008-214227. The work was mainly done while K.B.S. was a member of the International Max Planck Research School for Astronomy and Cosmic Physics at the University of Heidelberg (IMPRS-HD), Germany.
This work is based on observations taken by the 3D-HST Treasury Program with the NASA/ESA HST, which is operated by the Association of Universities for Research in Astronomy, Inc., under NASA contract NAS5-26555.
%

\begin{appendix}

\section{Parametrisation of merger morphology ($GM_{20}CAS$)}
\label{sec:GM20CAS}
To ease comparison between our visually selected merger candidates and the extensive literature selecting mergers based on empirically determined parametric classification schemes, we have estimated the Gini coefficient ($G$), the second-order moment of the brightest 20\% of the light ($M_{20}$), the concentration ($C$), the asymmetry parameter ($A$), and the smoothness parameter ($S$) \citep[e.g.,][]{Lotz:2004p12689,Lotz:2006p17552,Lotz:2008p24286,Conselice:2003p26534,Conselice:2008p22641,Conselice:2009p12404,Papovich:2005p26615,Scarlata:2007p26618} for the objects presented in this paper. We follow \cite{Lotz:2008p24286} when estimating the individual parameters.

\begin{figure*}
\centering{
\includegraphics[width=0.95\textwidth]{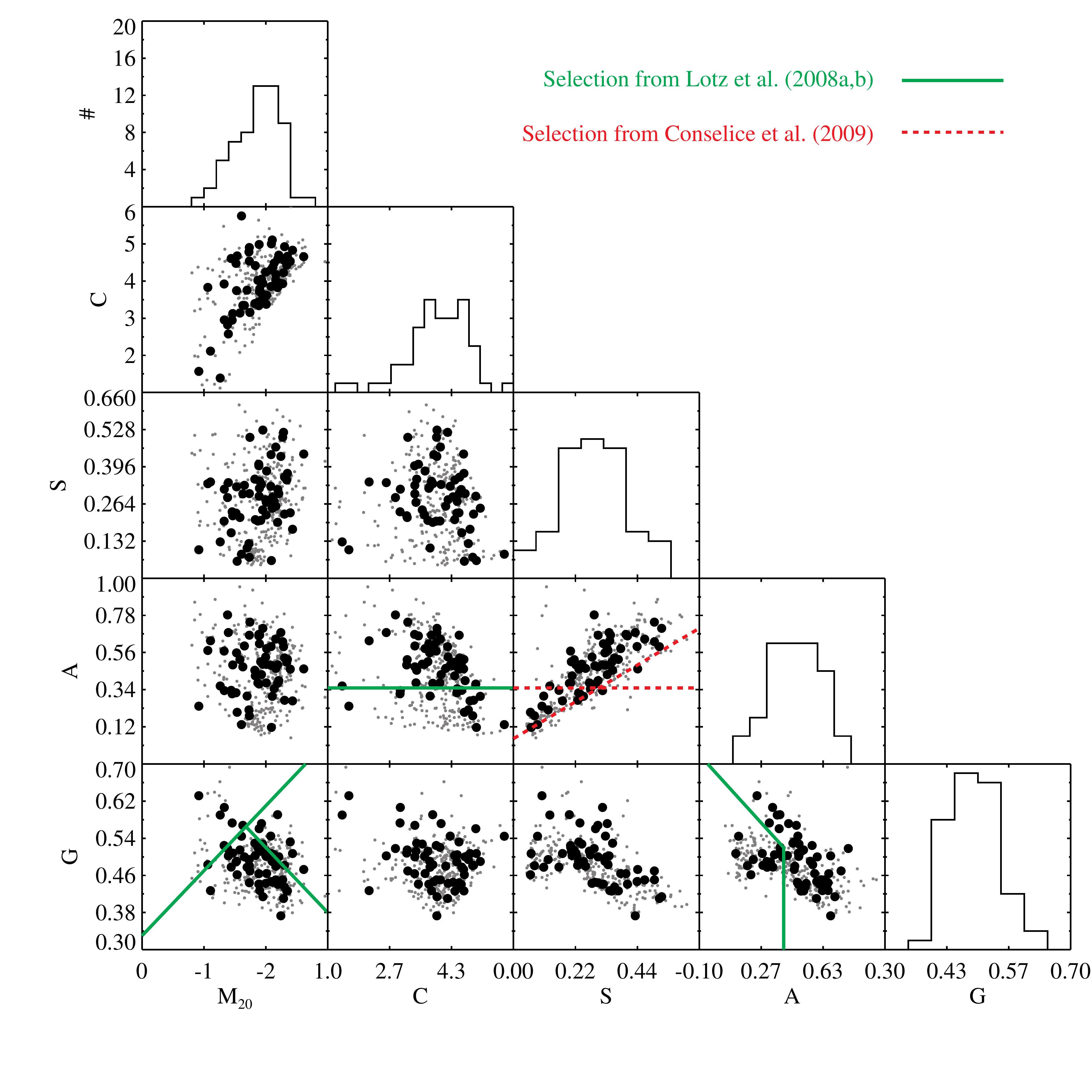}
\caption{The morphological parameters $G$, $M_{20}$, $C$, $A$, and $S$ for the 60 visually selected 3D-HST merger candidates (large black points) and the 292 visually discarded objects (small grey points). The histograms show the distribution of the parameters for the 60 merger candidates. The solid green lines in $A$--$C$, $G$--$M_{20}$, and $G$--$A$-space show the suggested merger selection from \protect\cite{Lotz:2008p12474,Lotz:2008p24286}. The dashed red lines in $A$--$S$-space are taken from \protect\cite{Conselice:2009p12404}. Parametrically selected mergers are believed to lie above and/or to the right of the solid and dashed lines.}
\label{fig:GM20CAS}}
\end{figure*}

In Figure~\ref{fig:GM20CAS} the estimated $GM_{20}CAS$ parameters are shown for the 60 visually selected 3D-HST merger candidates (large black points) and the 292 visually discarded objects (small grey points). 
The histograms show the distributions of the individual parameters of the 60 merger candidates.
The solid green and dashed red lines are from \cite{Lotz:2008p12474,Lotz:2008p24286} and \cite{Conselice:2009p12404}, respectively. Parametrically selected mergers have been shown to preferentially lie above and to the right of these lines. We see that the majority of the visually selected 3D-HST merger candidates indeed have $A>0.35$ as expected. The $G$--$M_{20}$ selection on the other hand seems to reject the majority of the merger candidates selected here. 
The selected merger candidates generally trace the parent population and does not distinct themselves clearly as would be expected for a `clean' merger selection. This is most probably due to the complications that clumpy and irregular morphology of high-$z$ objects and noisy images have on merger selections and in particular on `blind' parametric classification schemes.

\section{The 3D-HST Emission Line (Star Formation) Maps}
\label{sec:maps}

\begin{figure*}
\centering{
\includegraphics[width=0.13\textwidth]{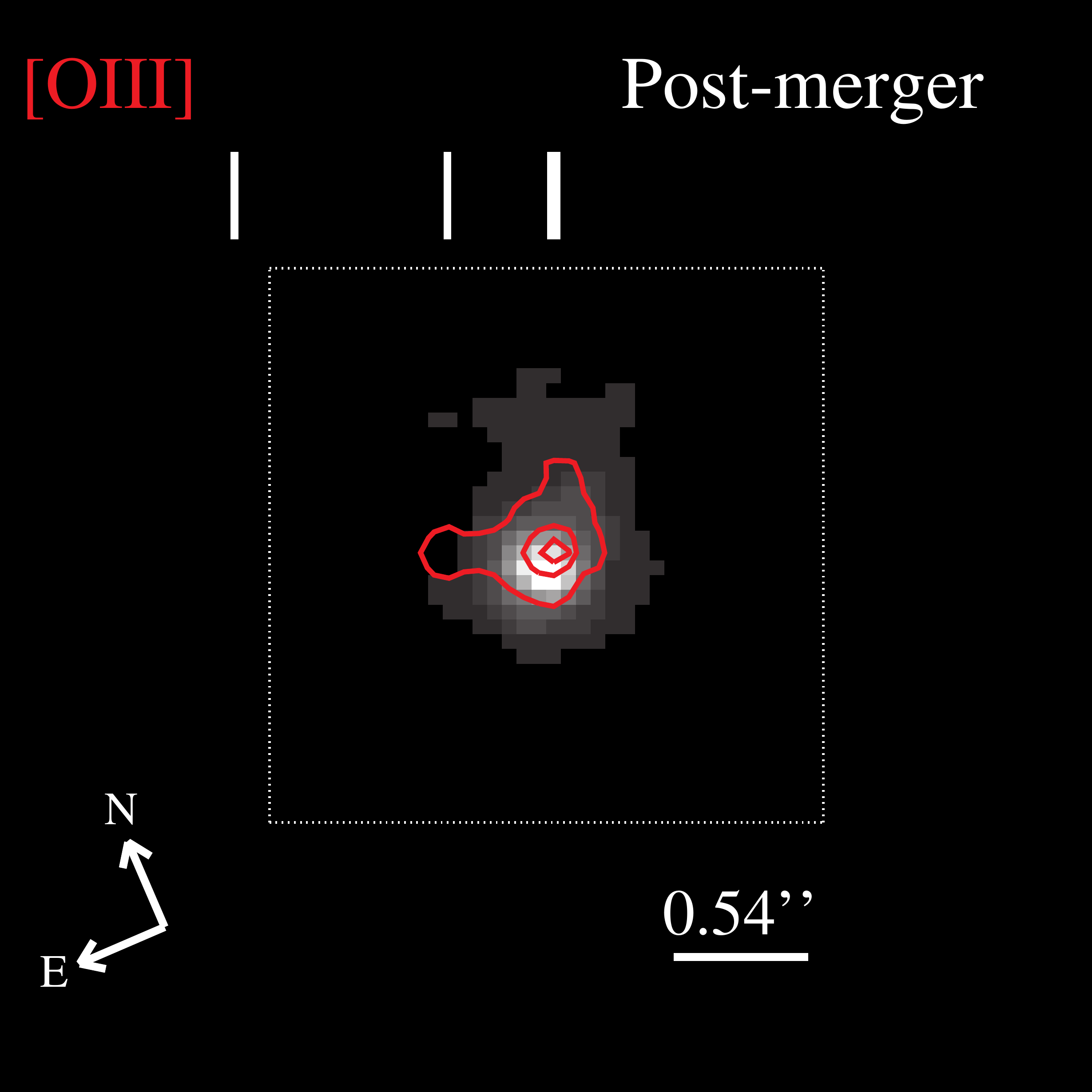}  
\includegraphics[width=0.13\textwidth]{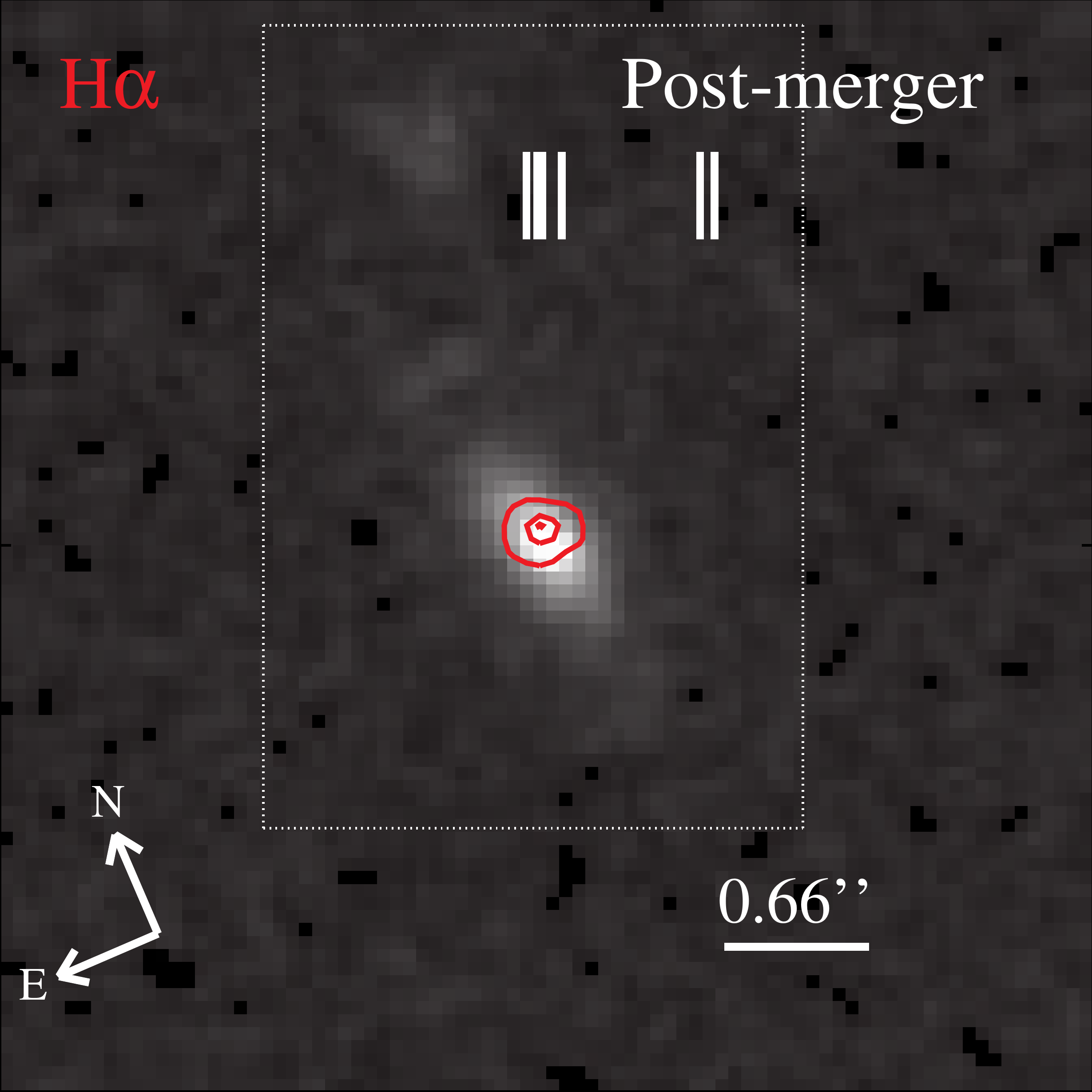}  
\includegraphics[width=0.13\textwidth]{COSMOS-12-G141_01119_lineimgoverlay_byhand.pdf}
\includegraphics[width=0.13\textwidth]{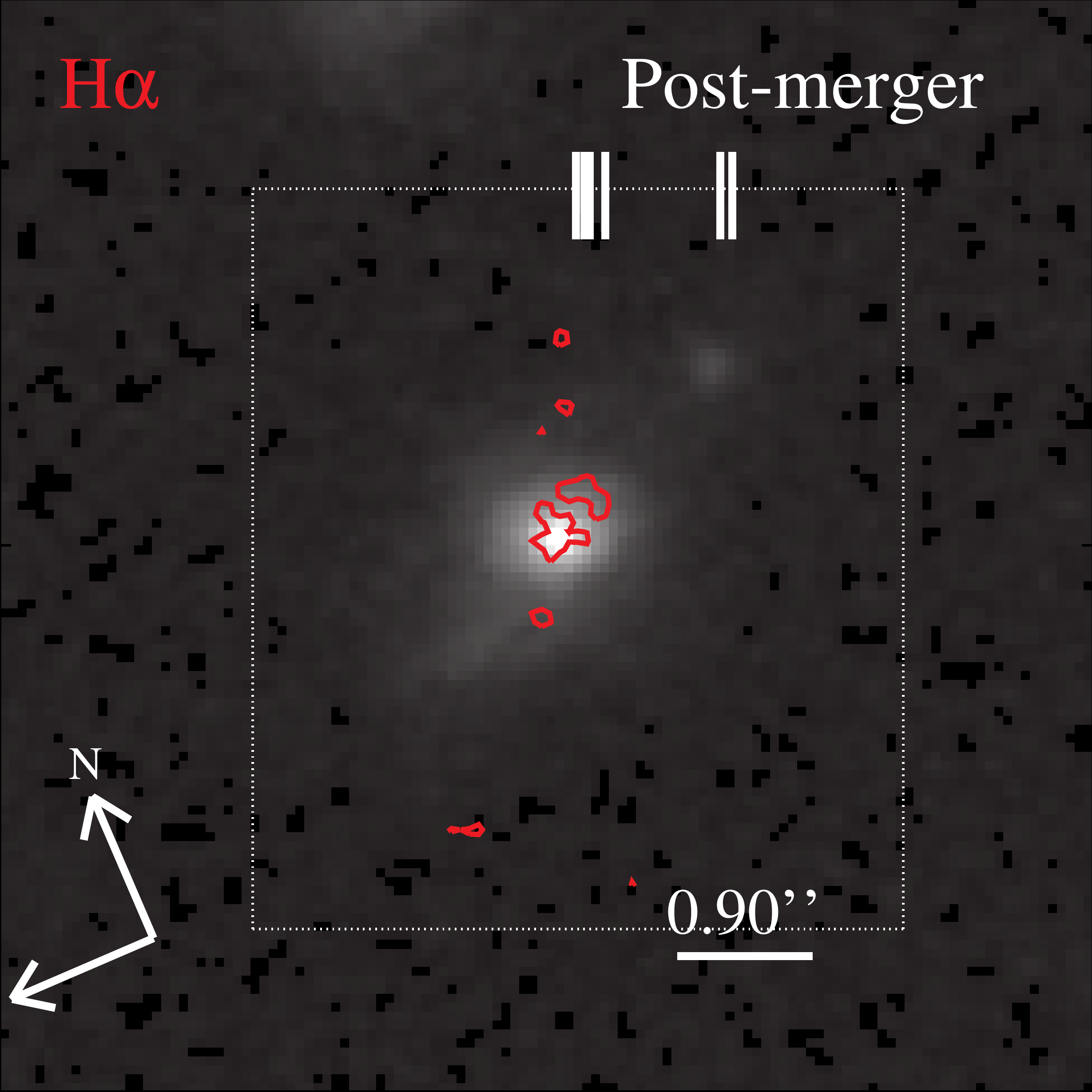}
\includegraphics[width=0.13\textwidth]{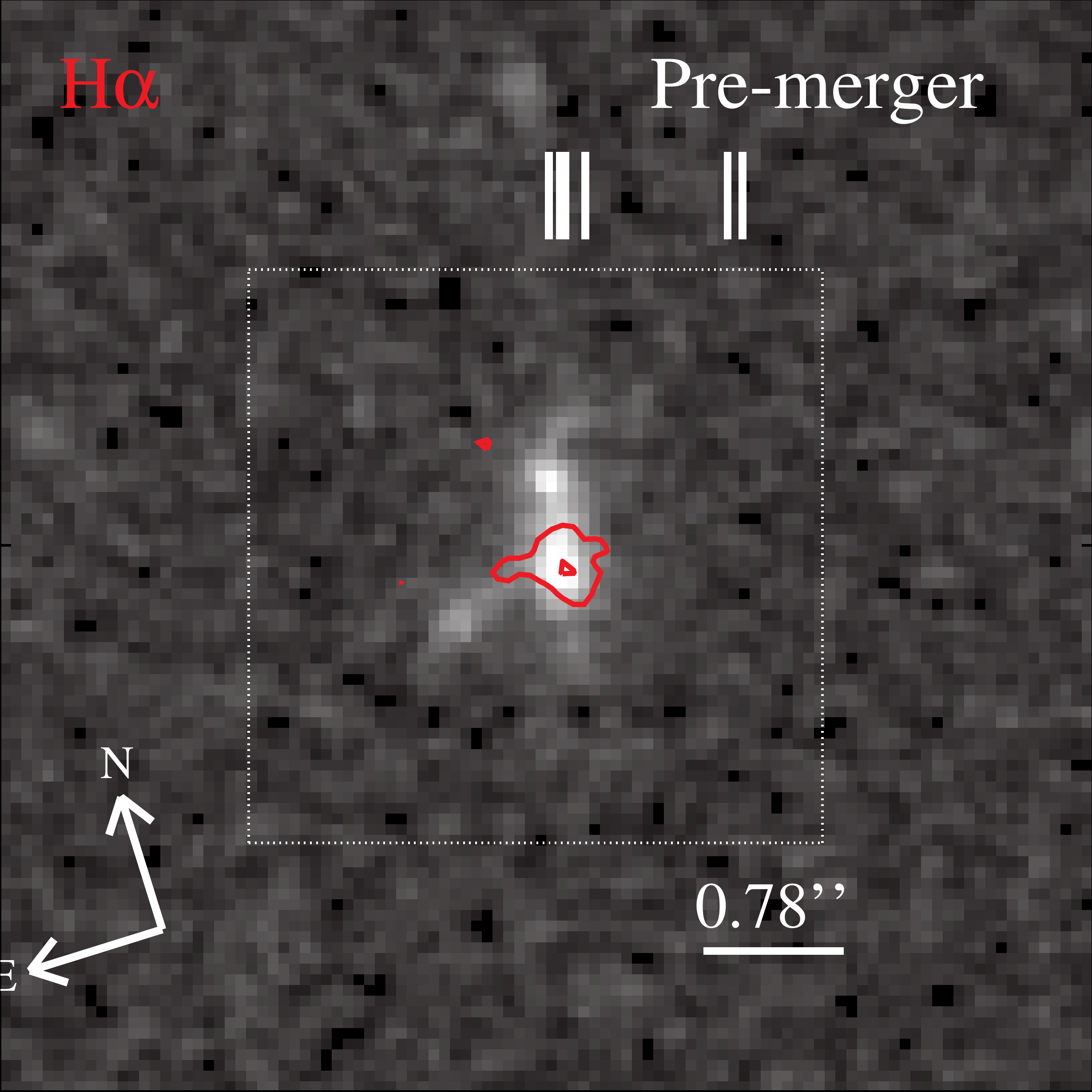}
\includegraphics[width=0.13\textwidth]{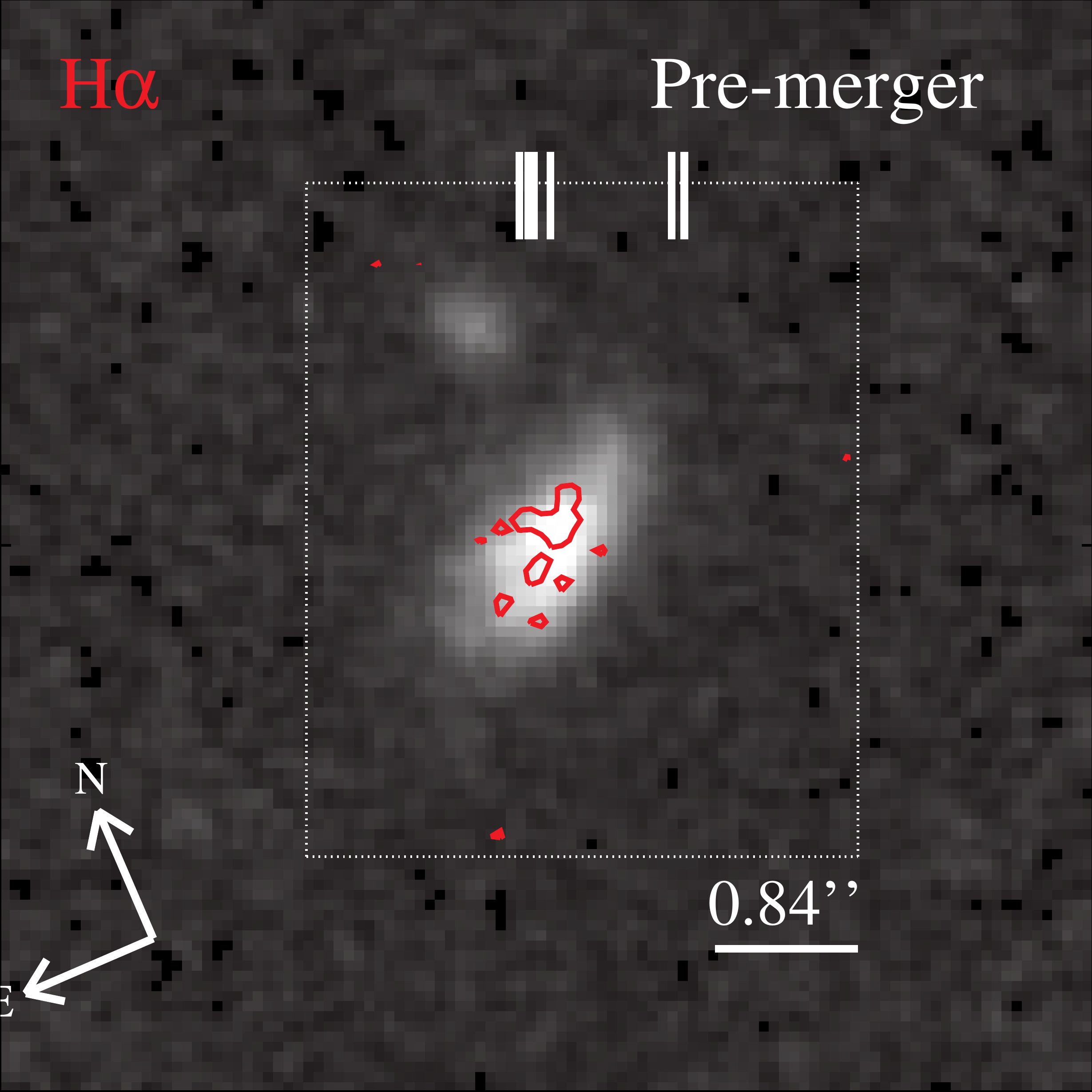}
\includegraphics[width=0.13\textwidth]{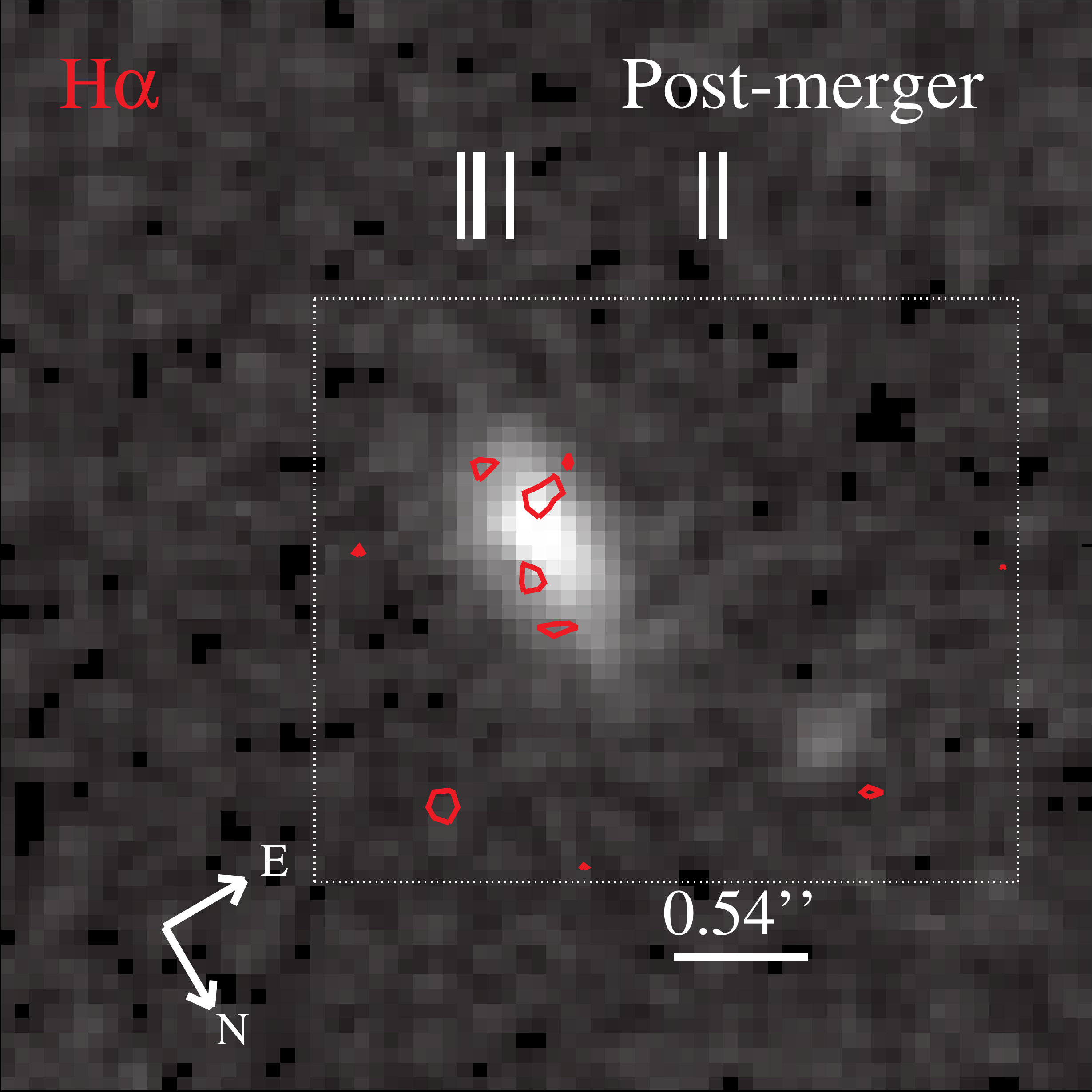}\\
\includegraphics[width=0.13\textwidth]{COSMOS-17-G141_01135_lineimgoverlay_byhand.pdf} 
\includegraphics[width=0.13\textwidth]{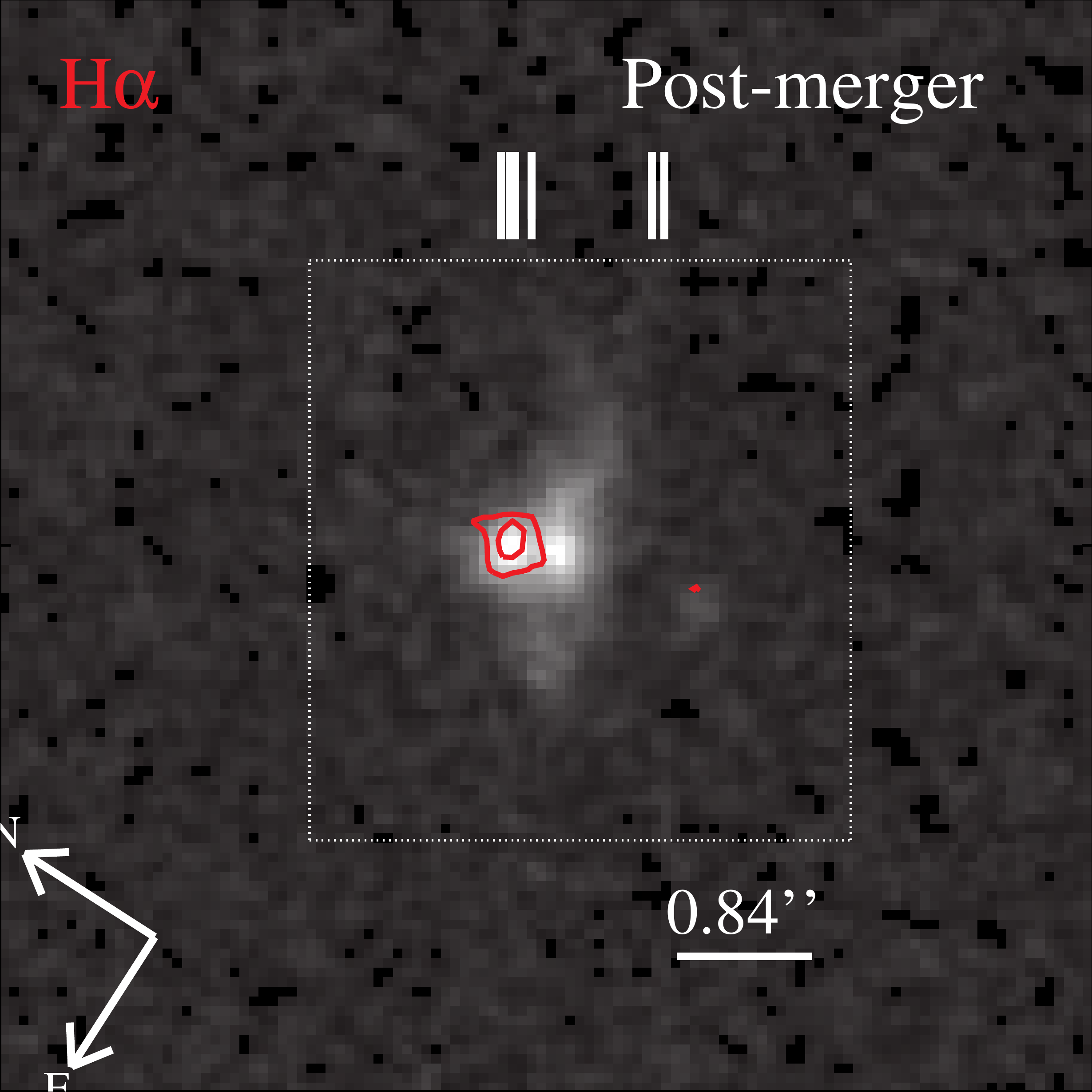} 
\includegraphics[width=0.13\textwidth]{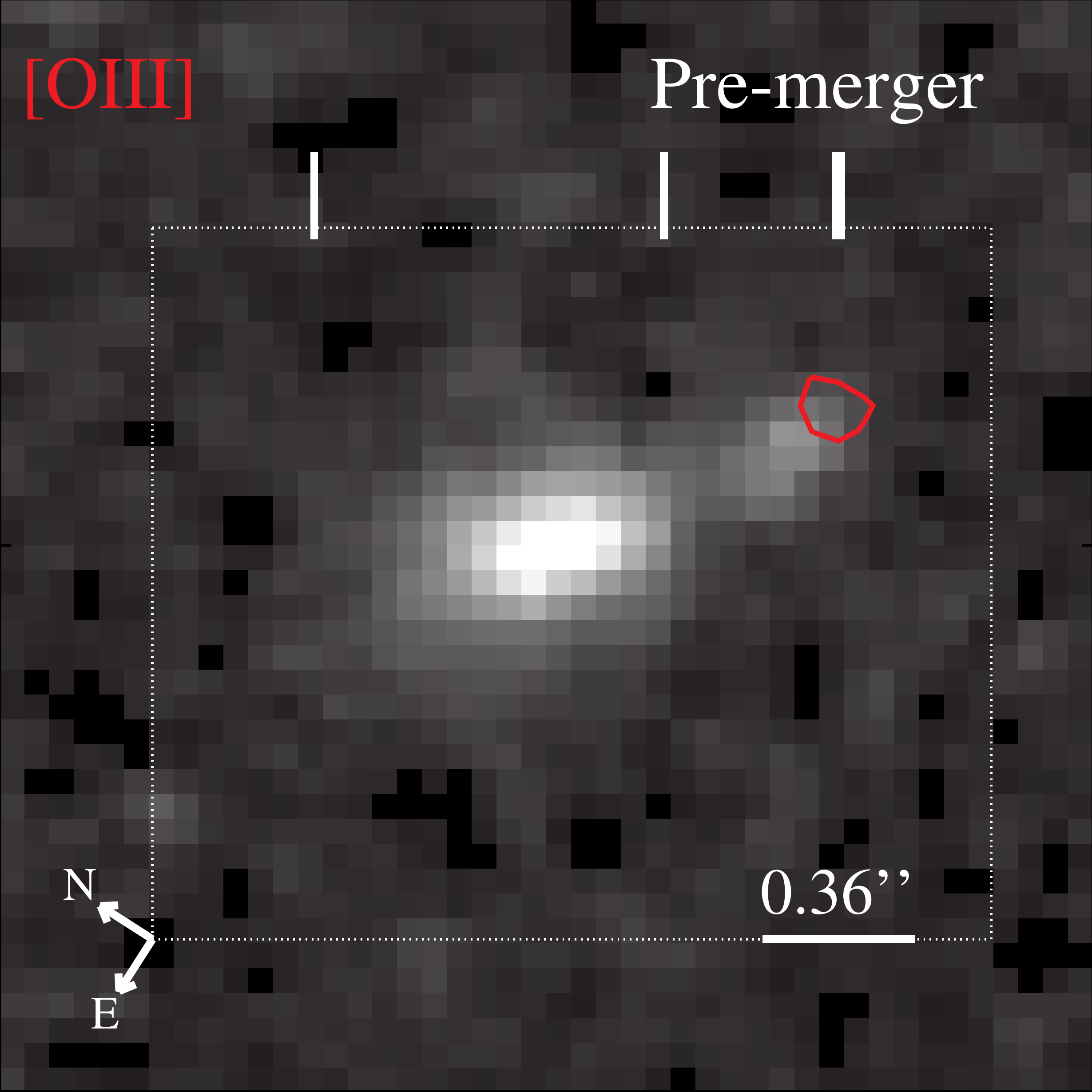} 
\includegraphics[width=0.13\textwidth]{COSMOS-20-G141_00150_lineimgoverlay_byhand.pdf} 
\includegraphics[width=0.13\textwidth]{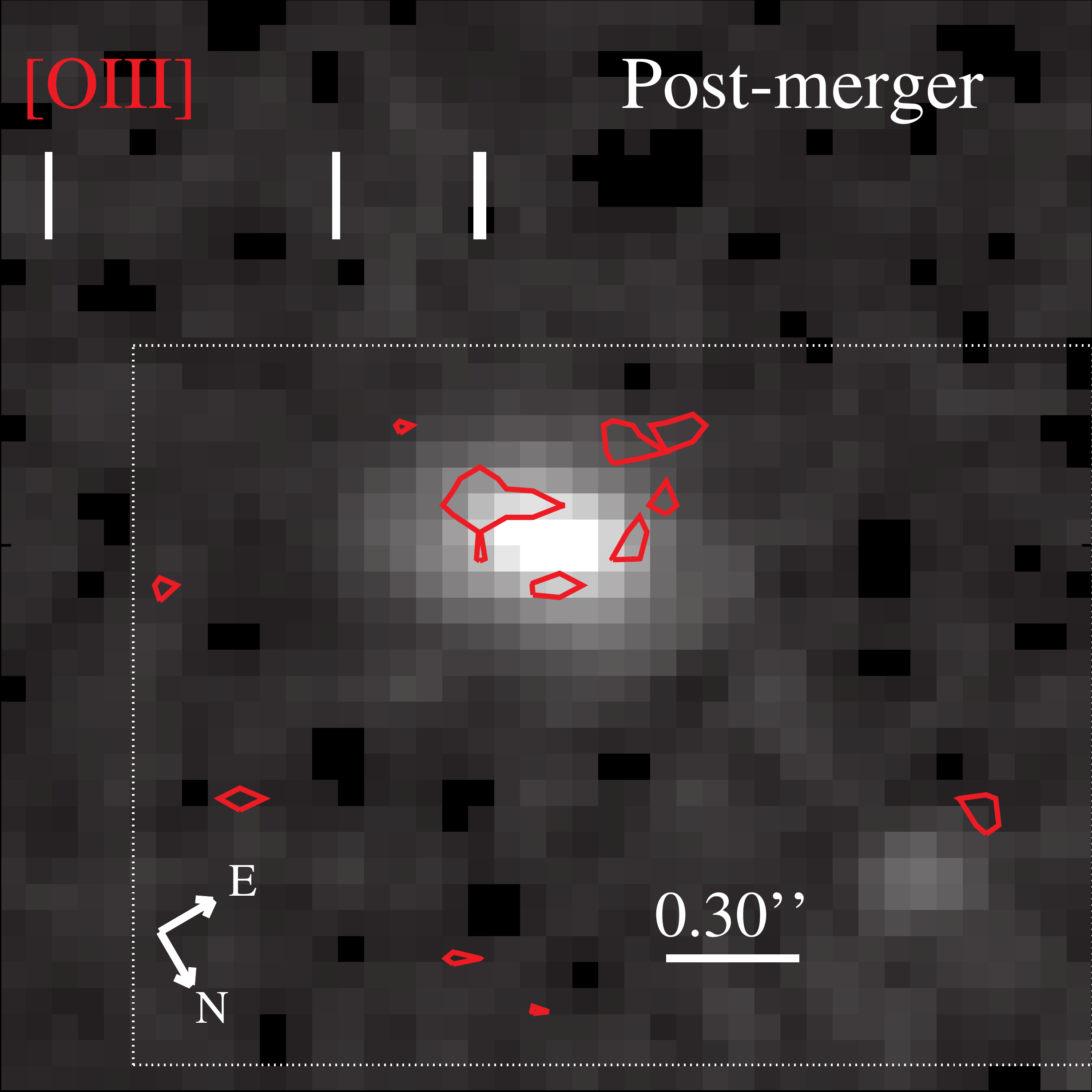} 
\includegraphics[width=0.13\textwidth]{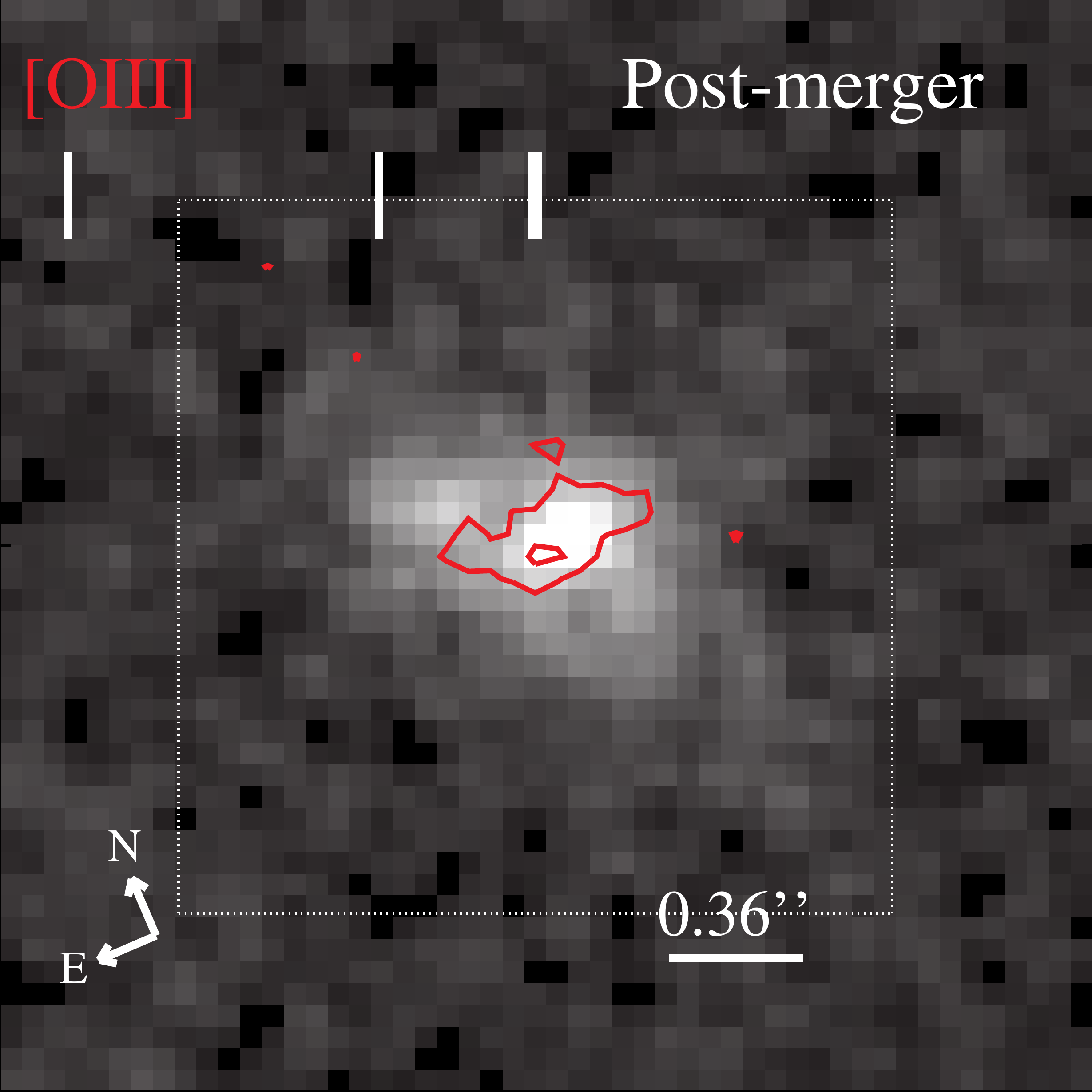} 
\includegraphics[width=0.13\textwidth]{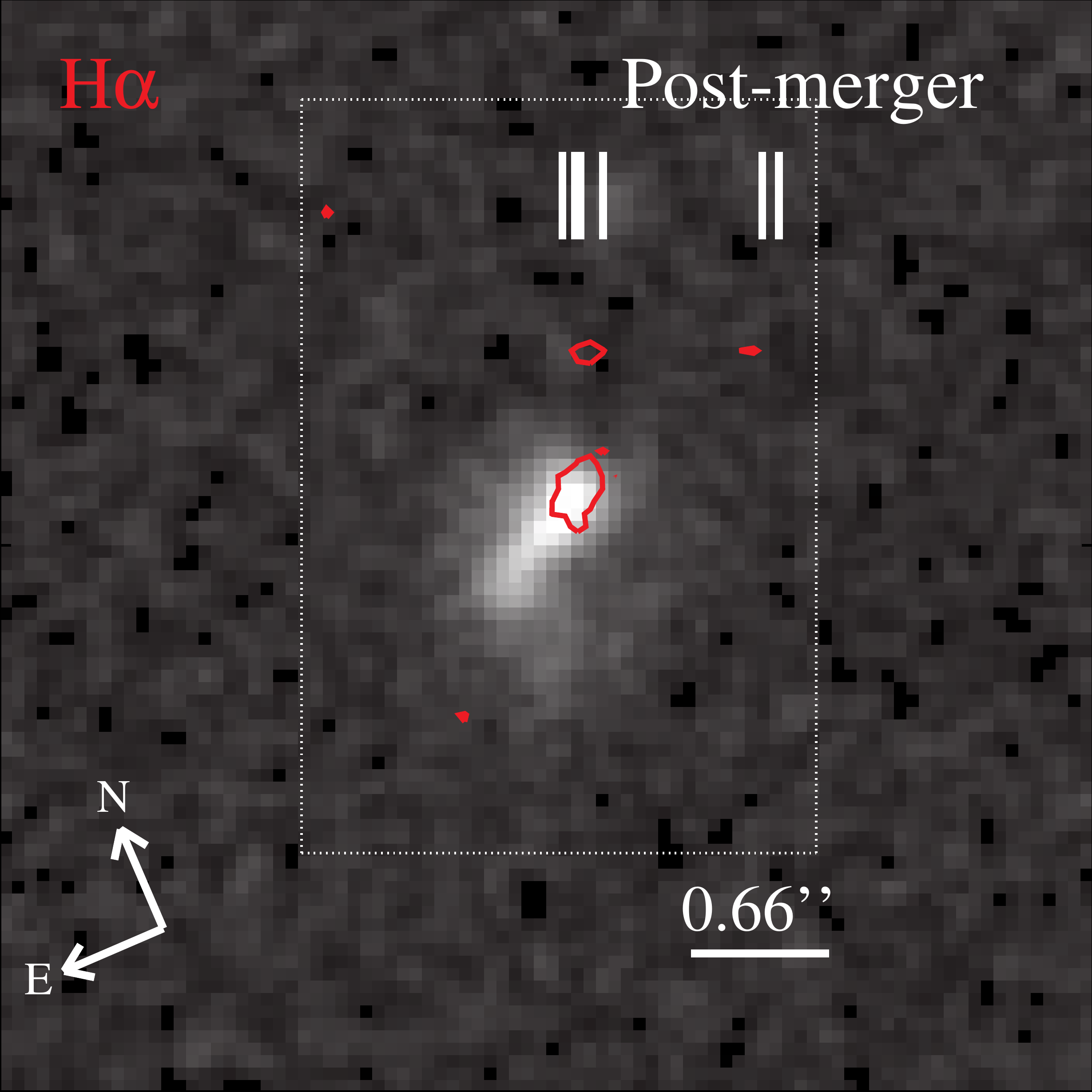}\\
\includegraphics[width=0.13\textwidth]{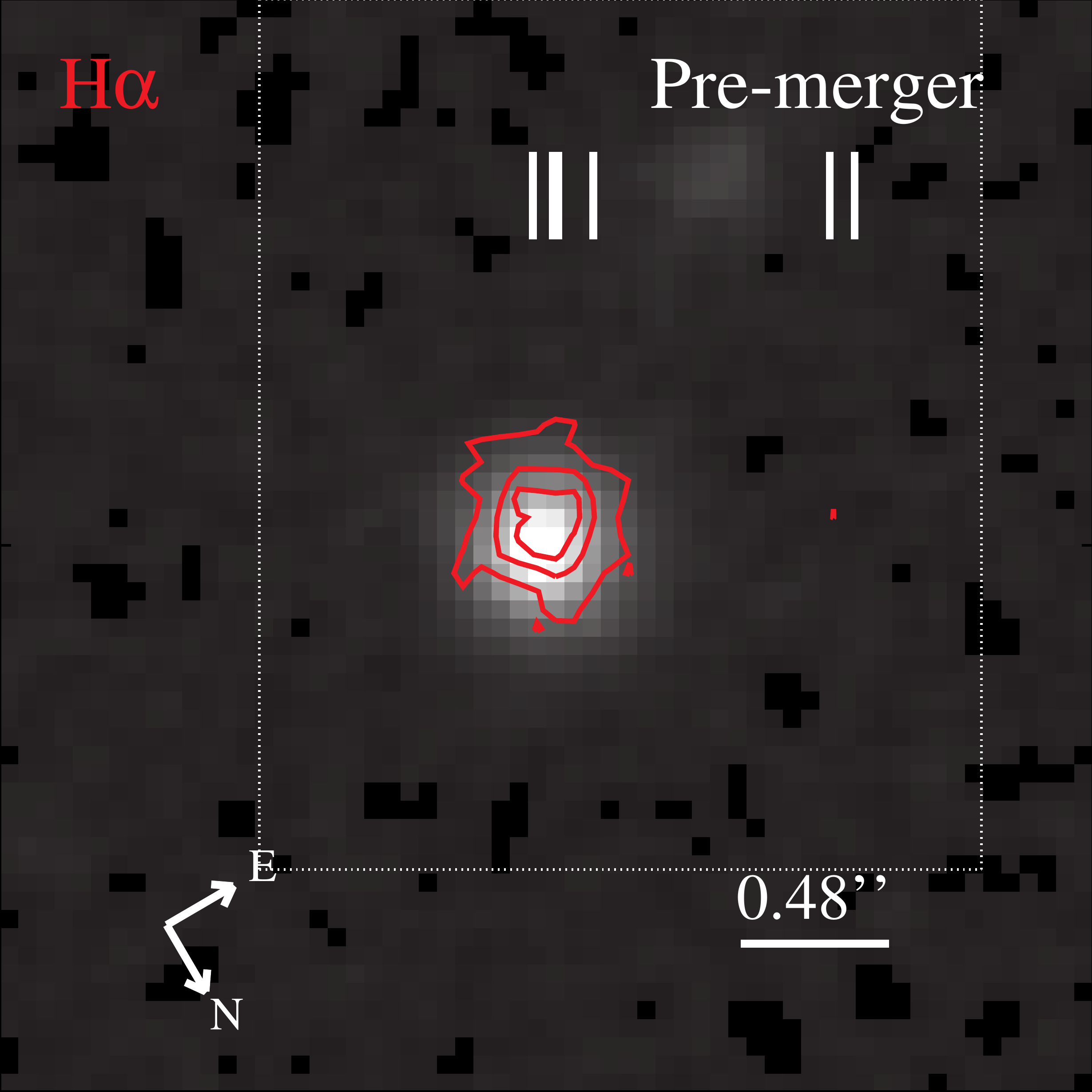}  
\includegraphics[width=0.13\textwidth]{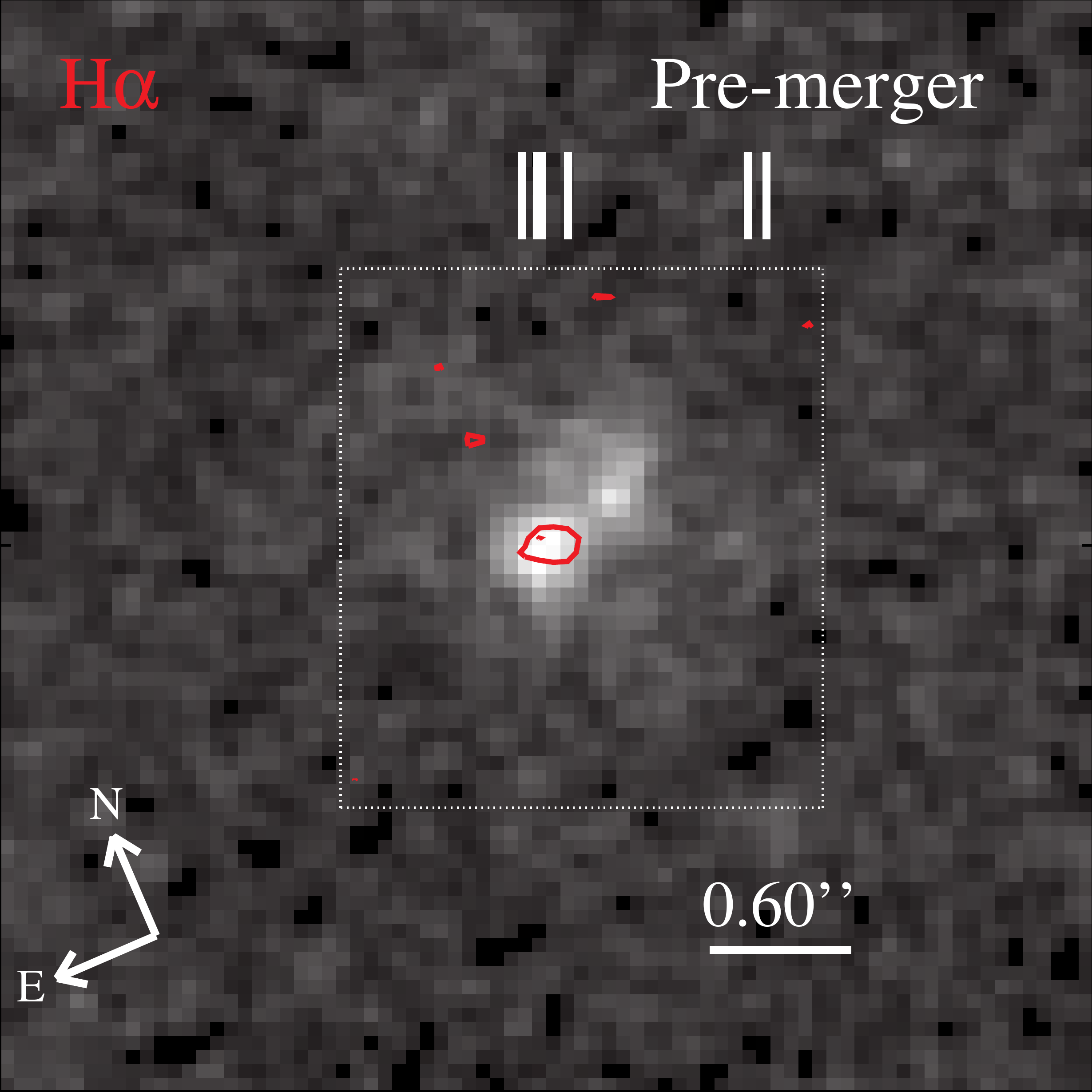}    
\includegraphics[width=0.13\textwidth]{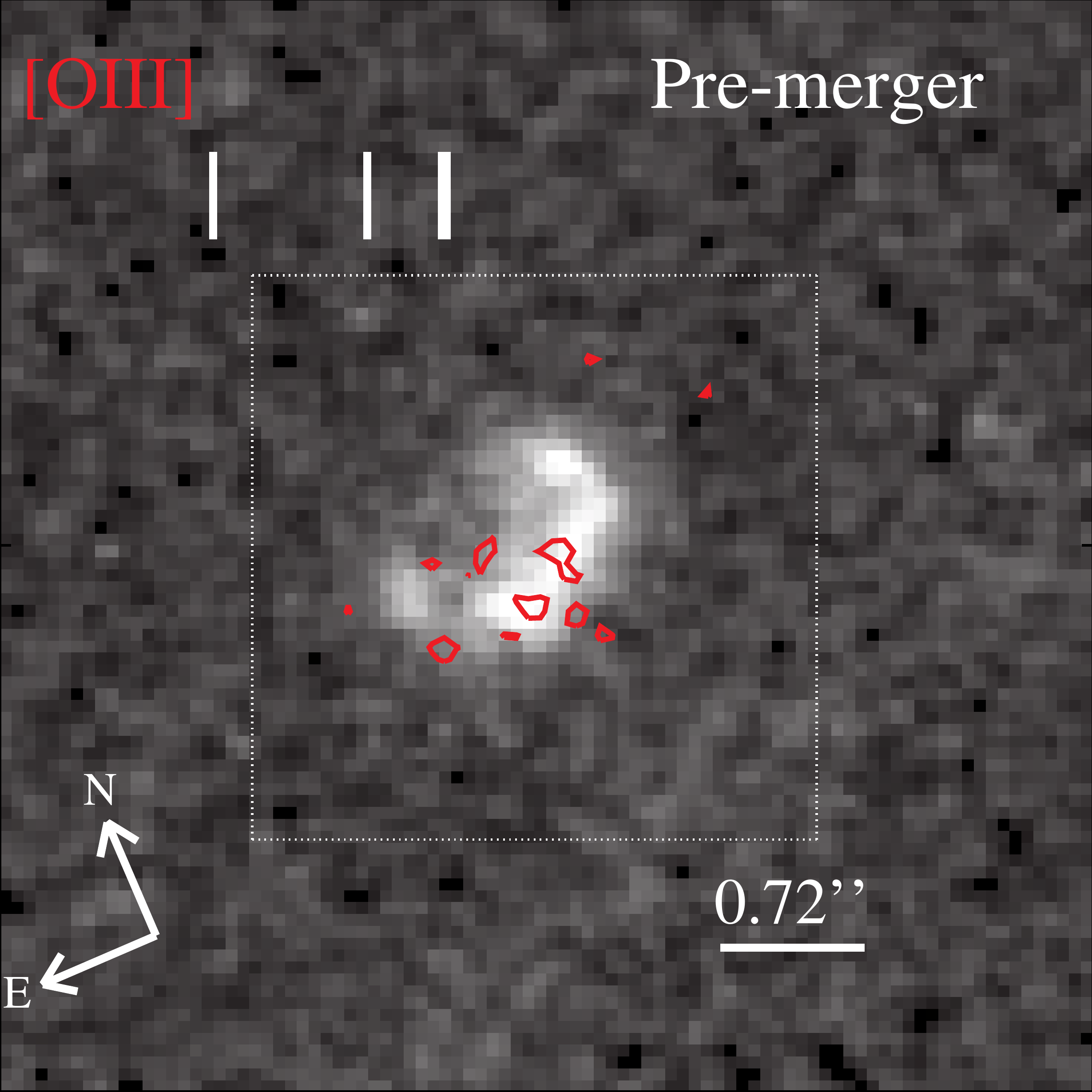}    
\includegraphics[width=0.13\textwidth]{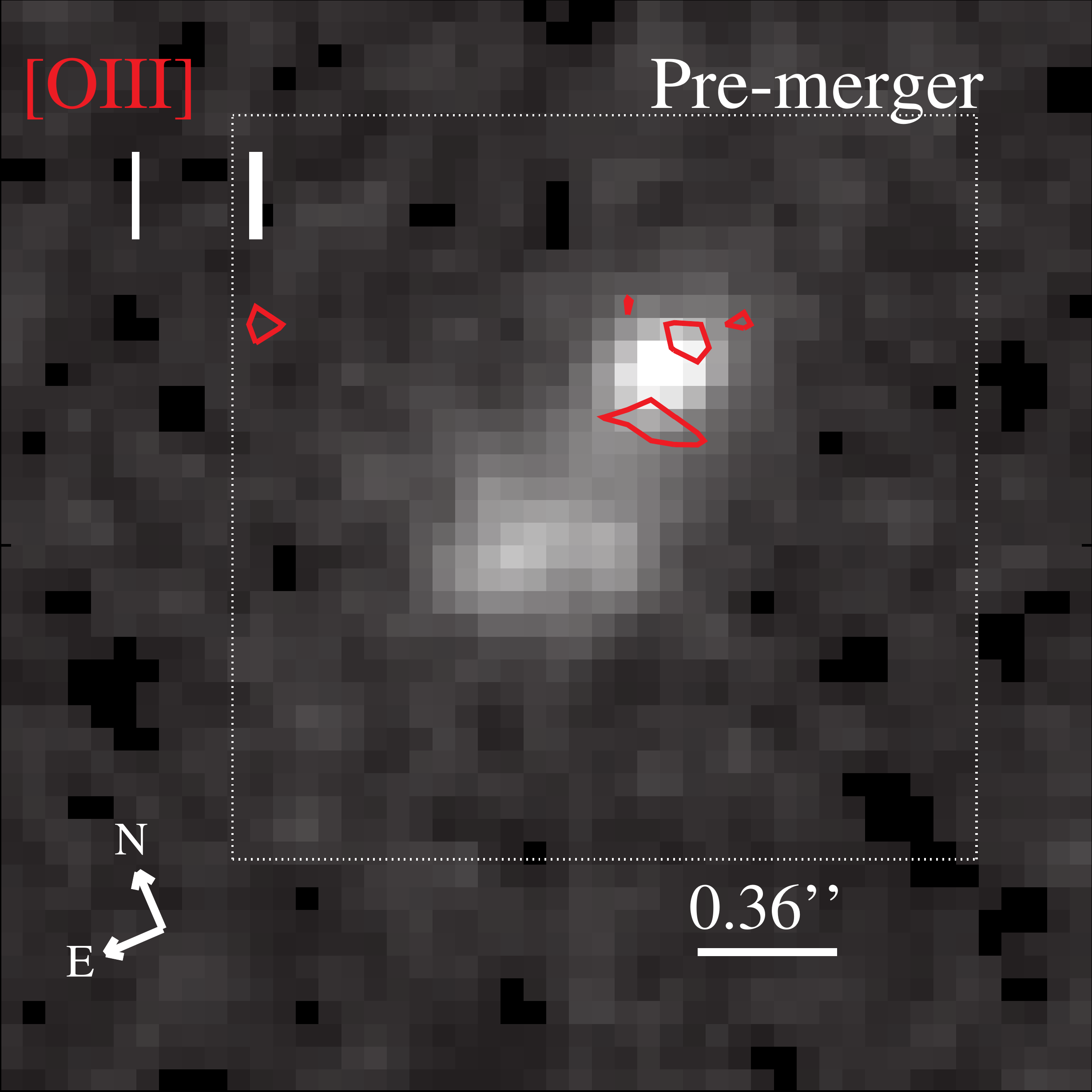}    
\includegraphics[width=0.13\textwidth]{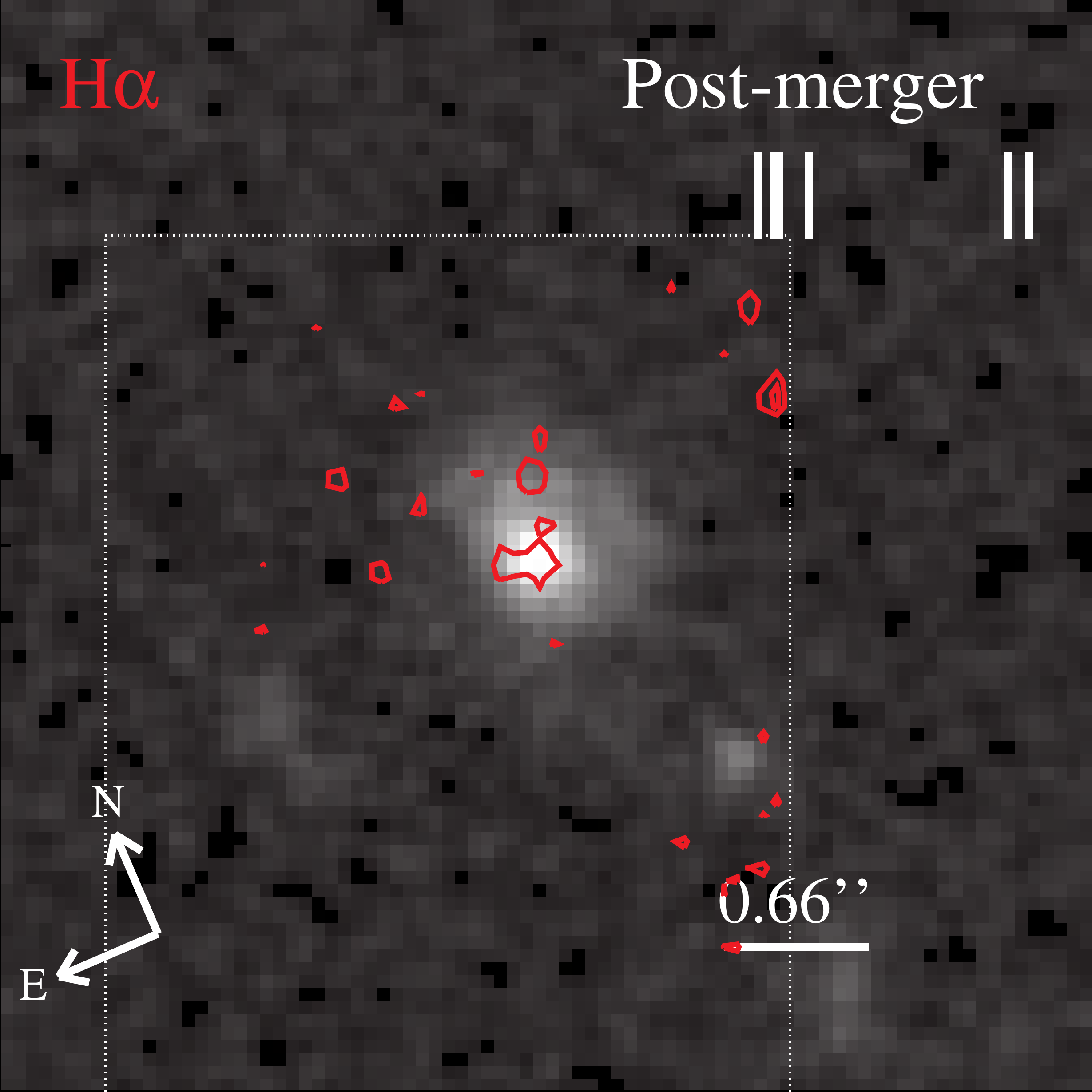}    
\includegraphics[width=0.13\textwidth]{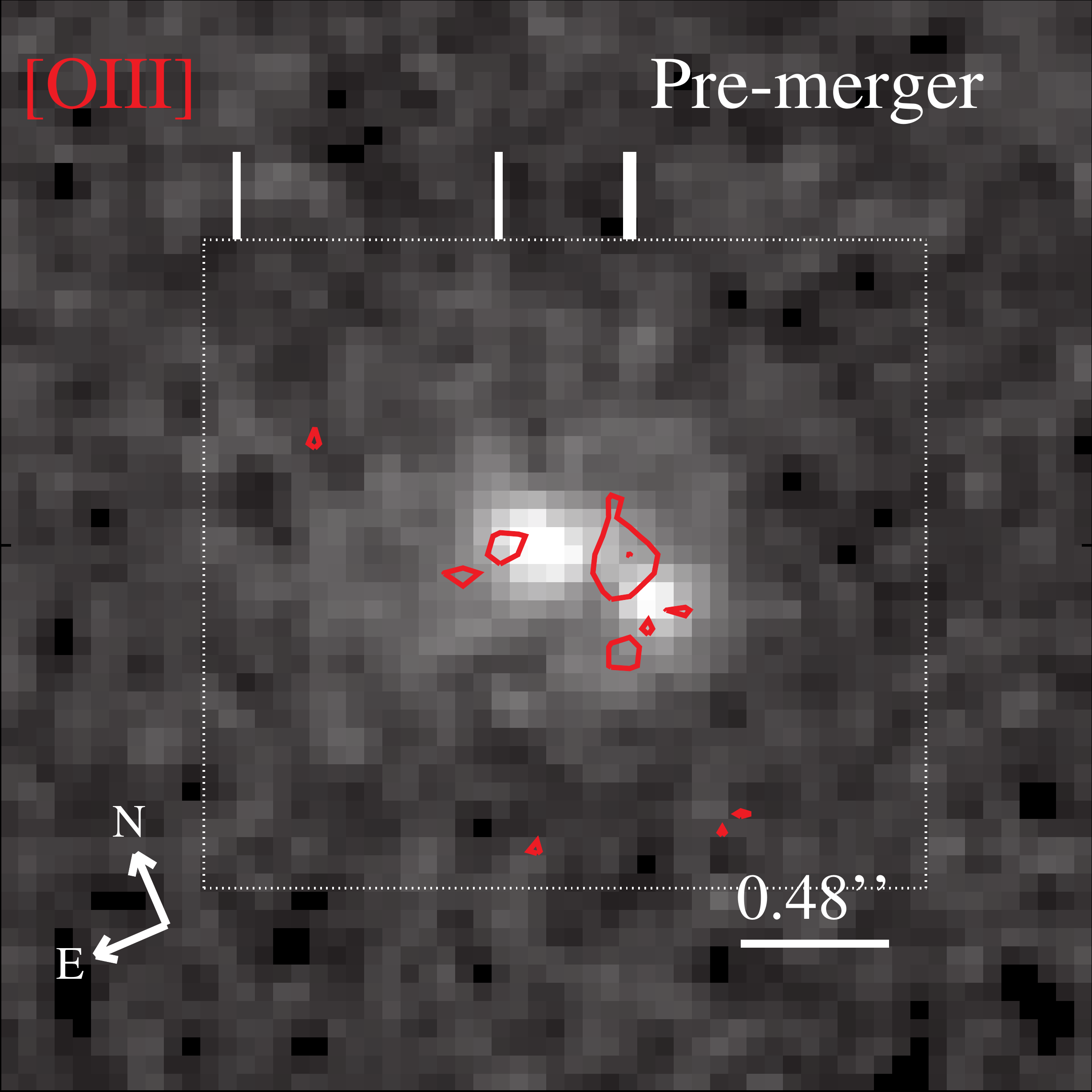}    
\includegraphics[width=0.13\textwidth]{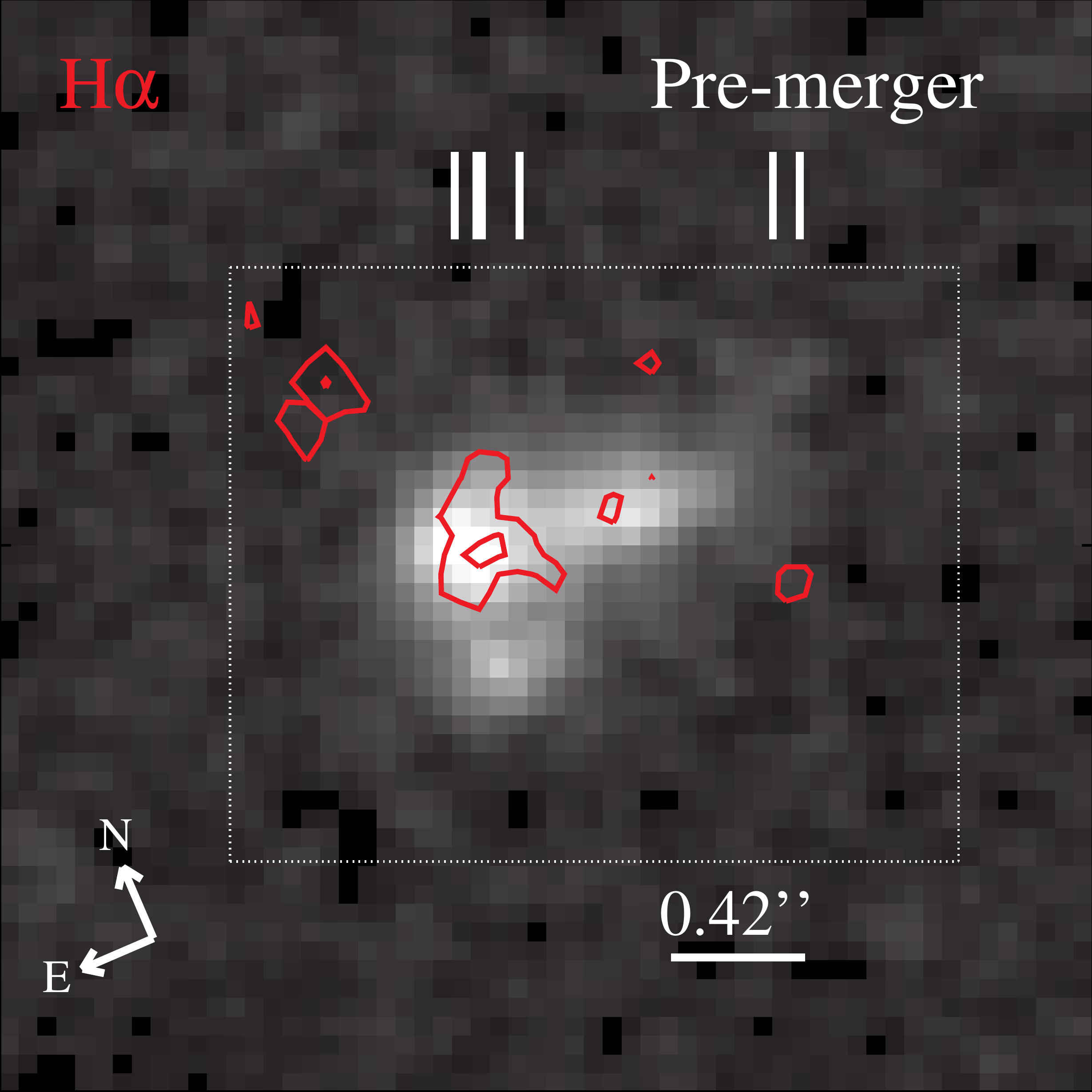}\\  
\includegraphics[width=0.13\textwidth]{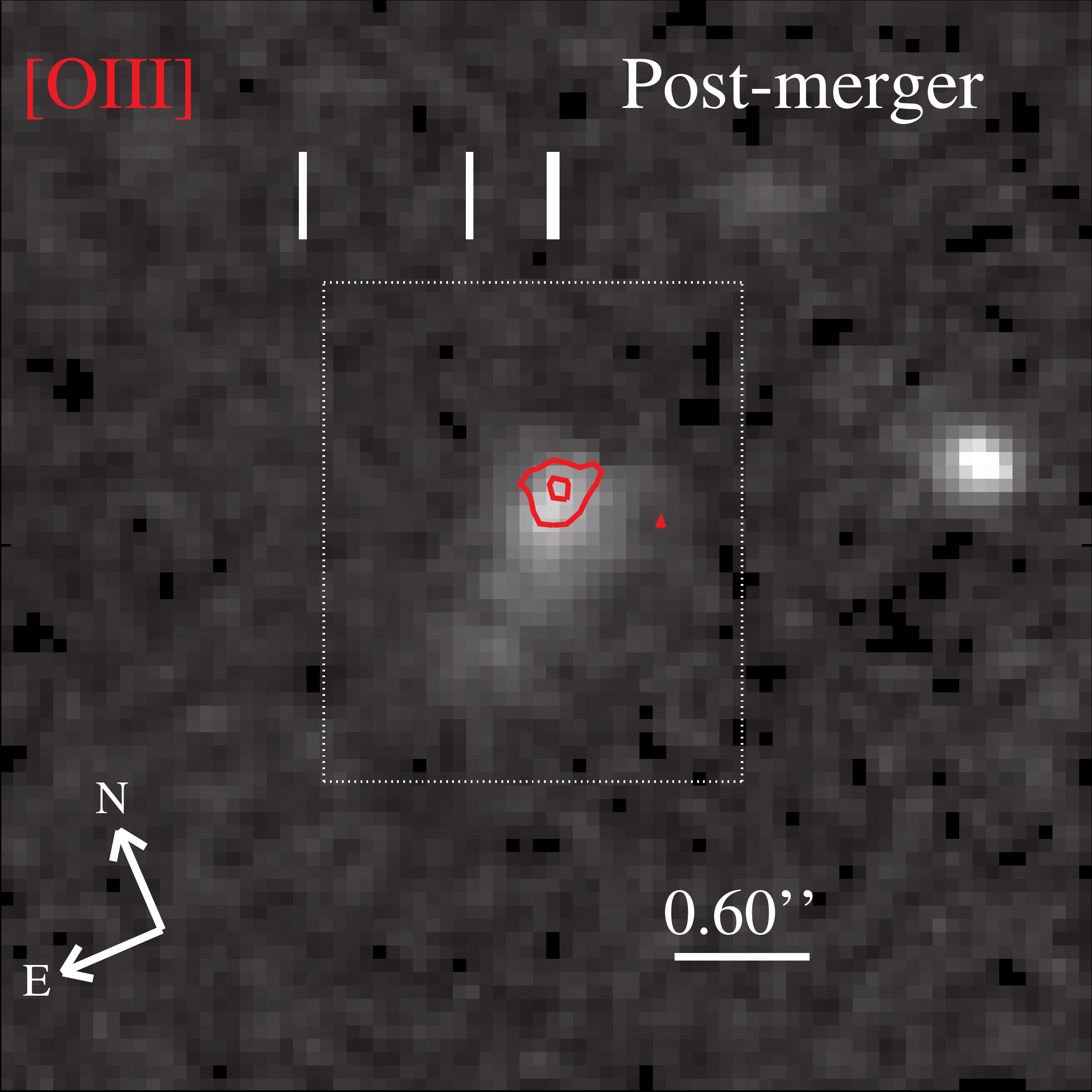}    
\includegraphics[width=0.13\textwidth]{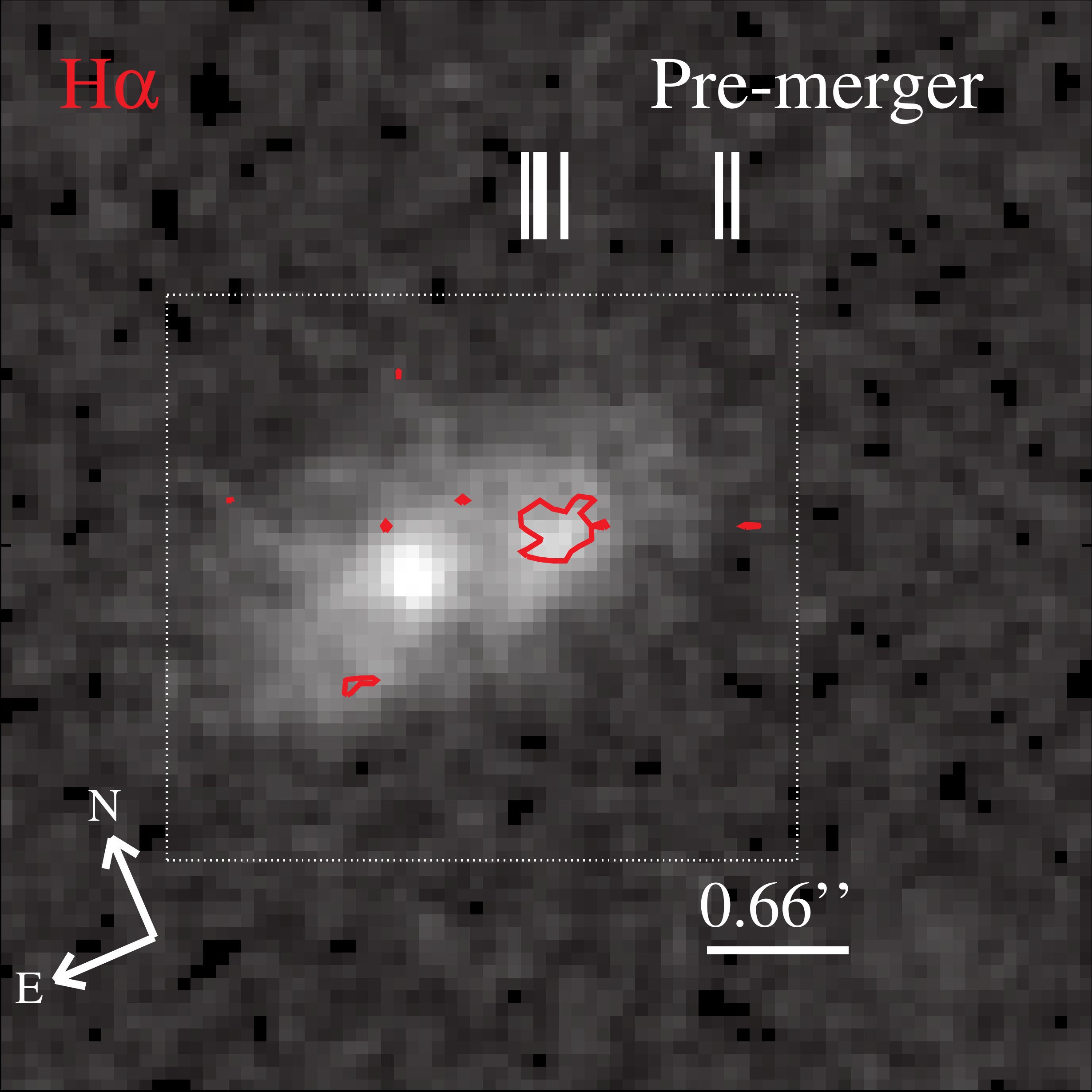}    
\includegraphics[width=0.13\textwidth]{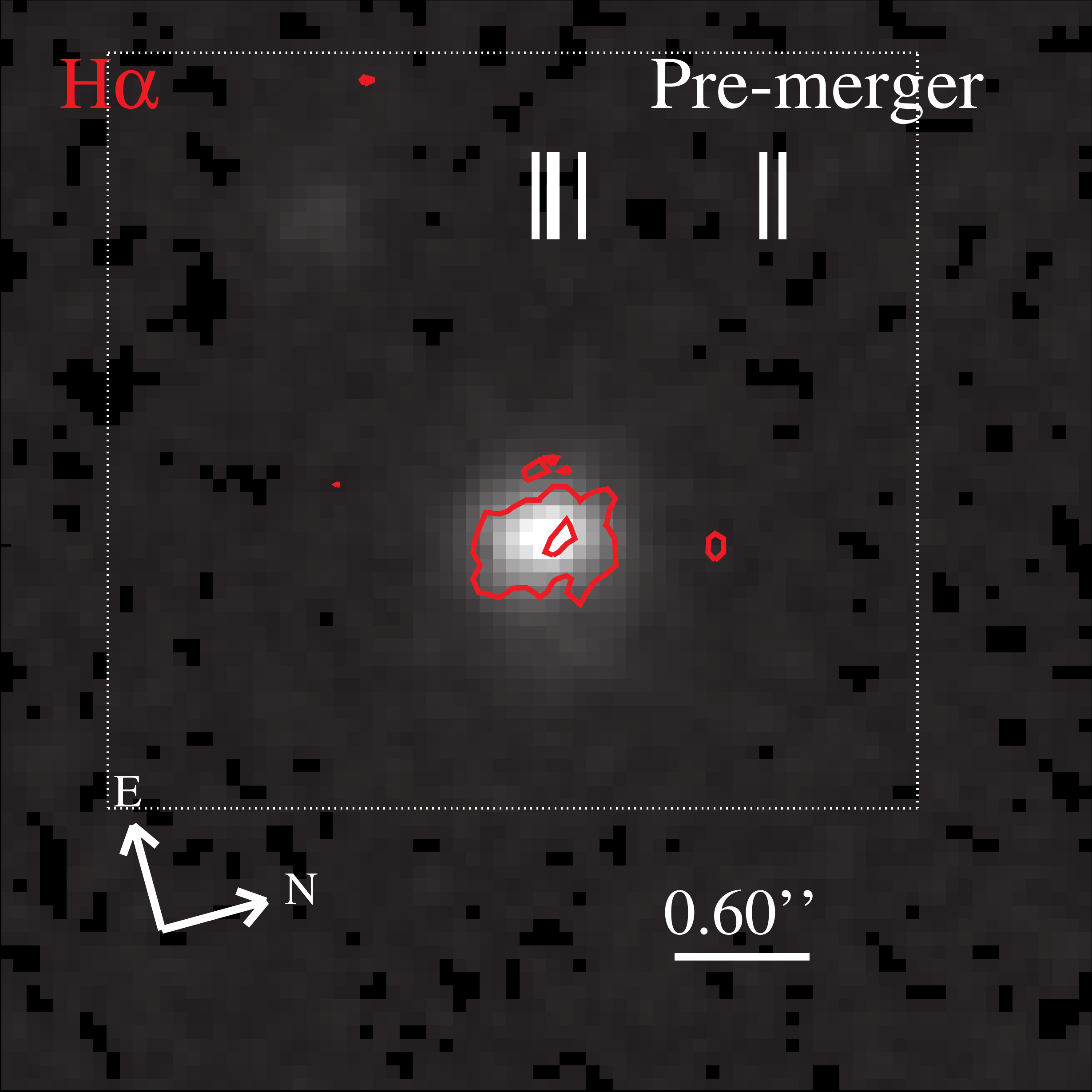} 
\includegraphics[width=0.13\textwidth]{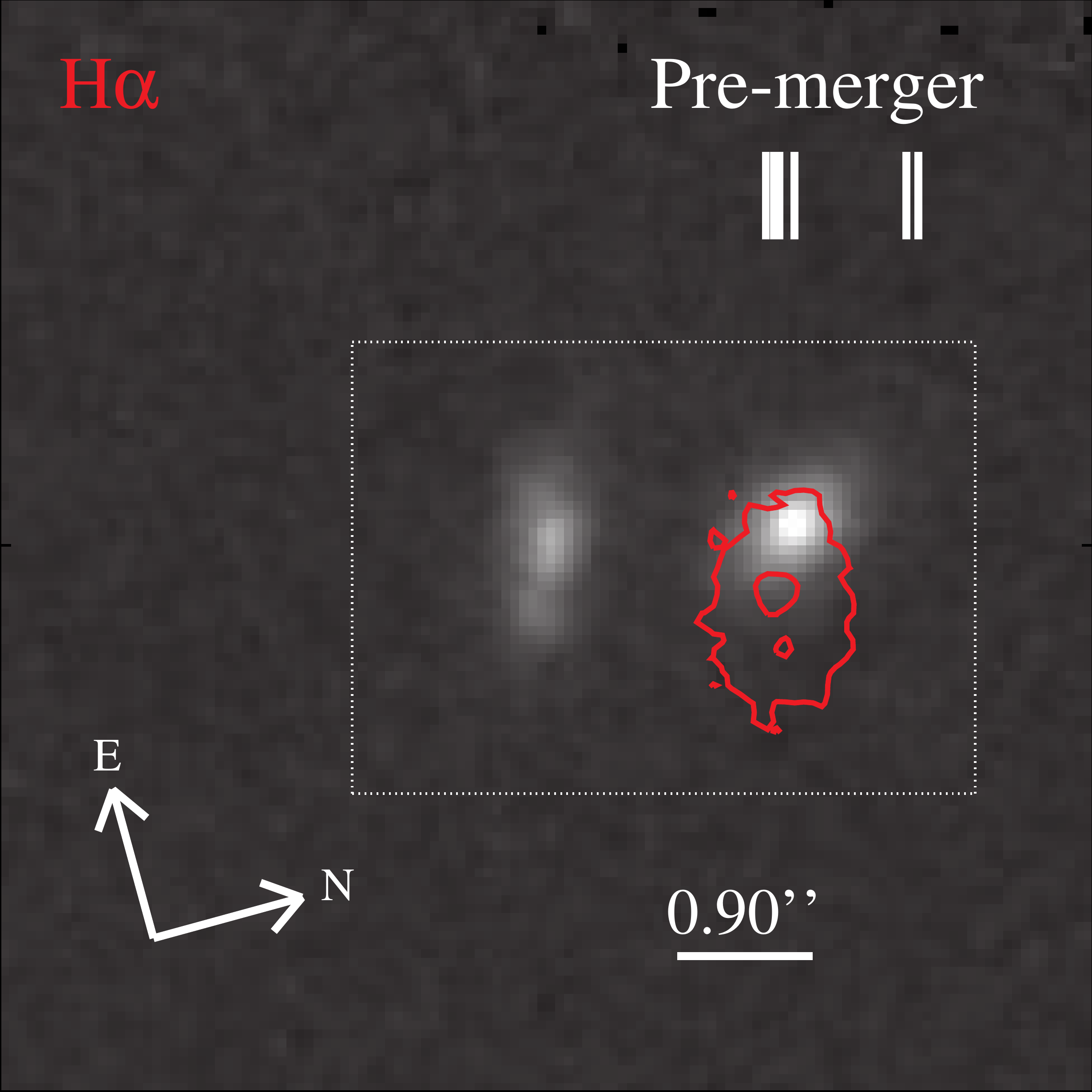} 
\includegraphics[width=0.13\textwidth]{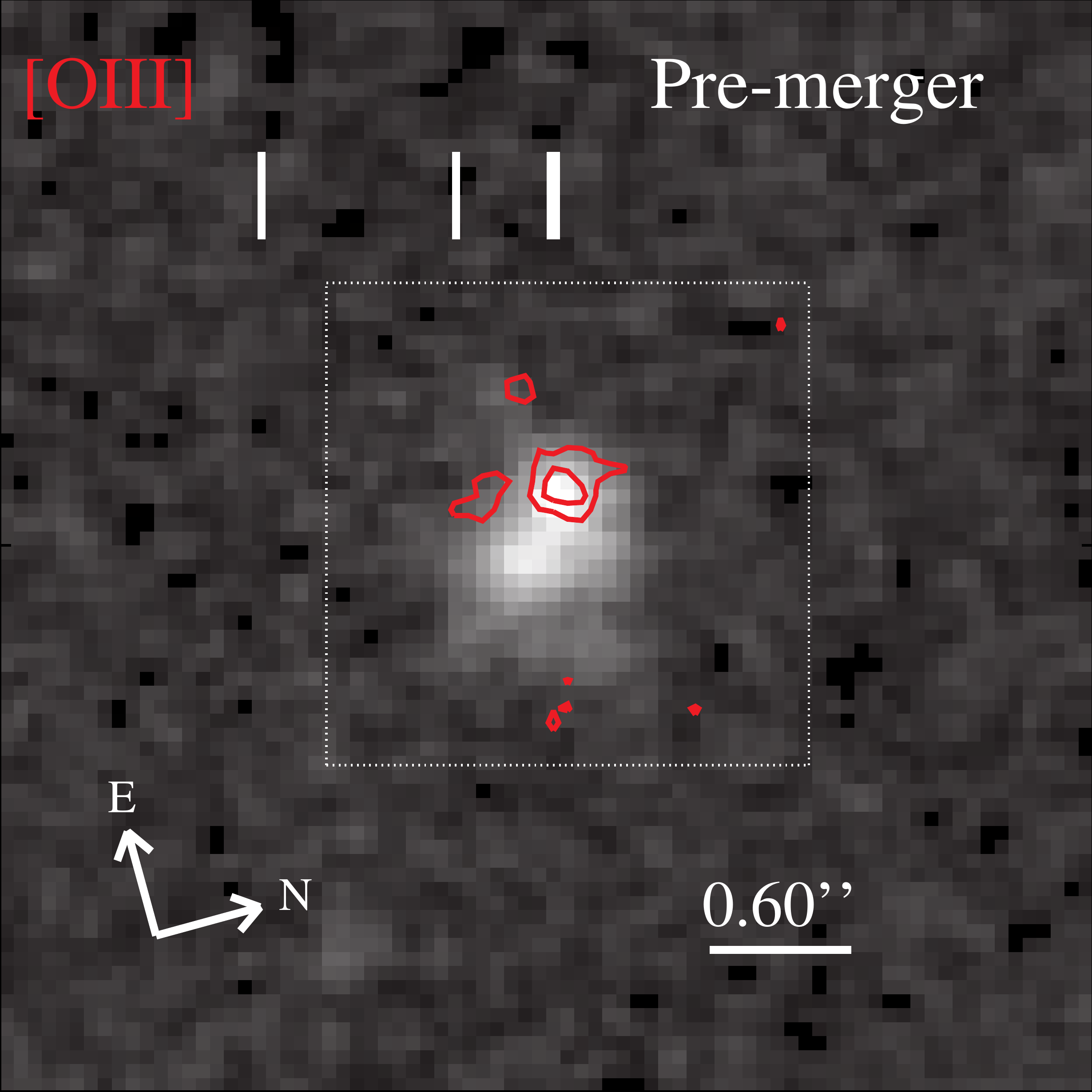} 
\includegraphics[width=0.13\textwidth]{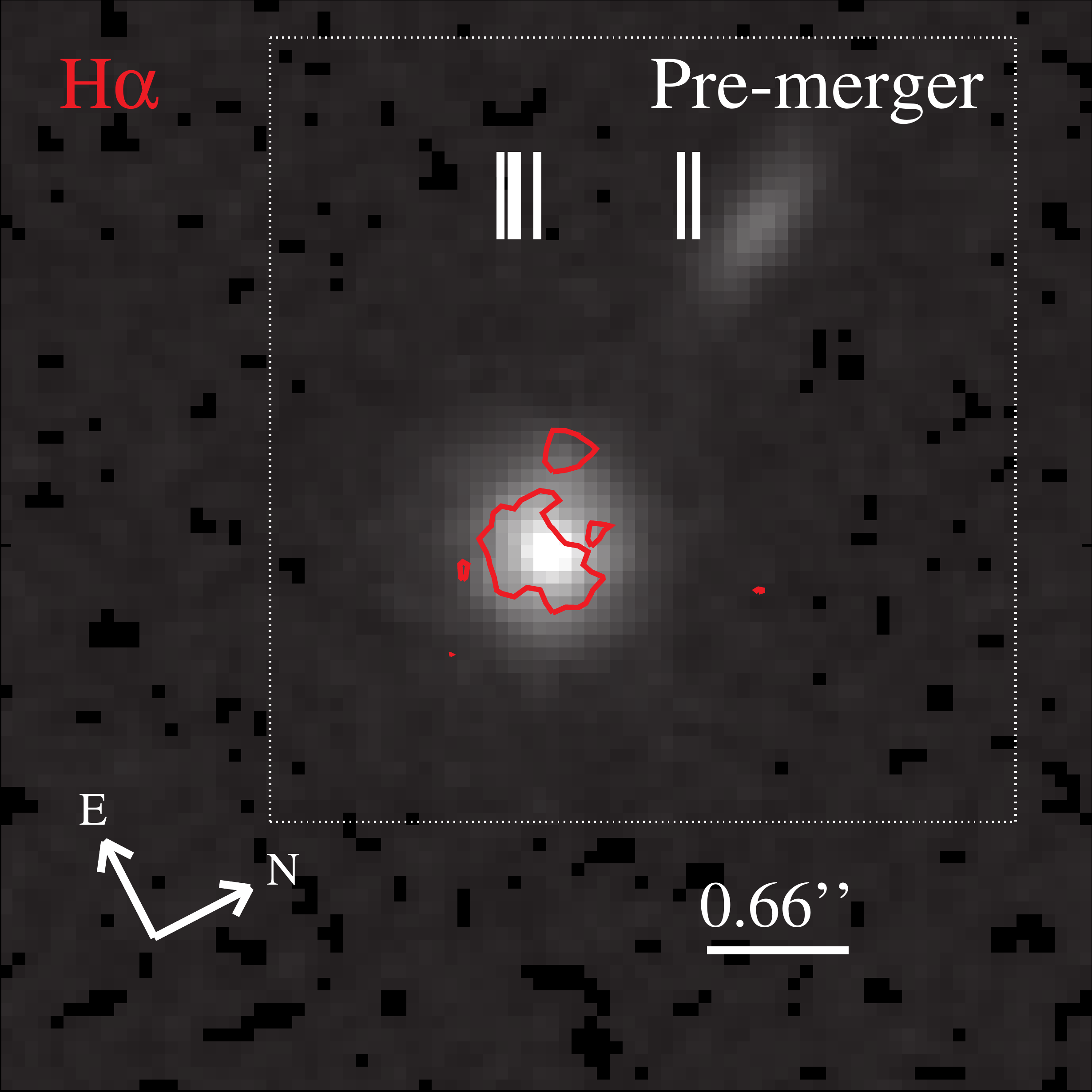} 
\includegraphics[width=0.13\textwidth]{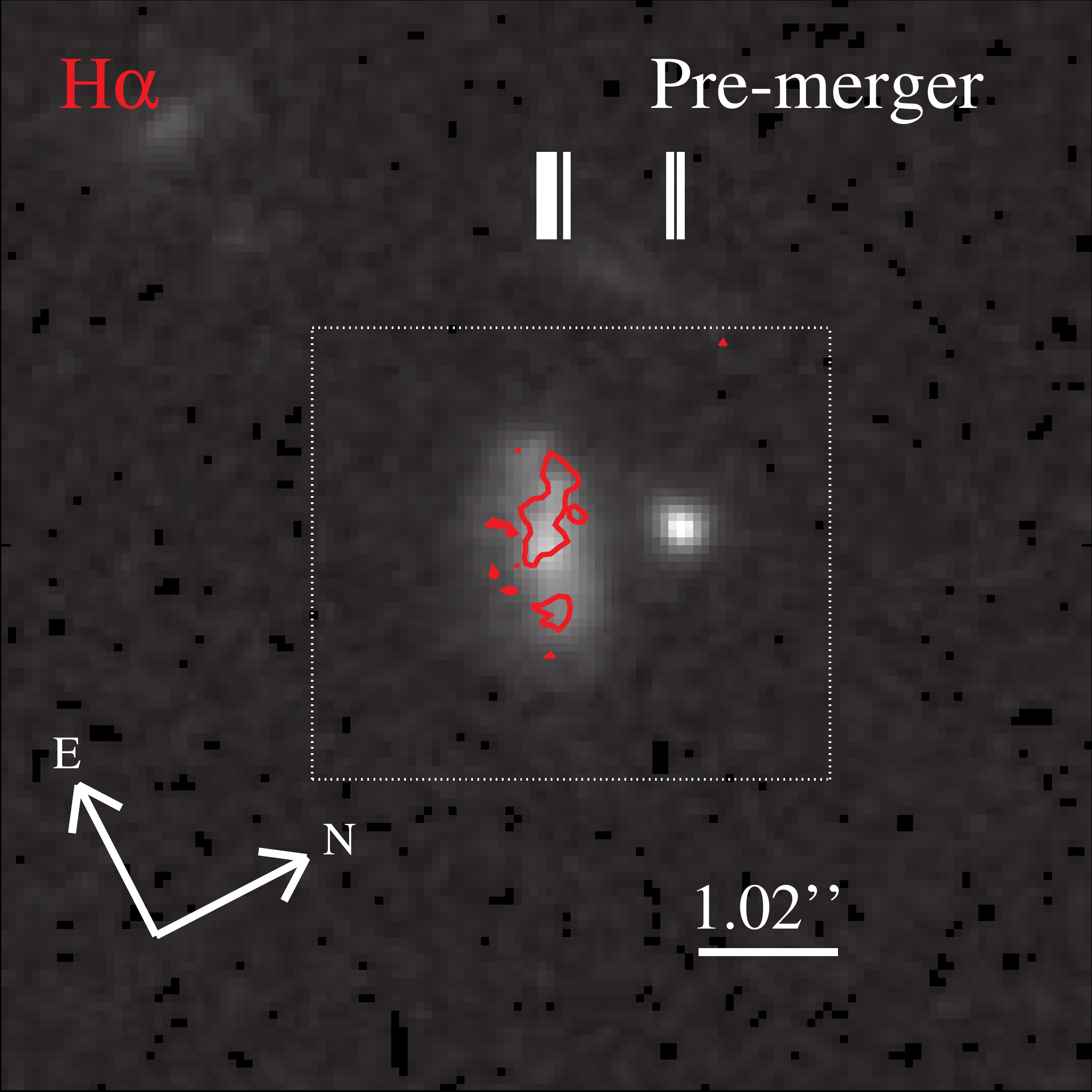}\\
\includegraphics[width=0.13\textwidth]{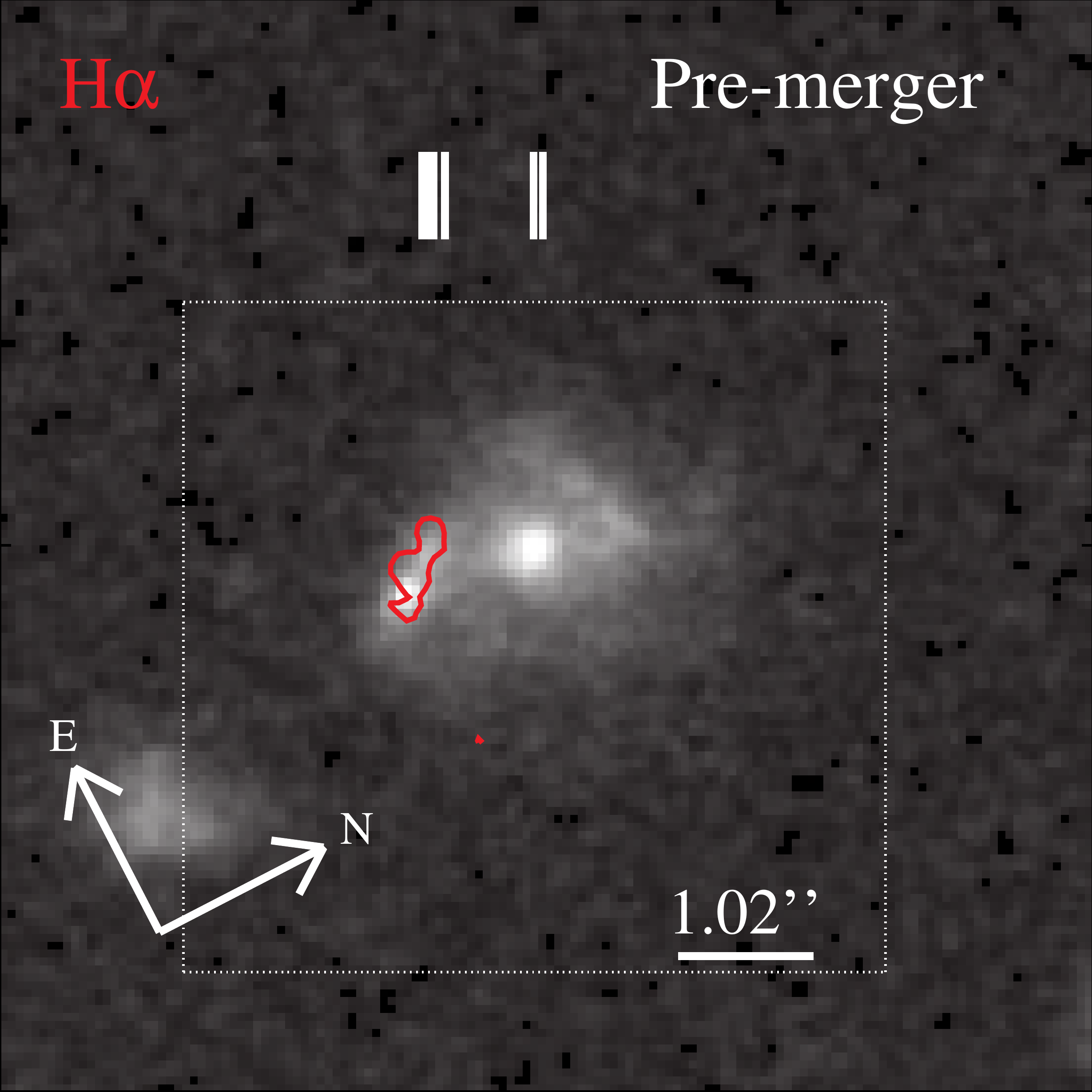}  
\includegraphics[width=0.13\textwidth]{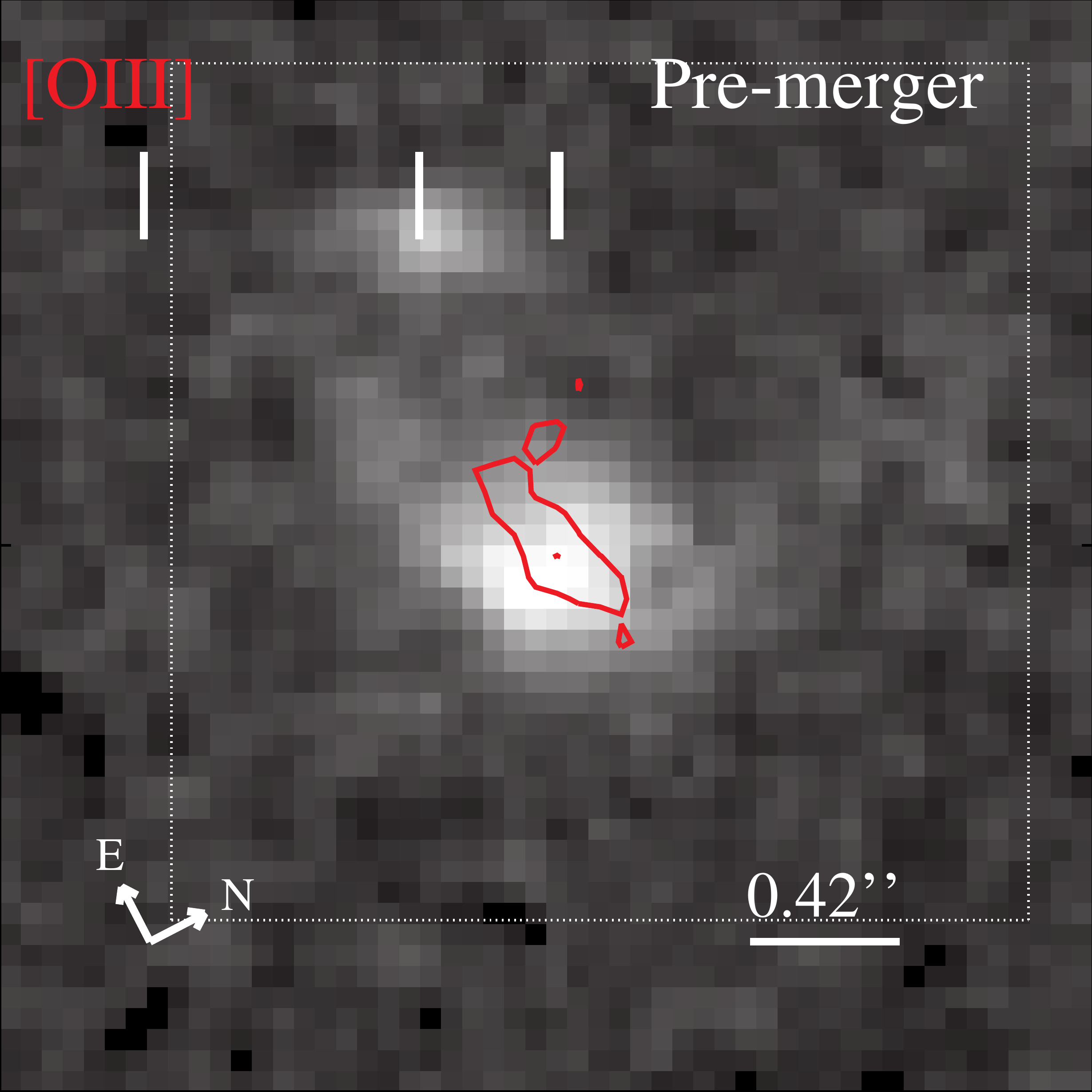}  
\includegraphics[width=0.13\textwidth]{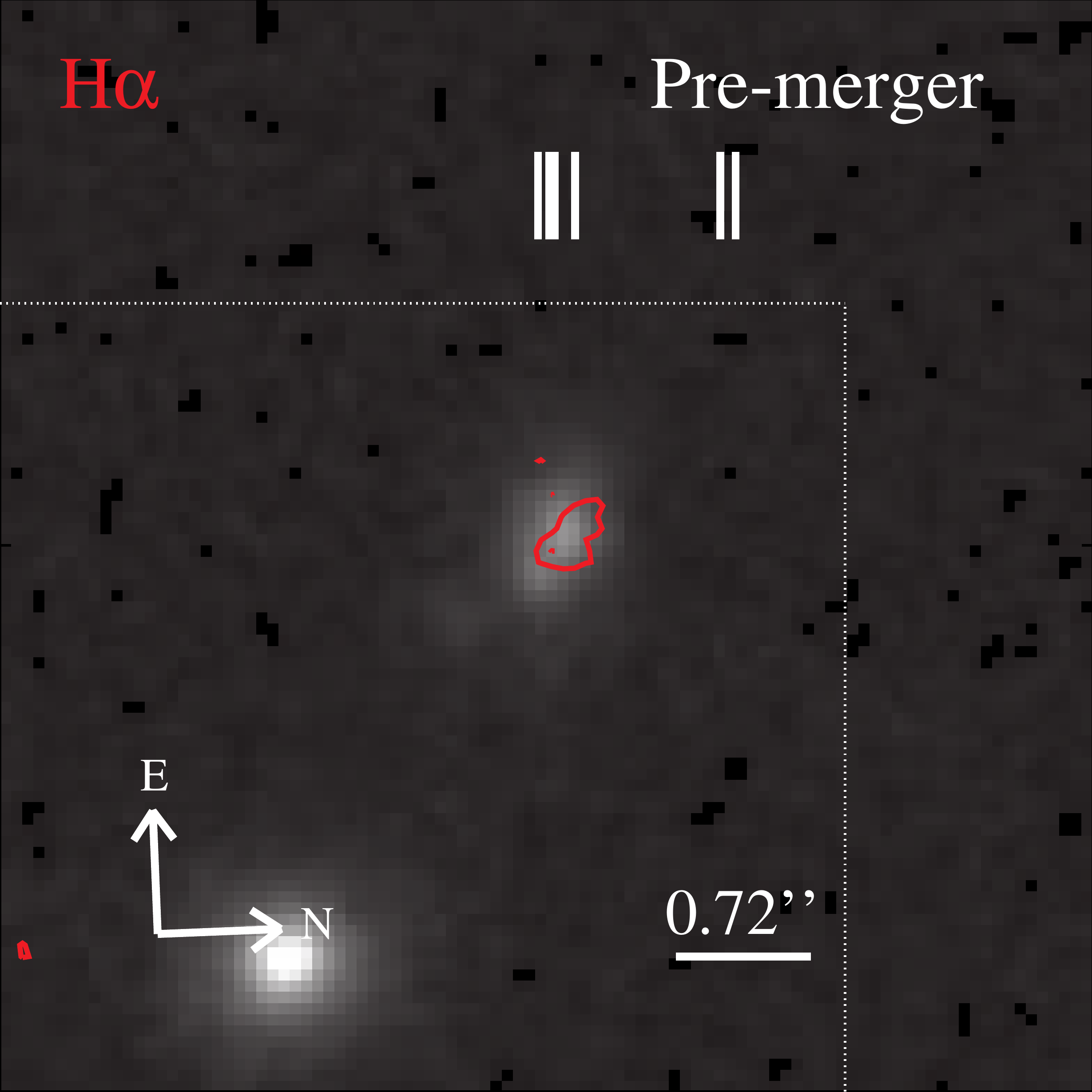}  
\includegraphics[width=0.13\textwidth]{GOODS-S-27-G141_00714_lineimgoverlay_byhand.pdf}  
\includegraphics[width=0.13\textwidth]{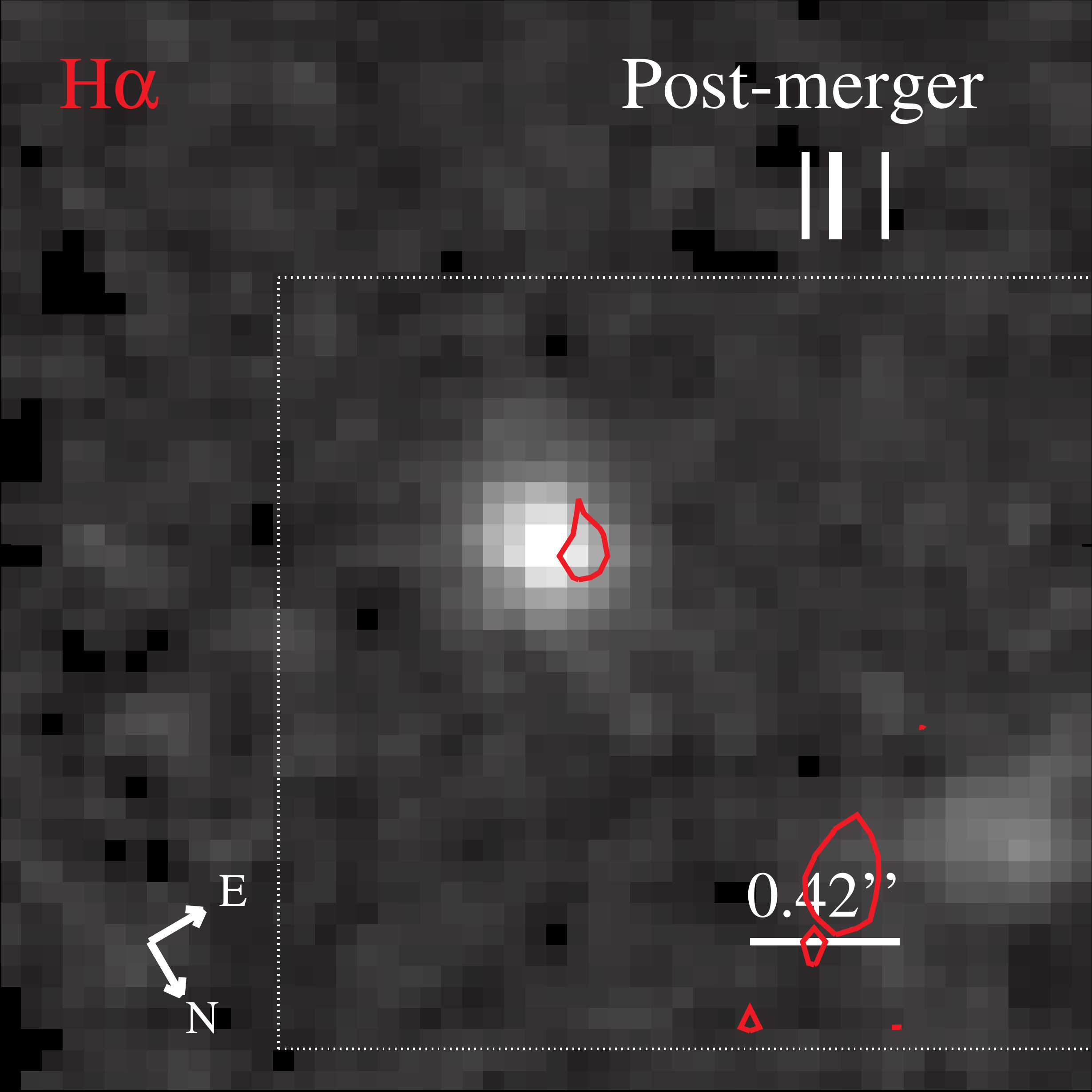}   
\includegraphics[width=0.13\textwidth]{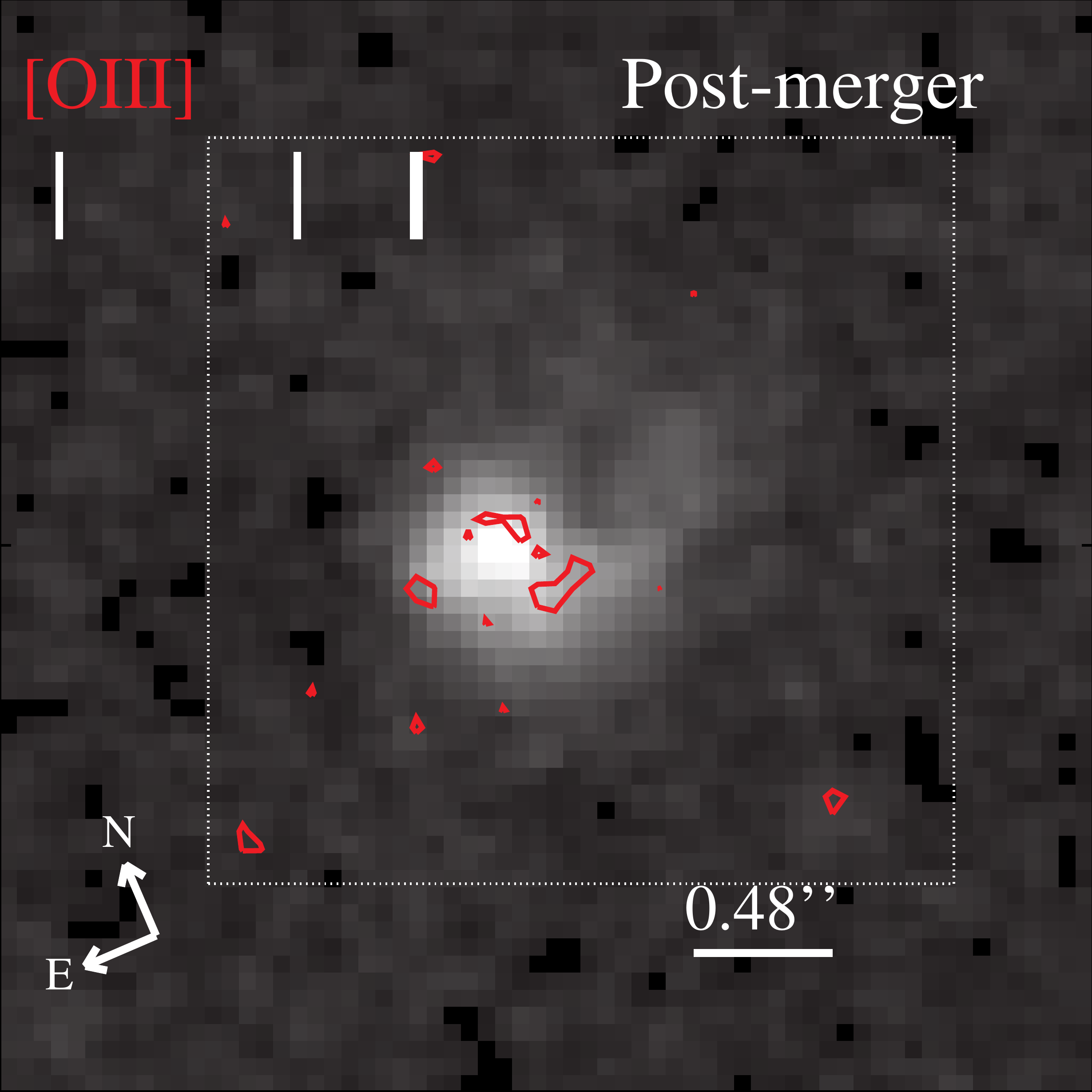}   
\includegraphics[width=0.13\textwidth]{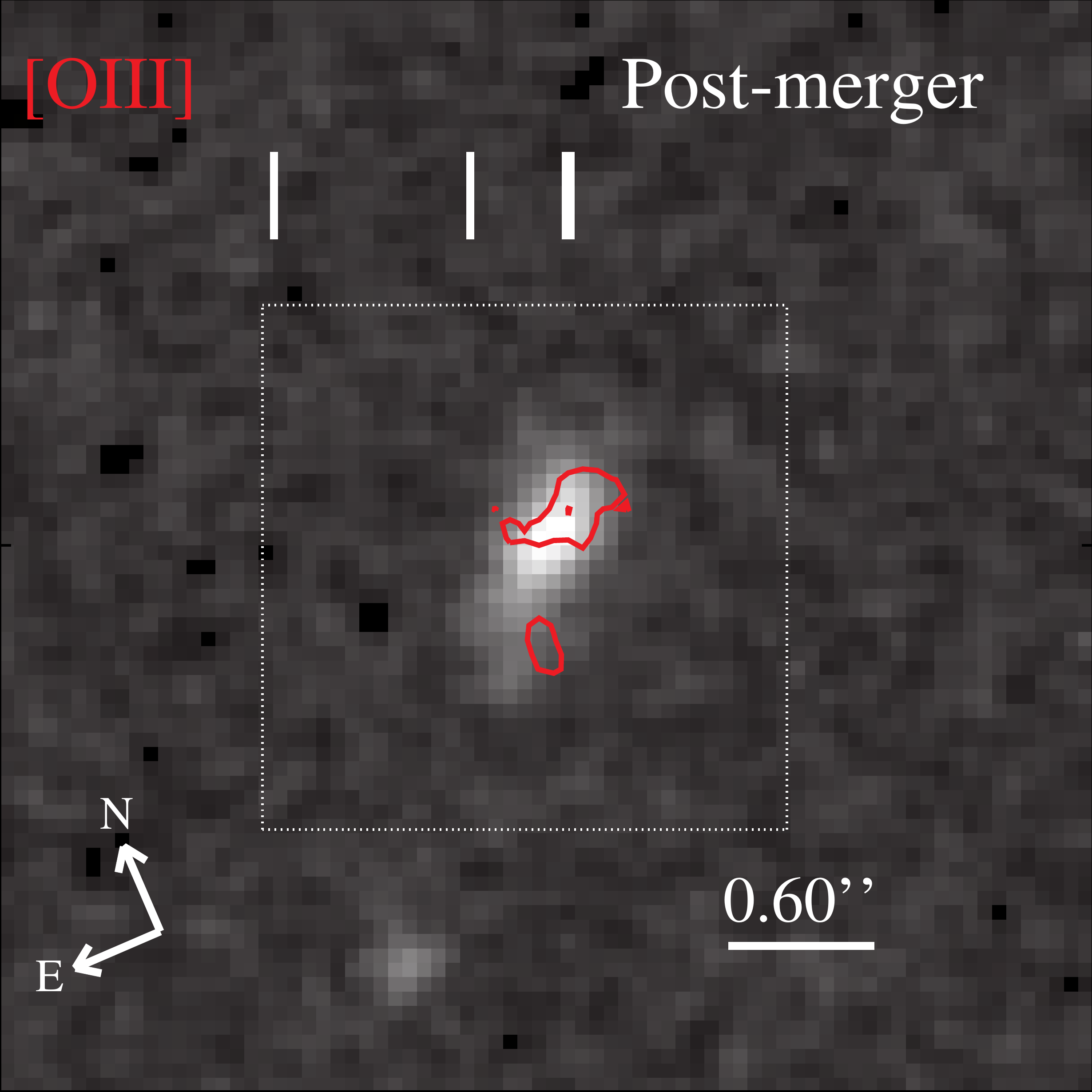}\\ 
\vspace{0.5cm}
\includegraphics[width=0.13\textwidth]{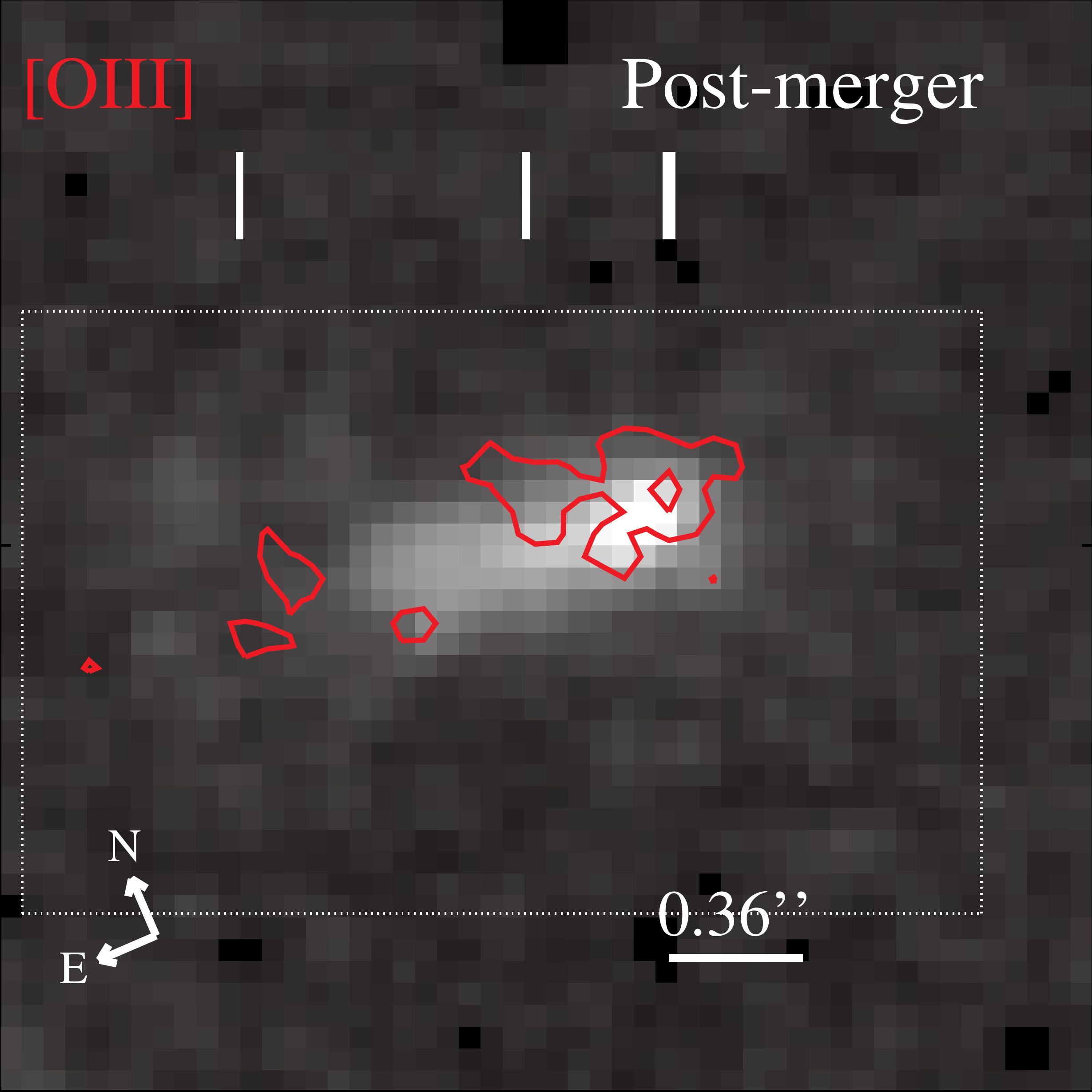}   
\includegraphics[width=0.13\textwidth]{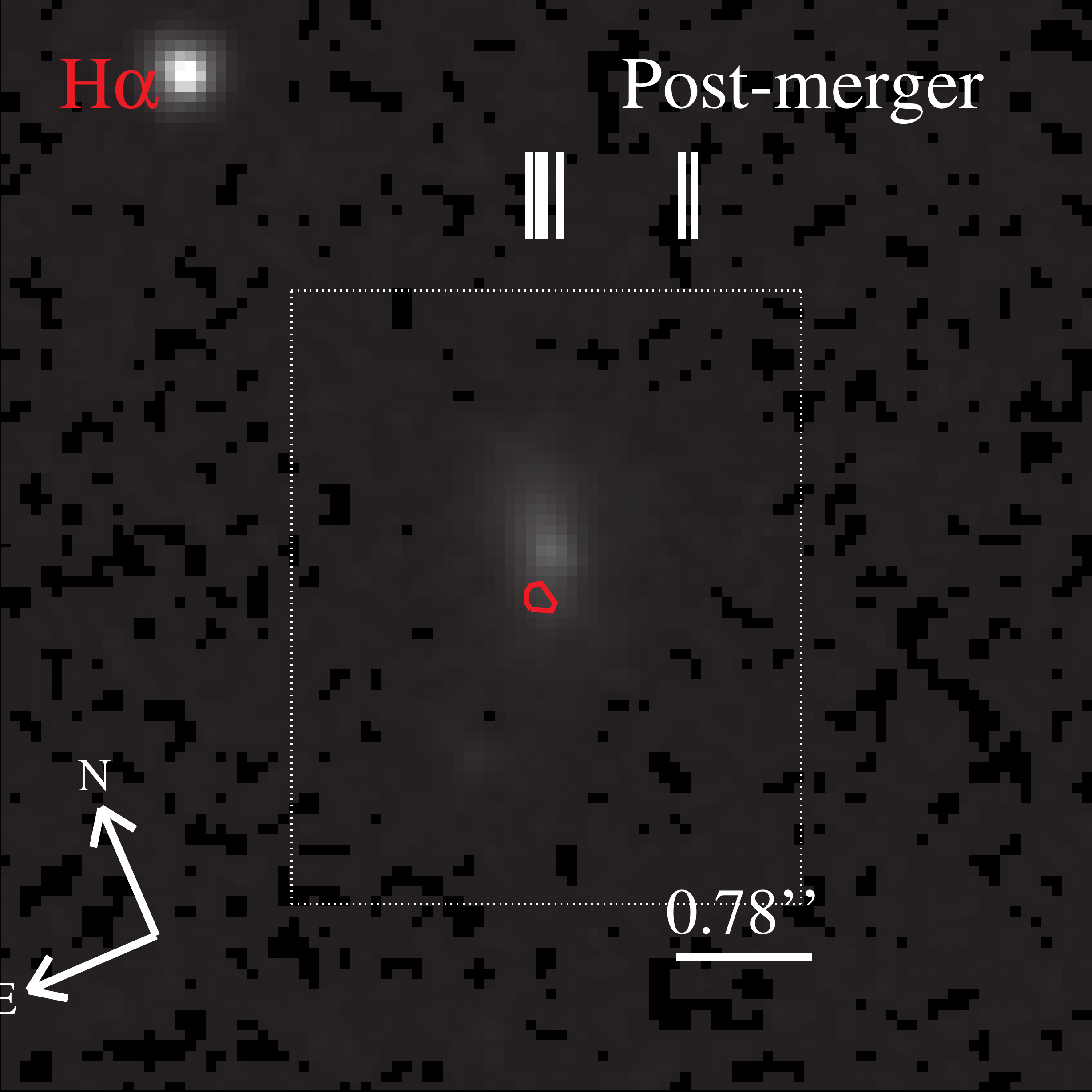}   
\includegraphics[width=0.13\textwidth]{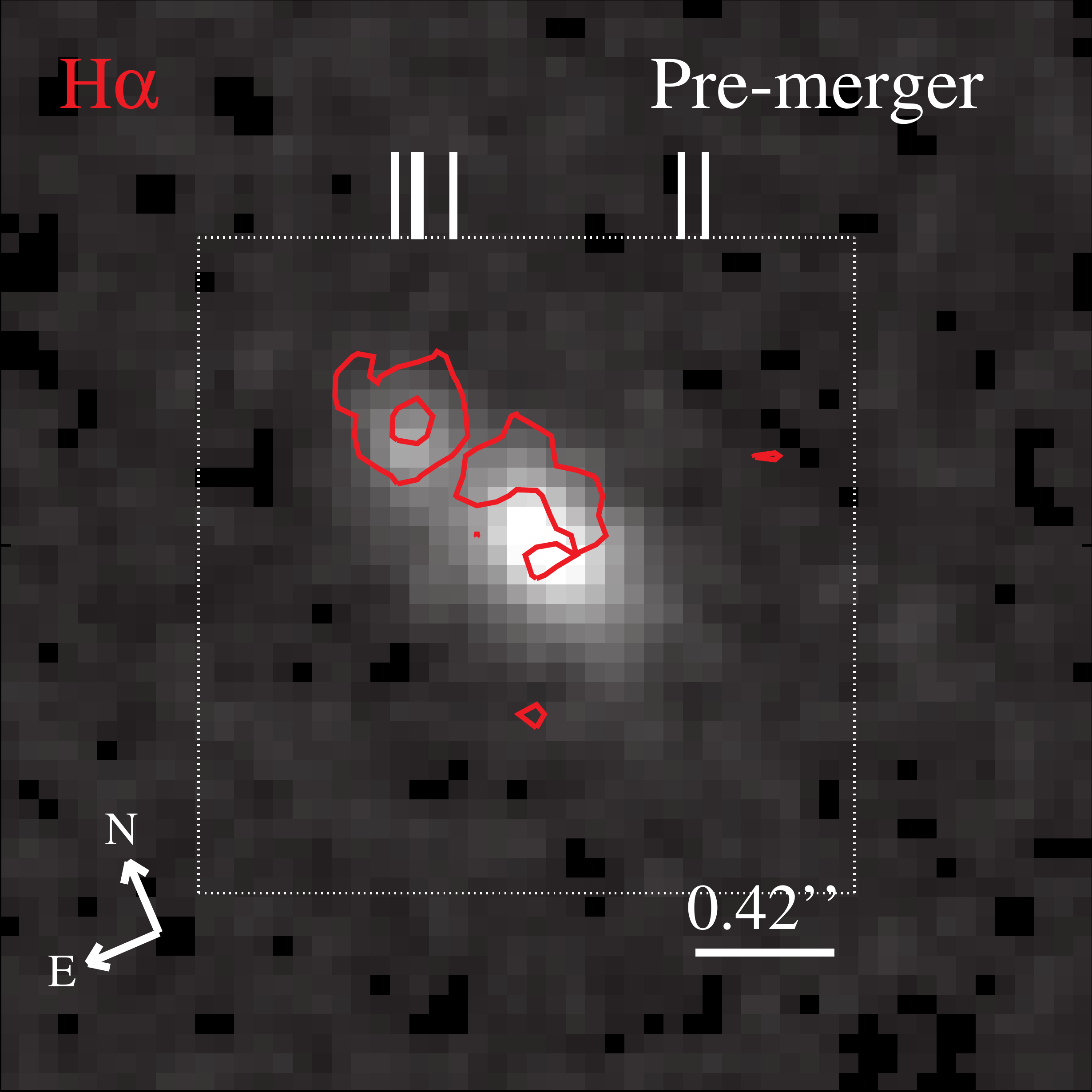}   
\includegraphics[width=0.13\textwidth]{COSMOS-1-G141_01186_lineimgoverlay_byhand.pdf}   
\includegraphics[width=0.13\textwidth]{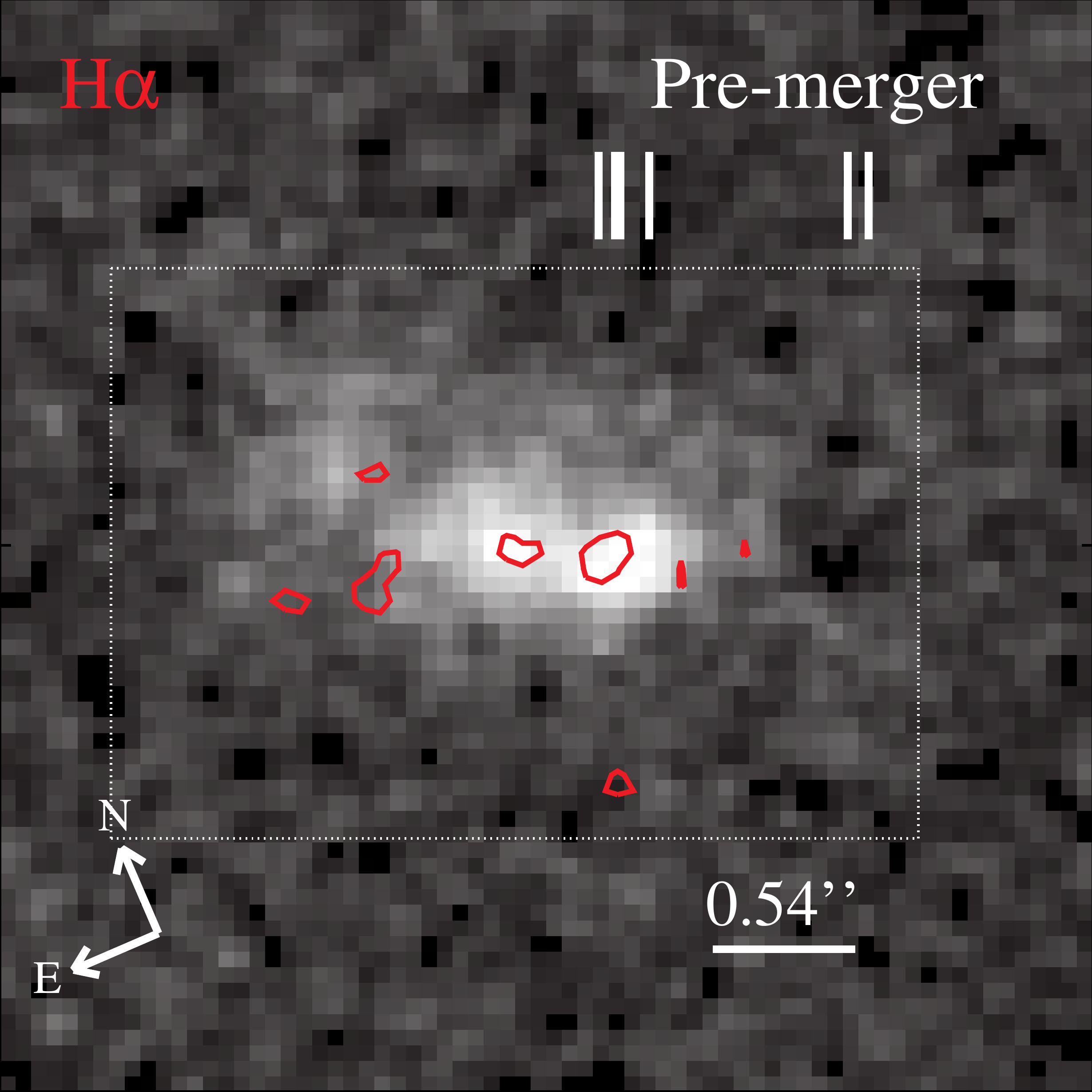} 
\includegraphics[width=0.13\textwidth]{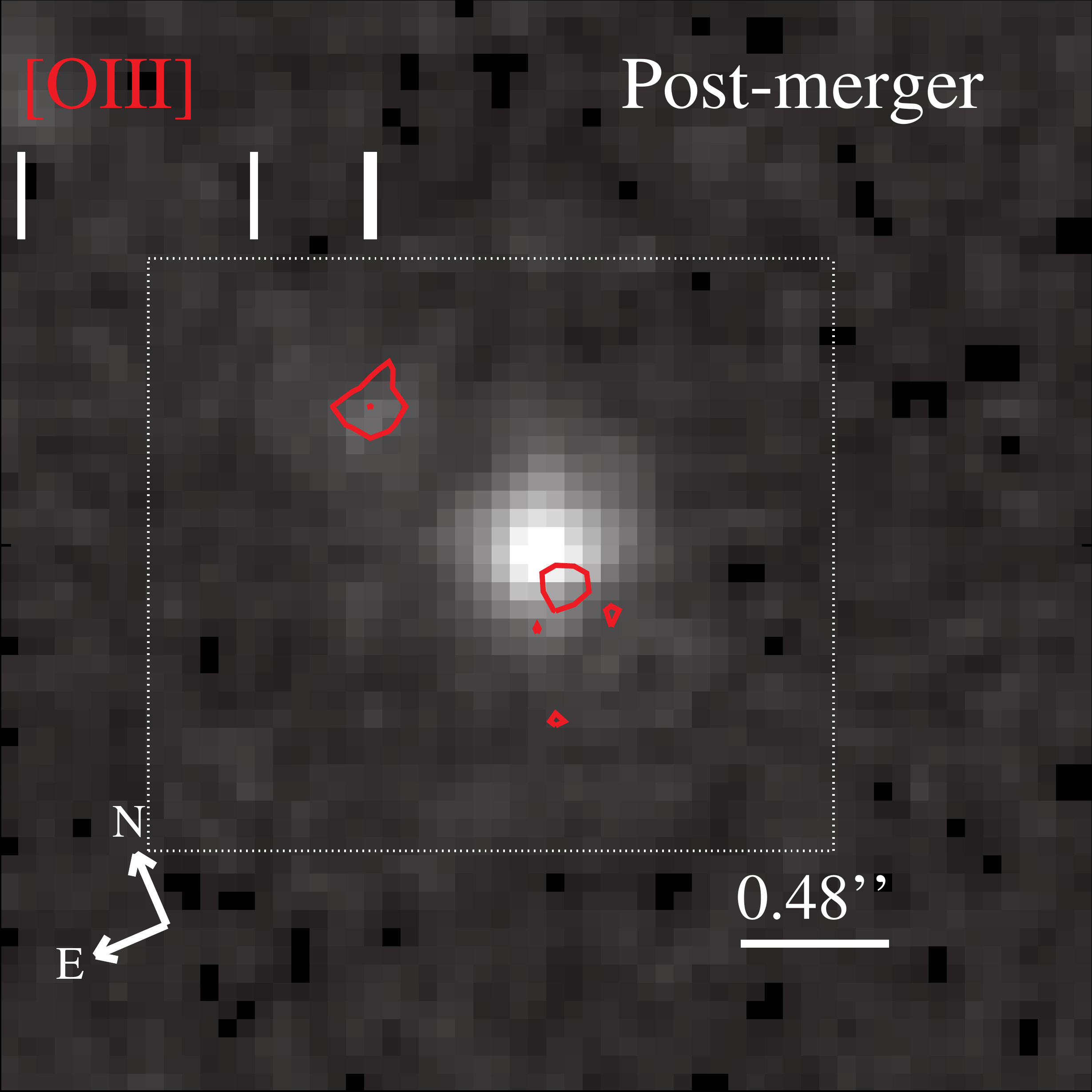} 
\includegraphics[width=0.13\textwidth]{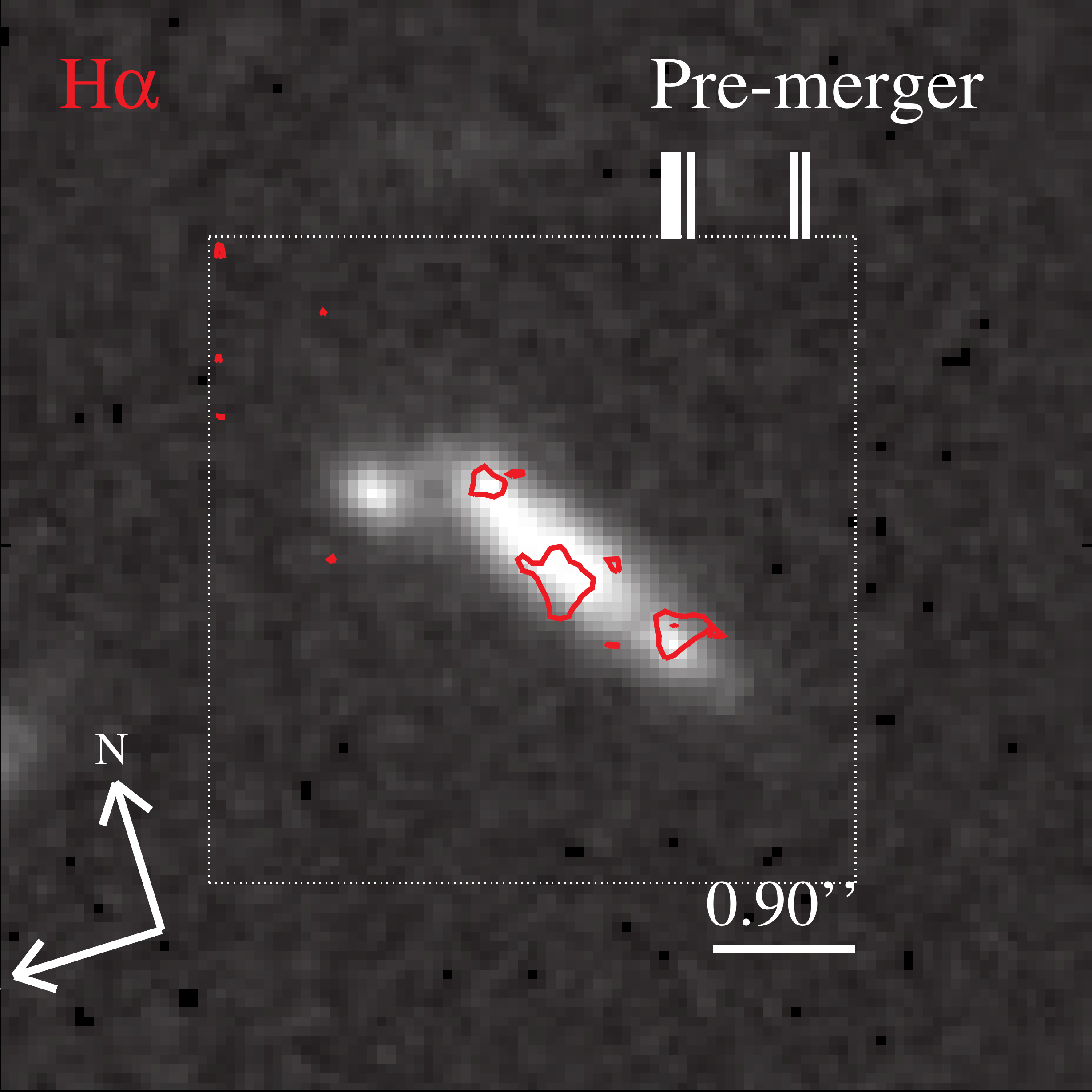}\\
\includegraphics[width=0.13\textwidth]{COSMOS-17-G141_00408_lineimgoverlay_byhand.pdf}  
\includegraphics[width=0.13\textwidth]{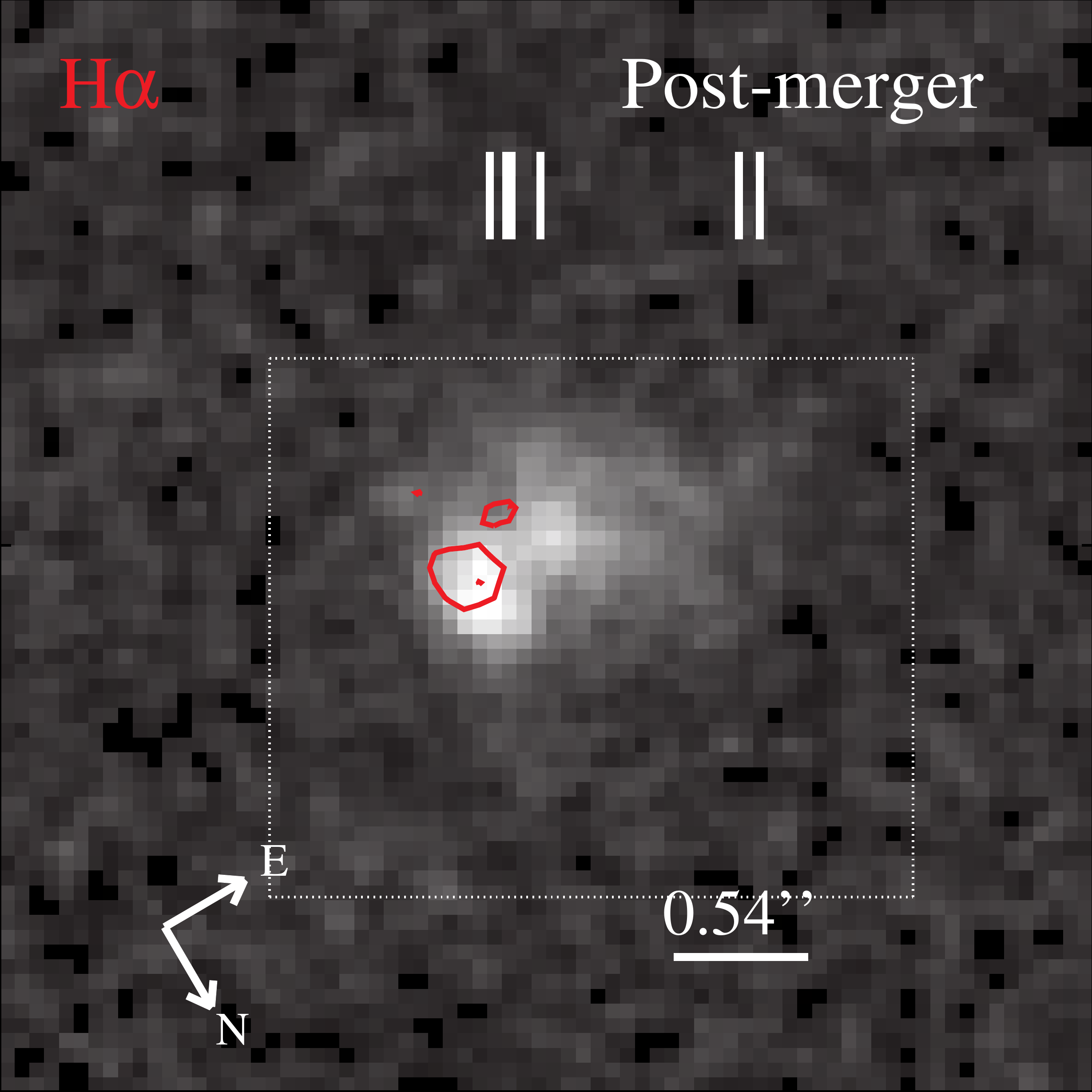}  
\includegraphics[width=0.13\textwidth]{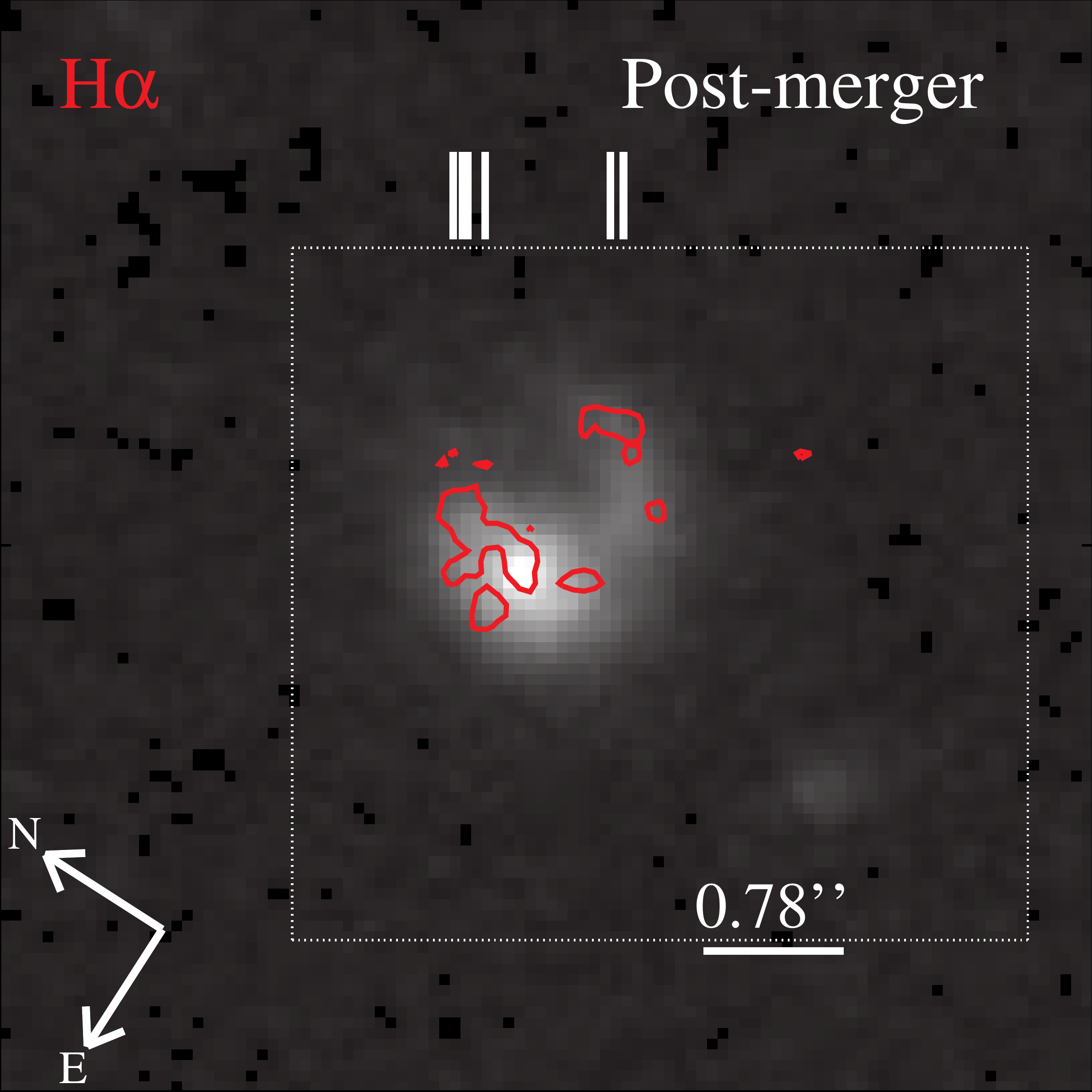}  
\includegraphics[width=0.13\textwidth]{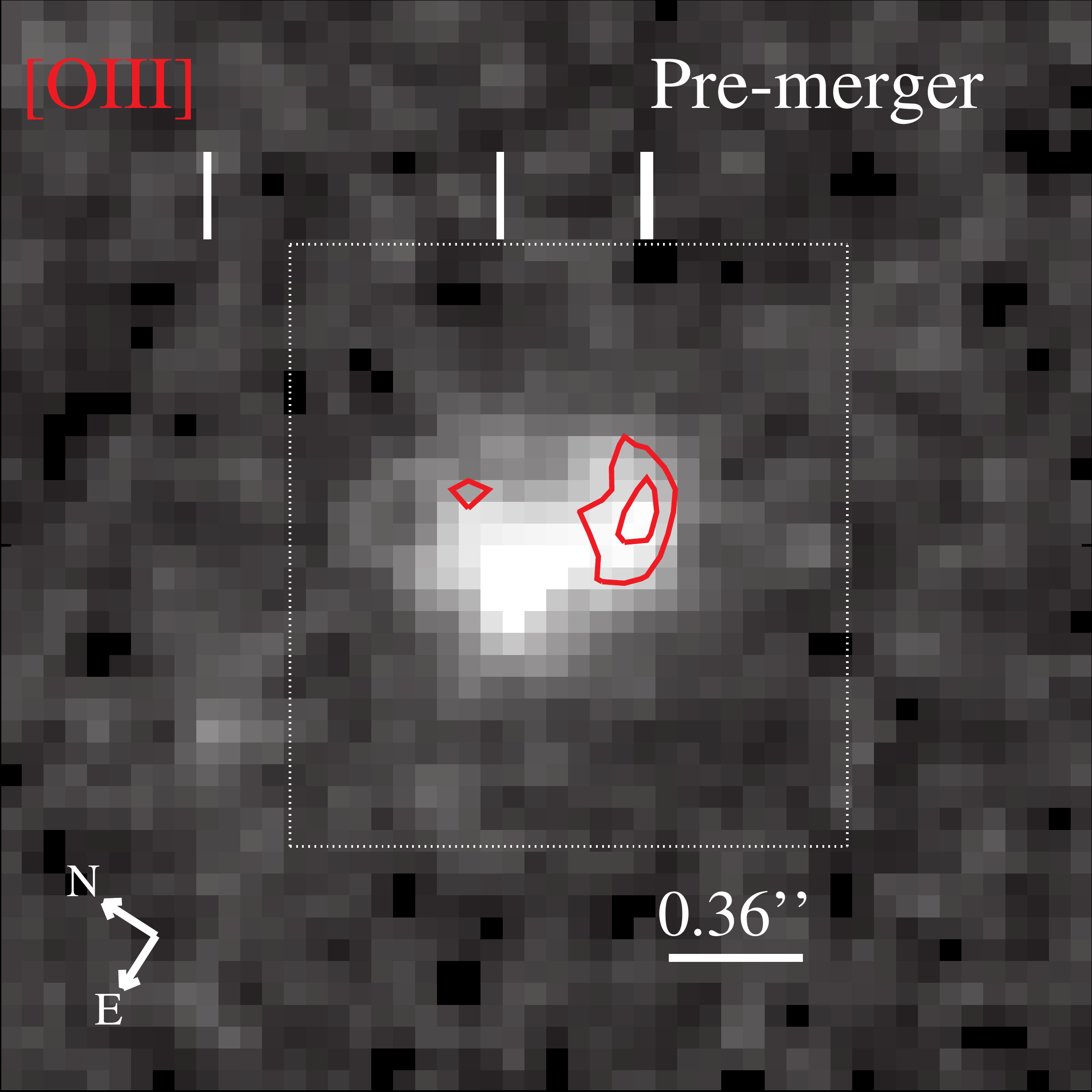}  
\includegraphics[width=0.13\textwidth]{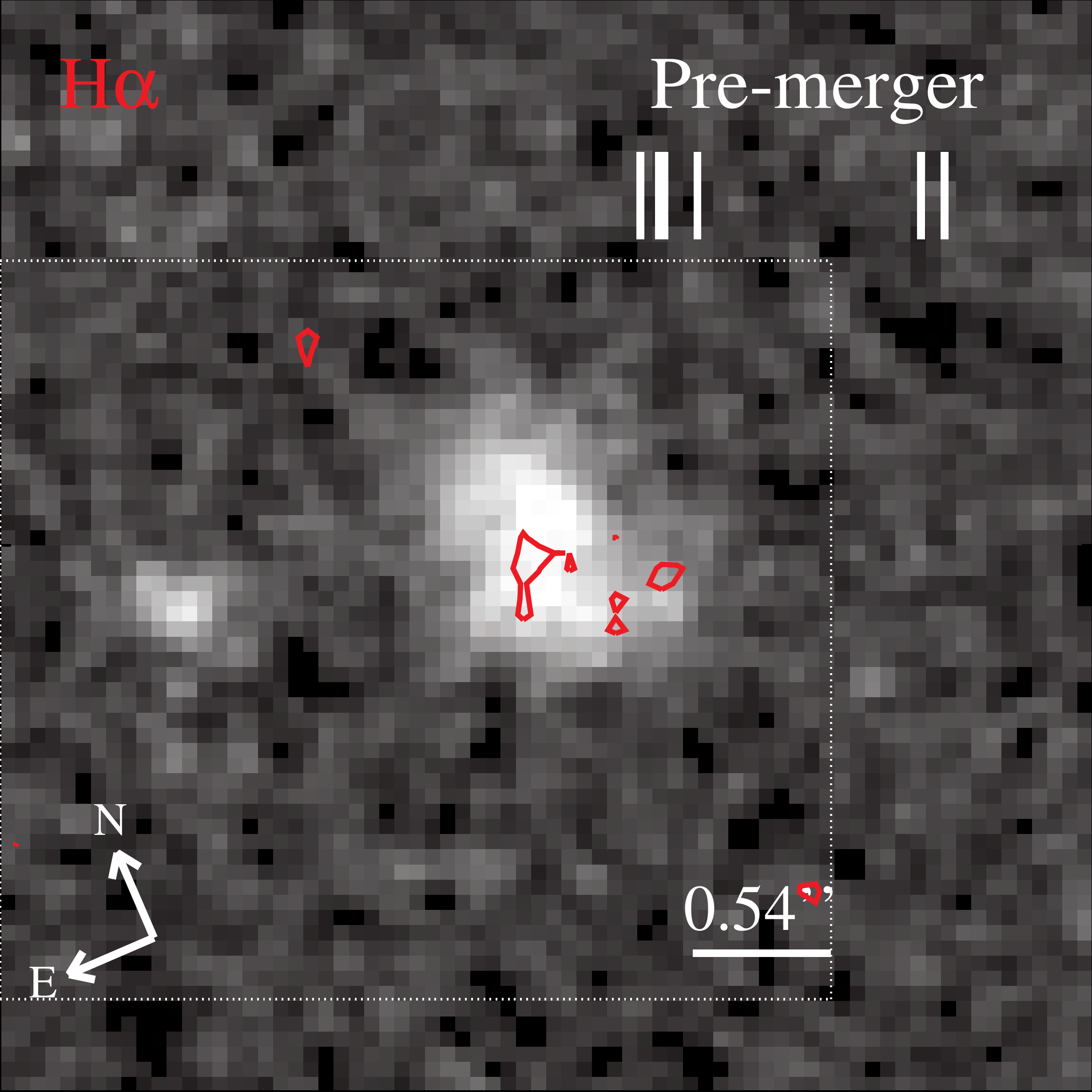}    
\includegraphics[width=0.13\textwidth]{COSMOS-23-G141_00308_lineimgoverlay_byhand.pdf}  
\includegraphics[width=0.13\textwidth]{GOODS-S-23-G141_00734_lineimgoverlay_byhand.pdf}\\
\includegraphics[width=0.13\textwidth]{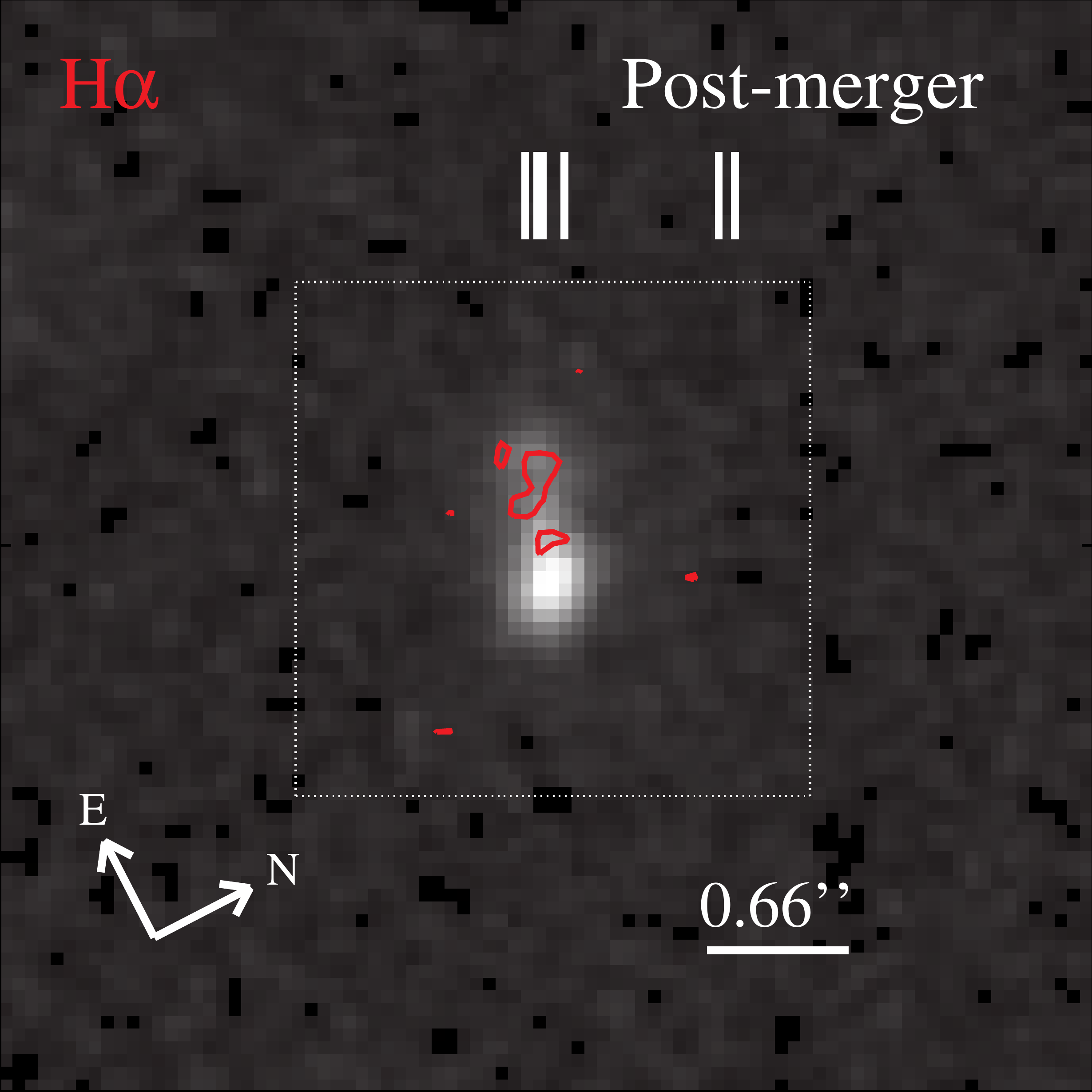}  
\includegraphics[width=0.13\textwidth]{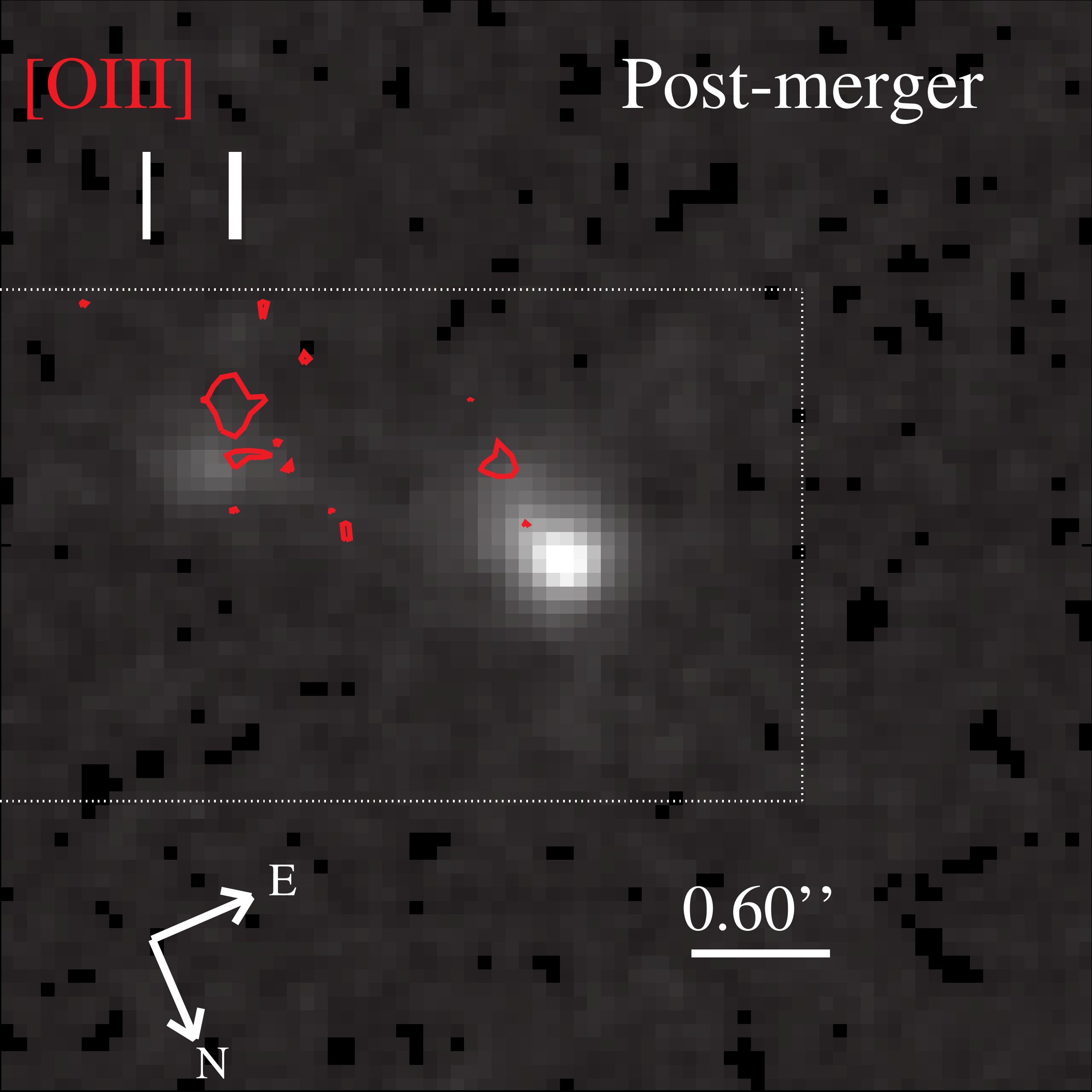}   
\includegraphics[width=0.13\textwidth]{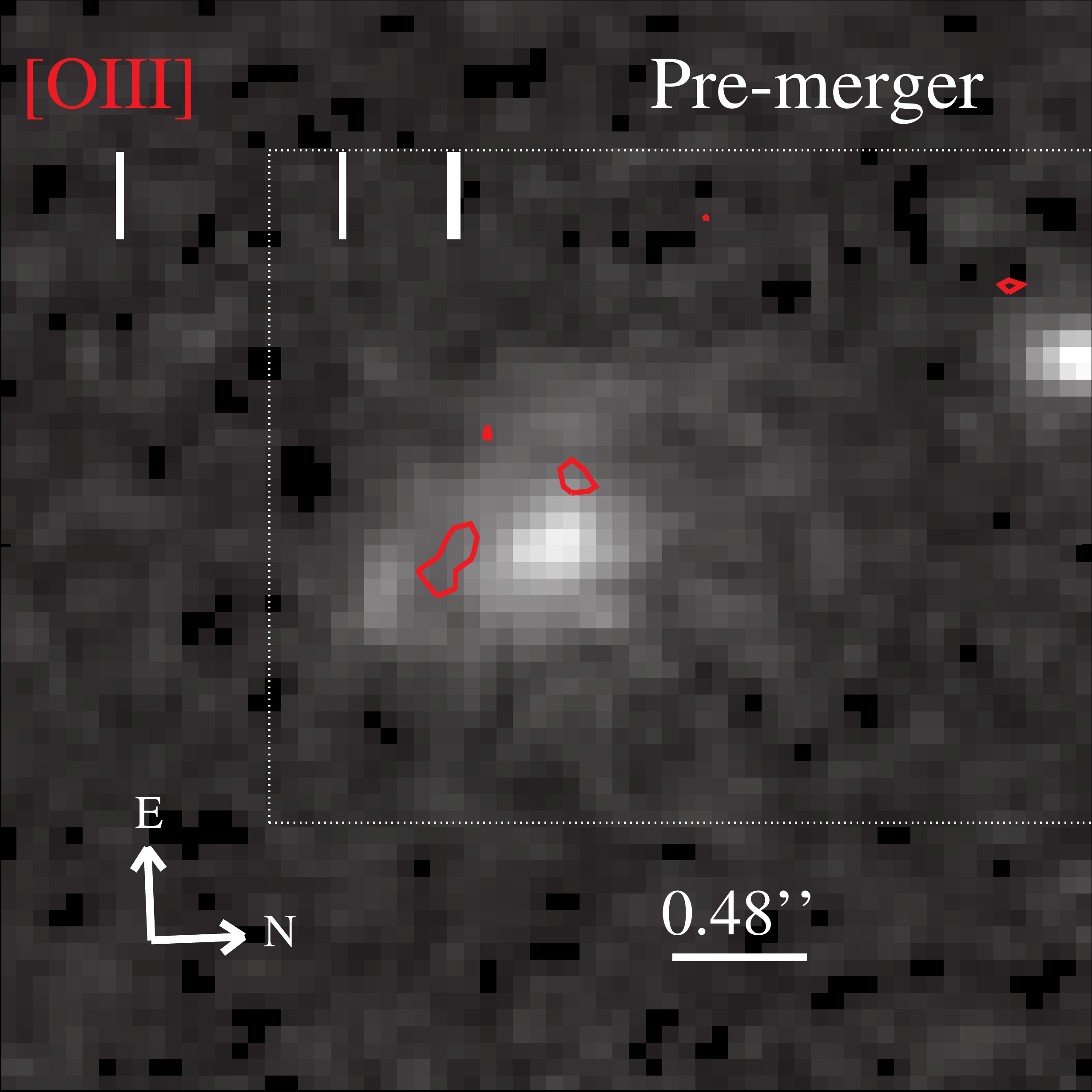}  
\includegraphics[width=0.13\textwidth]{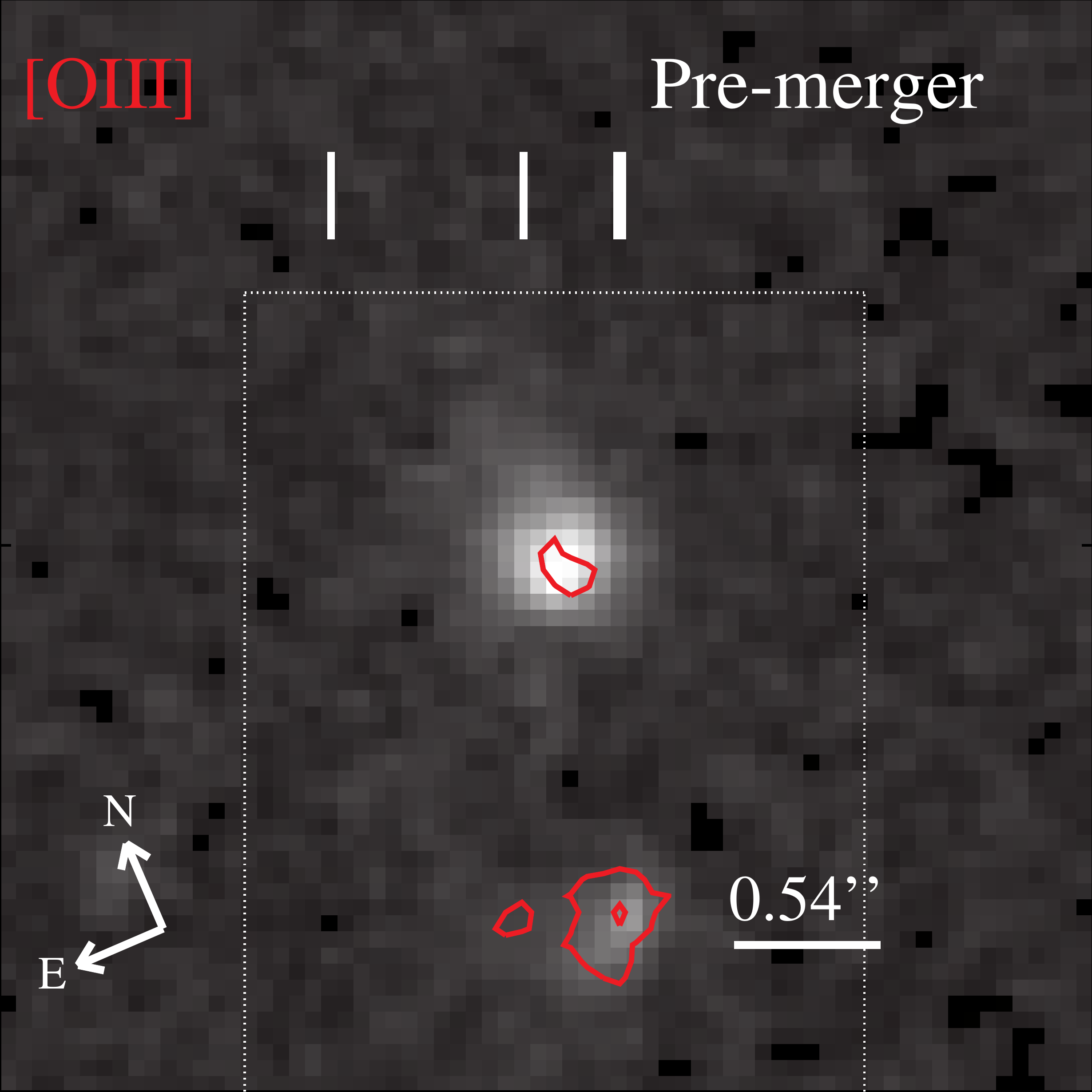}    
\includegraphics[width=0.13\textwidth]{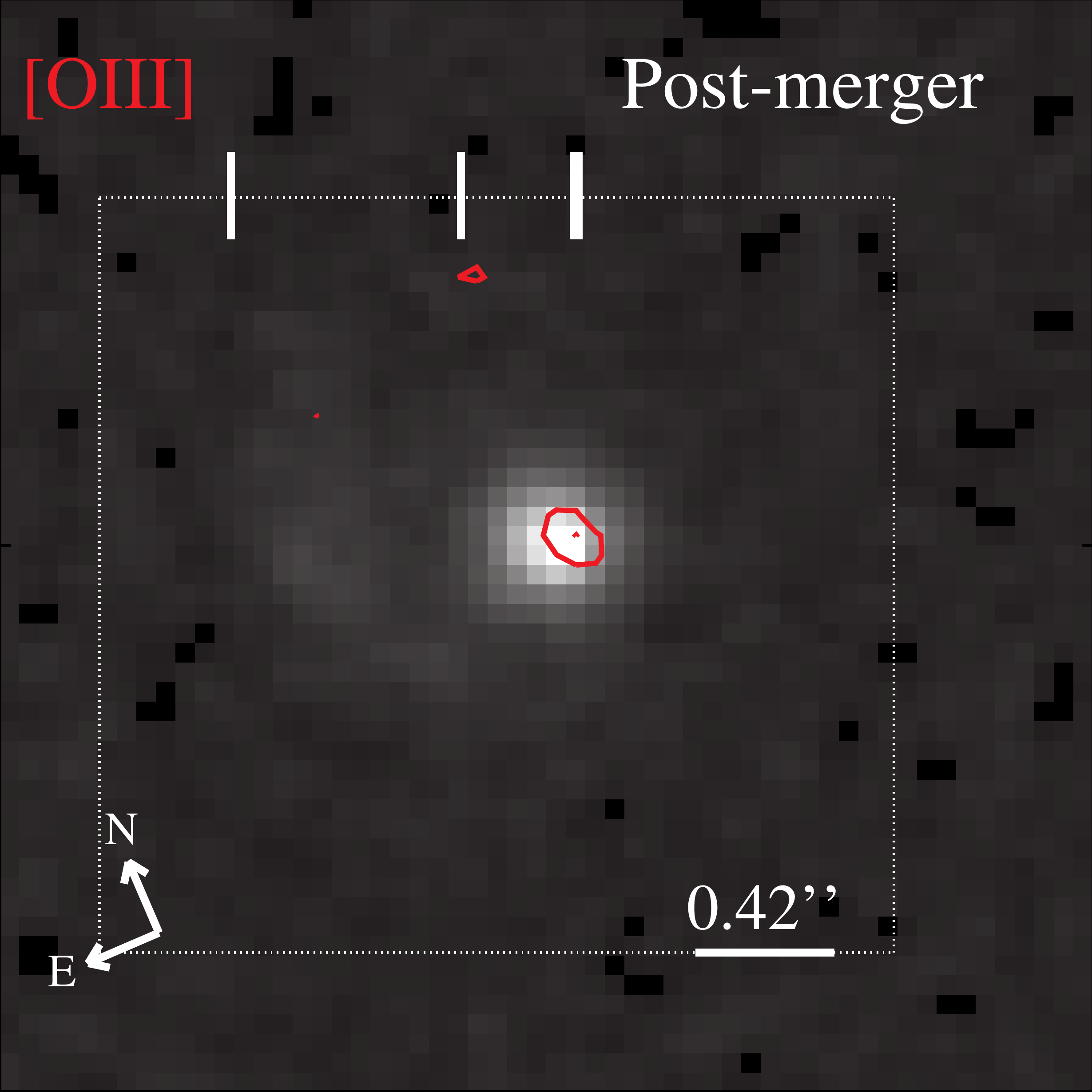}\\  
\vspace{0.5cm}
\includegraphics[width=0.13\textwidth]{COSMOS-3-G141_01186_lineimgoverlay_byhand.pdf}    
\includegraphics[width=0.13\textwidth]{COSMOS-1-G141_01290_lineimgoverlay_byhand.pdf}    
\hspace{0.5cm}
\includegraphics[width=0.13\textwidth]{COSMOS-23-G141_00743_lineimgoverlay_byhand.pdf}  
\includegraphics[width=0.13\textwidth]{GOODS-S-24-G141_00325_lineimgoverlay_byhand.pdf} 
\includegraphics[width=0.13\textwidth]{COSMOS-26-G141_00350_lineimgoverlay_byhand.pdf}  
\includegraphics[width=0.13\textwidth]{COSMOS-9-G141_00895_lineimgoverlay_byhand.pdf}    
\caption{Figure similar to Figure~\ref{fig:ELmap} for the full sample of emission line (star formation) maps obtained from the 60 morphologically selected 3D-HST mergers. The maps are sorted according to SF type (Section~\ref{sec:resultsEL}) with the 35 systems of SF type 1 at the top, the 19 SF type 2 systems in the centre, and the SF type 3 and 4 mergers, also shown in Figure~\ref{fig:ELmap}, at the bottom.}
\label{fig:ELmapALL}}
\end{figure*}

In Figure~\ref{fig:ELmapALL} we show the full sample of the 60 3D-HST emission line (star formation) maps. The objects have been sorted according to the estimated morphology of the star formation (SF type) described in Section~\ref{sec:resultsEL}.

\end{appendix}

\bibliographystyle{mn2eFIX}
\bibliography{ms.bib}

\label{lastpage} 
\end{document}